\documentclass[english,aps,pre,twocolumn]{revtex4}
\usepackage[T1]{fontenc}
\usepackage[latin9]{inputenc}
\usepackage{graphicx}
\usepackage{amssymb}

\makeatletter

\providecommand{\tabularnewline}{\\}

\@ifundefined{textcolor}{}
{%
 \definecolor{BLACK}{gray}{0}
 \definecolor{WHITE}{gray}{1}
 \definecolor{RED}{rgb}{1,0,0}
 \definecolor{GREEN}{rgb}{0,1,0}
 \definecolor{BLUE}{rgb}{0,0,1}
 \definecolor{CYAN}{cmyk}{1,0,0,0}
 \definecolor{MAGENTA}{cmyk}{0,1,0,0}
 \definecolor{YELLOW}{cmyk}{0,0,1,0}
 }

\makeatother

\usepackage{babel}

\begin{document}

\title{Simulations of Ground State Fluctuations in Mean-Field Ising Spin
Glasses}

\author{Stefan Boettcher}
\email{www.physics.emory.edu/faculty/boettcher}
\affiliation{Physics Department, Emory University, Atlanta, Georgia 30322; USA}
\begin{abstract}
The scaling of fluctuations in the distribution of ground-state energies
or costs with the system size $N$ for Ising spin glasses is considered
using an extensive set of simulations with the Extremal Optimization
heuristic across a range of different models on sparse and dense graphs.
These models exhibit very diverse behaviors, and an asymptotic extrapolation
is often complicated by higher-order corrections. The clearest picture,
in fact, emerges from the study of graph-bipartitioning, a combinatorial
optimization problem closely related to spin glasses. Aside from two-spin
interactions with discrete bonds, we also consider problems with Gaussian
bonds and three-spin interactions, which behave differently to a significant
degree. 
\end{abstract}

\pacs{75.10.Nr , 02.60.Pn , 05.50.+q }

\maketitle

\section{Introduction\label{sec:Introduction}}

Ising spin glasses are the paradigmatic model for disorder not only in
materials~\citep{Edwards75}, but provide the archetype for complex
behavior in many contexts, such as hard combinatorial
problems~\citep{MPV,Mezard06}, information theory~\citep{Nishimori01}, and
learning~\citep{Hopfield82,Schneidman06}. An Ising spin glass is generally
described by the Hamiltonian
\begin{eqnarray}
H & = &
-\sum_{<i,j>} J_{i,j}\sigma_{i}\sigma_{j}
\label{eq:SK-Ham}
\end{eqnarray}
with $N$ Ising spins $\sigma_{i}\in\left\{ \pm1\right\} $ and quenched
random bonds $J_{i,j}$. The bonds are chosen from a distribution, here,
of zero mean and width $\sqrt{\langle J^2\rangle}= J_0$ that is either
bimodal ($J_{i,j}\in\{\pm J_0\}$) or Gaussian. The sum parses over all
extant bonds $<i,j>$ between any pair of spins $\sigma_i$ and
$\sigma_j$. The Sherrington-Kirkpatrick model
(SK)~\citep{Sherrington75} is the mean-field limit of Ising spin
glasses, in which each spin is coupled to every other spin in the
system. To keep the energies $E$ of an equilibrium spin configuration
in SK extensive, we set $J_0=1/\sqrt{N}$; the other models discussed
here are defined on sparse graphs and require $J_0=1$.

It is surprising to find still new features of such a well-studied
model after some 30 years study. Therefore, the unusual behavior of
the distribution of ground-state energies $E_{0}$ over the bond
disorder in SK has raised significant interest in recent years. This
interest is further elevated by its close connection with the
statistics of extremely rare events, that in many jammed, disordered
systems can become the controlling feature of the dynamics. It was
found~\citep{Palassini03,Bouchaud03,Andreanov04,EOSK,Palassini08} that
the fluctuations in $E_{0}$ behave in a highly non-normal fashion and
rather resemble distributions found in extremal-value
statistics~\citep{Bouchaud97}.  This connection becomes rigorous for
the fluctuations of the random-energy model (REM)~\citep{derrida:80},
for which a Gumbel distribution can be derived exactly. The shape in
SK shares similarities with a higher-order Gumbel
distribution~\citep{Palassini03,EOSK,koerner:06} but its precise
functional form remains unknown. There exists a high degree of
universality in the extreme-value statistics of intrinsically
uncorrelated states, but in long-range connected systems with quenched
disorder such an assumption may fail. A general discussion of the ordinary
versus extreme fluctuations in the SK and other disordered models is
provided in Ref.~\citep{Monthus10}.

Of special interest for the energy fluctuations is the variation in
width of the distribution with the system size, as it provides
important clues to the structural properties of the ground states (or
low temperature states
generally)~\citep{Bouchaud03,Aspelmeier08,aspelmeier08b}. A number of
arguments regarding the system-size scaling of the standard deviation
for the ground-state energy densities $e_{0}=E_{0}/N$,
\begin{eqnarray}
\sigma\left(e_{0}\right) & = & \sqrt{\left\langle
  e_{0}^{2}\right\rangle -\left\langle e_{0}\right\rangle ^{2}}
\label{eq:rho}\\
 & \sim & A\, N^{-\rho}+B\,N^{-a}+\ldots\quad\left(a>\rho\right),\nonumber
\end{eqnarray}
for $N\to\infty$ have been put forward, leading to values of
either~\citep{Bouchaud03,Aspelmeier03} $\rho=\frac{3}{4}$
or~\citep{crisanti:92,Aspelmeier07,Parisi08,Parisi09,Rizzo09b,Rizzo09}
$\rho=\frac{5}{6}$. Both conjectures predict decay that is faster than
for normal fluctuations $\rho=\frac{1}{2}$, which would be obtained
from the central limit theorem under the assumption of negligible
correlations between individual terms of the spin glass Hamiltonian in
Eq.~(\ref{eq:SK-Ham}). While a bound of $\rho\geq\frac{3}{4}$ has been
shown~\citep{Aspelmeier08,aspelmeier08b} for the SK, Gaussian behavior
is indeed found for spin glasses on a finite-dimensional
lattice~\citep{Wehr90}, and has also been observed on sparse random
graphs~\citep{Bouchaud03}. Exact values different from any of these
are known for the replica symmetric spherical spin glass
$\left(\rho=\frac{2}{3}\right)$ and the $m$-vector spin glass for
$m\to\infty$ $\left(\rho=\frac{4}{5}\right)$~\citep{Aspelmeier10}.

Numerically, both conjectures for $\rho$ on the SK have proven
difficult to distinguish with any certainty, and while most initial
predictions~\citep{Palassini03,Bouchaud03,Andreanov04,EOSK} seems to
favor a value close to $\rho=\frac{3}{4}$, more recently a trend
towards $\rho=\frac{5}{6}$ was found at larger system
sizes~\citep{Palassini08}. That would support the current consensus in
the theoretical work~\citep{Aspelmeier07,Parisi08,Parisi09}. Such a
larger value is also desirable for consistency between relations
connecting $\rho$ to the exponent describing domain wall
excitations~\citep{Aspelmeier03} as found in high
dimensions~\citep{Boettcher05d}.

Here, we report on extensive simulations to clarify this important
question regarding the low-temperature properties of spin glasses.
The results are at best marginally consistent with any of the
theoretical predictions. Since the data analysis proves to be
complicated by transient behavior, we have widened the scope of our
investigation to incorporate a large number of related models for
comparison. For instance, we provide corresponding data for spin
glasses on random regular graphs ({}``Bethe lattices'') of sparse
degree, both for two- and three-spin interactions and discrete ($\pm
J$) as well as Gaussian bonds. Two-spin coupled spin glasses on Bethe
lattices of degree $r$ provide a convenient one-parameter family of
models with smoothly extrapolates to SK for increasing degree,
$r\to\infty$~\citep{Boettcher03a,Boettcher03b}. The results are quite
similar to those we find in SK, thus, providing a likely trend for the
SK behavior itself, but they are equally beset with strong
transients. The three-spin data with discrete bonds draws a different
picture, consistent with a recent study~\citep{Nakajima09}, and also
highlights the question of universality of the results when continuous
bonds are used. As we have noted before~\citep{Boettcher10a}, on
sparse graphs finite-size corrections may already depend on details of
the bond distribution; this dependence appears to extend also to
ground-state deviations.

Alternatively, we study the graph bipartitioning problem (GBP) on
these Bethe lattices. We find much diminished transients and a
consistent extrapolation for the value of $\rho$. But that value is
between -- and likely distinct -- from $\frac{3}{4}$ and
$\frac{5}{6}$.  Although it was recently predicted~\citep{Zdeborova10}
that GBP in the thermodynamic limit is equivalent to the corresponding
spin glass  at $T=0$ on those Bethe lattices, it is of course less
clear whether such relation would hold at finite size. For instance,
finite-size corrections to the average ground state energies already
differ significantly between GBP and spin glasses.

This paper is structured as follows: We start with a few remarks about
the optimization heuristic used in all of our simulations in
Sec.~\ref{sec:Optimization-Methods}. Since the clearest case is
provided by GBP, we present our data for this problem first in
Sec.~\ref{sec:Graph-Bi-Partitioning}.  In
Sec.~\ref{sec:Sherrington-Kirkpatrick-Model}, we discuss the SK data
at length. In Sec.~\ref{sec:Spin-Glasses-on-Bethe}, we present the
data for the corresponding spin glass problem on Bethe lattices,
followed by a similar study on ordinary random graphs in
Sec.~\ref{sec:Spin-Glasses-RG}.  In
Sec.~\ref{sec:Three-Spin-Interactions-on}, we supplement our
investigation with a study of three-spin interactions on a Bethe
lattice. We summarize with a few conclusions in
Sec.~\ref{sec:Conclusions}.

\section{Means and Methods\label{sec:Methods}}

In this section, we introduce the simulation methods and techniques by
which the resulting data was analyzed.

\subsection{Optimization Methods\label{sec:Optimization-Methods}}

We have employed the Extremal Optimization heuristic
(EO)~\citep{Boettcher00,Boettcher01a,Dagstuhl04} in the implementation
described previously for the SK spin glass~\citep{EOSK} and those on
Bethe lattices~\citep{Boettcher03b}, with various improvements to
attain an order of magnitude in speed-up %
\footnote{A sample code of this implementation of EO for SK can be
  found at
  http://www.physics.emory.edu/\break
faculty/boettcher/Research/EO\_demo/demoSK.c.%
}. Resorting to a simple bimodal bond distribution whenever possible
allows further an efficient use of integer arithmetic. Not only do
discrete bonds provide computational advantages over continuous ones;
Gaussian bonds pose significant entropic barriers in addition to the
usual complexities faced by local search in a multi-modal energy
landscape~\citep{Bauke04}. To find approximations to GBP ground
states, we use EO as described in Ref.~\citep{Percus08,Zdeborova10}.

Unlike, for instance, the parallel tempering Monte Carlo technique
used at larger system sizes for SK in Ref.~\citep{Palassini08}, which
operates at small but finite temperatures, EO generally performs its
local search for minima in the landscape formed by the internal energy
itself, with activated spin flips as elementary
moves~\citep{Boettcher01a}.  The only free parameter controlling EO
was set to $\tau=1.25$ for discrete problems and somewhat higher at
$\tau=1.5$ for problems with Gaussian weights, and about $0.1\,N^{3}$
spin flips were executed in each run, restarting form a random initial
configuration. At least 2 such restarts were performed for each
instance, and the number of runs is doubled on-the-fly whenever a new
optimum is found in the latter half of all runs. For some rare
instances, more than 10 runs were required, always ensuring that the
latter half of all runs merely confirms the previously found optimum
but does not exceed it. In each section, we have listed the number of
instances treated at each $N$. Besides trying to reach large sizes, we
have often emphasized simulating a very large number of instances on a
relatively dense set of intermediate $N$. We have conducted extensive
tests for each model to ensure the accuracy of the results. Testing on
testbeds of exactly-solved instance (using branch-and-bound) typically
results in perfect agreement, but system sizes are small
($N\approx50-70$). On larger systems, we have done sample runs with
ten-times more updates and found only few inaccuracies, which lead to
systematic errors far below statistical errors (unless otherwise
noted).

Although all of the problems in this project could be classified as
spin glasses, they stand in for a large class of combinatorial
problems. It should be noted that the EO heuristic provides data of
great detail with only small changes in the implementation. Only the
input for the various graph types needed to be changed. (For GBP, we
have to impose the additional constraint of vanishing magnetization.)
While we treat only mean-field problems here, pertinent results have
been obtained previously for structured instances, such as
finite-dimensional lattices~\citep{Boettcher04b,Boettcher04c}.

\subsection{Data Analysis\label{data}}

To obtain insights into the finite-size scaling of our observables, we
study fits of the data generated in our simulations to a number of
asymptotic forms. These forms seem to be the most nature candidates to
describe corrections to the leading asymptotic behavior. But there can
never be an exhaustive list of all possibilities without any
theoretical knowledge a-priori. Therefore, we list as much as possible
the data points obtained in the simulations in tables that would allow
the readers to pursue their own hypotheses.

Any asymptotic ($N\to\infty$) fit bares considerable risks: Not only
may a presumed asymptotic form be insufficient and may, for instance,
miss logarithmic corrections, etc. But even if correct, it would
certainly fit data for larger system sizes $N$ better than any
transients, raising the question of how many data points for lower $N$
to include. Any of these uncertainties can introduce potentially
sizable systematic errors, even if the heuristic had been perfectly
accurate.  Such errors can only be eliminated fully with a theory to
compare to, which at present does not exist. The obtained
qualities-of-fit $Q$ have to be seen in this context, and a poor
$Q$-value should not be taken immediately as disproving a hypothesis,
unless it is seen in relation to alternative fits and a discussion of
how much transient data has been included. In fact, to obtain any
reasonable $Q$-values, in each fit we have -- somewhat arbitrarily --
increased the error bars \emph{uniformly} by a factor of four from the
purely statistical errors given as a bracketed uncertainty in the last
digit of the simulation data listed in the tables. Such a uniform
allowance for systematic errors, either from heuristic inaccuracies or
functional uncertainty in the fit, is definitely inadequate and would
deserve better consideration in the future.

Thus, we fit the asymptotic extrapolation of the finite-size
data towards the thermodynamic limit ($N\to\infty$) for the presumed
ground-state energy $\left\langle e_{r}\right\rangle_{N}$ or cost
densities $\left\langle c_{r}\right\rangle _{N}$ to the following
forms, abbreviating $x\in\{c,e\}$:
\begin{eqnarray}
\left\langle x_{r}\right\rangle _{N} & \sim & \left\langle x_{r}\right\rangle _{\infty}+
A\, N^{-\omega}+\dots
\label{eq:Escaling}
\end{eqnarray}
for just the first-order correction, or
\begin{eqnarray}
\left\langle x_{r}\right\rangle _{N} & \sim & \left\langle x_{r}\right\rangle _{\infty}+
A\, N^{-\omega}+B\, N^{-\omega_{1}}+\dots
\label{eq:Cscaling}
\end{eqnarray}
as a second order correction, where we have to fix the first-order
exponent $\omega$ and require $\omega_1>\omega$ to achieve a stable
fit. An alternative second-order, logarithmic fit,
\begin{eqnarray}
 \left\langle x_{r}\right\rangle _{N}\sim\left\langle
x_{r}\right\rangle _{\infty}+A\, N^{-\omega}+B\,\frac{\ln N}{N}+\ldots,
\label{eq:logscaling}
\end{eqnarray}
with arbitrary $\omega<1$ also seems conceivable.

\begin{table*}[!ht]
\caption{\label{tab:costGBP} 
Average cost per spin $\left\langle c_{r}\right\rangle _{N}$ for
approximate ground states of GBP on Bethe lattices of degree
$r=3,\dots,10$. The given errors, in parentheses after each average,
denote the uncertainty in the last given digit of that average. This
uncertainty is solely based on the statistical error.  Fits of this
data and their plots were discussed previously in
Ref.~\citep{Zdeborova10} (see Fig.~2 there).}
\begin{centering}
\begin{tabular}{r|l|l|l|l|l|l|l|l}
\hline 
$N$ & $\left\langle c_{3}\right\rangle _{N}$  & $\left\langle c_{4}\right\rangle _{N}$  & $\left\langle c_{5}\right\rangle _{N}$  & $\left\langle c_{6}\right\rangle _{N}$  & $\left\langle c_{7}\right\rangle _{N}$  & $\left\langle c_{8}\right\rangle _{N}$  & $\left\langle c_{9}\right\rangle _{N}$  & $\left\langle c_{10}\right\rangle _{N}$ \tabularnewline
\hline 
32  & 0.19495(3)  & 0.35652(4)  & 0.53893(5)  & 0.71445(5)  & 0.91552(6)  & 1.09781(6)  & 1.31056(6)  & 1.49700(7) \tabularnewline
40  & 0.18161(3)  & 0.34104(3)  & 0.51730(4)  & 0.69348(4)  & 0.88800(5)  & 1.07190(5)  & 1.27774(5)  & 1.46657(6) \tabularnewline
50  & 0.16627(2)  & 0.32820(3)  & 0.49487(3)  & 0.67581(4)  & 0.85919(4)  & 1.05003(4)  & 1.24320(5)  & 1.44071(5) \tabularnewline
64  & 0.15822(2)  & 0.31632(2)  & 0.48312(3)  & 0.65972(3)  & 0.84438(3)  & 1.02997(4)  & 1.22578(4)  & 1.41715(4) \tabularnewline
80  & 0.15042(1)  & 0.30750(2)  & 0.47119(2)  & 0.64753(3)  & 0.82899(3)  & 1.01495(3)  & 1.20743(3)  & 1.39936(3) \tabularnewline
100  & 0.14406(2)  & 0.30016(2)  & 0.46138(3)  & 0.63749(3)  & 0.81632(3)  & 1.00232(4)  & 1.19222(4)  & 1.38449(4) \tabularnewline
128  & 0.13825(2)  & 0.29350(1)  & 0.45240(2)  & 0.62822(2)  & 0.80481(3)  & 0.99082(3)  & 1.17837(3)  & 1.37085(3) \tabularnewline
160  & 0.13411(1)  & 0.28858(2)  & 0.44582(2)  & 0.62133(2)  & 0.79628(2)  & 0.98220(2)  & 1.16812(3)  & 1.36075(3) \tabularnewline
200  & 0.13064(1)  & 0.28450(1)  & 0.44037(2)  & 0.61553(2)  & 0.78920(2)  & 0.97498(2)  & 1.15965(2)  & 1.35227(3) \tabularnewline
256  & 0.12753(1)  & 0.28087(1)  & 0.43545(1)  & 0.61037(2)  & 0.78281(2)  & 0.96852(2)  & 1.15195(2)  & 1.34449(2) \tabularnewline
320  & 0.12526(1)  & 0.27818(1)  & 0.43185(2)  & 0.60652(2)  & 0.77813(2)  & 0.96370(2)  & 1.14630(2)  & 1.33874(3) \tabularnewline
512  & 0.12170(1)  & 0.27395(1)  & 0.42607(1)  & 0.60054(2)  & 0.77073(2)  & 0.95605(2)  & 1.13734(2)  & 1.32968(2) \tabularnewline
1024  & 0.11852(1)  & 0.26993(1)  & 0.42075(1)  & 0.59497(2)  & 0.76407(4)  & 0.94948(4)  & 1.12965(5)  & 1.32211(6) \tabularnewline
2048  & 0.11670(4)  &  &  &  &  &  &  & \tabularnewline
\hline
\end{tabular}
\par\end{centering}
\end{table*}

The asymptotic scaling for the respective deviation exponent $\rho$
proceeds either through a direct fit (to first- or second order) to
Eq.~(\ref{eq:rho}), or by the corresponding extrapolated form
\begin{eqnarray}
-\frac{\log\sigma}{\log N} & \sim & \rho-\frac{\log C}{\log
  N}+O\left(\frac{N^{-\left(a-\rho\right)}}{\log  N}\right).
\label{eq:rhoextra}
\end{eqnarray}
Either method has its advantages and disadvantages. A  direct
fit, plotted on a double-logarithmic scale, provides easier
convergence when higher-order terms are included. But the logarithmic
scale smoothes out variability or transient in a data set.  In
turn, the extrapolated form in Eq.~(\ref{eq:rhoextra}) is far less
forgiving and provides significantly more insight into fluctuations in
the data set. When plotted on an $1/\log N$-scale, data that extends
over decades is squashed into a relatively small space. But if a solid
linear regime emerges, it provides stronger evidence for the existence
and value of an exponent than a double-logarithmic plot. The slope of
this extrapolation is arbitrary, though, as we could multiply
$\sigma$ by any constant factor without affecting the scaling. We may choose to
fix this factor in a way (such as for the spin glasses below) that the
$\log C$-term in Eq.~(\ref{eq:rhoextra}) vanishes, or (such as for the
GBP below) to splay out several plots into one graph. Clearly, such a choice
does not alter any linear regime, if it exists. Without such a linear
regime, as we will see, it is very difficult to extract satisfactory
information from such a fit, even when higher-order
(and strongly non-polynomial) corrections are taken into account.

As we have argued in Ref.~\citep{Boettcher10a}, under certain
circumstances it is necessary to consider the cost of frustration (sum
of all violated bonds) in a ground state, instead of its energy
(difference between all violated and all unviolated bonds). We will
switch in our discussion between cost and energy repeatedly. There is
a simple linear relation between their average densities,
\begin{equation}
\left\langle e_{0}\right\rangle_N =2\left\langle c_{0}\right\rangle_N
-\frac{r}{p}\left\langle \left|J\right|\right\rangle_N,
\label{eq:cost_energy}
\end{equation}
where $r$ is the (average or fixed) degree and $p$ refers to the
number of spins coupled via a single bond. The average absolute
bond-weight for the Gaussian distribution of zero mean and standard
width ($J_0^2=1$ on sparse graphs, $J_0^2=1/N$ for SK) is given by
$\lim_{N\to\infty}\left\langle \left|J\right|\right\rangle_N
=\sqrt{\frac{2}{\pi}}$, whereas it is simply $\left\langle
\left|J\right|\right\rangle_N =1$ for a bimodal distribution on a
sparse graph. While there is no difference between cost and energy at
the level of averages, the corresponding relation of the
(co-)variances mixes the fluctuations in cost and energy with that of
the absolute bond weights. If in each instance the absolute sum of all
bond weights is the same, i.~e. a $\delta$-peak without fluctuations,
cost and energy fluctuations are proportional. But if the absolute
bond weight sum fluctuates (independently) from instance to instance,
either energy- or cost-fluctuations will become equally
normal-distributed, which leaves at most one of them
non-trivial. Typically, for SK cost-fluctuations are trivial
(i.~e. normal) and energy-fluctuations are interesting, while for
sparse graphs with variable degree and/or continuous bonds the role
reverses. Only for Bethe lattices (of fixed degree) with bimodal
bonds, cost and energy-fluctuations are proportional. When $r\gg1$ and
each spins acquires a nearly extensive number of neighbors, matters get
more complicated.

\section{Graph Bi-Partitioning on Bethe Lattices\label{sec:Graph-Bi-Partitioning}}

In GBP, a graph of (even) $N$ vertices and a certain number of edges
is divided into two sets of equal size $N/2$ such that the number of
edges connecting both sets, the {}``cutsize'', is minimized.  The
global constraint of an equal division places the GBP among the
hardest problems in combinatorial optimization, since determining the
\emph{exact} solution with certainty would in general require a
computational effort growing faster than any power of $N$~\citep{G+J}.
Applications of graph partitioning reach from the design of integrated
circuits (VLSI)~\citep{A+K} to the partitioning of sparse
matrices~\citep{HL}, leading to very different requirement regarding
the mix of speed and accuracy in a heuristic~\citep{CISE}. In
Refs.~\citep{Boettcher99a,Boettcher01b}, we have considered a range of
different graph ensembles, which can affect characteristics of GBP
drastically; here, as in Ref.~\citep{Zdeborova10}, we merely focus on
Bethe lattices, which are locally tree-like such that some analytical
results have been derived~\citep{Banavar87b,Mezard87,Wong87,Wong88}.
These Bethe lattices are also known as $r$-regular graphs, and are
generated by fixing the degree $r$ at each vertex, which in turn is
connected to other vertices at random.

\subsection{Average Ground State Costs\label{sub:Ground-State-Costs}}

In Tab.~\ref{tab:costGBP}, we have obtained approximate optima in the
cutsize per vertex $\left\langle c_{r}\right\rangle _{N}$ on Bethe
lattices for degrees $r$ between 3 and 10, and graph sizes between
$N=32$ and 2048. Statistical errors of our averages have been kept
small by generating a large number of instances for each $N$ and $r$,
typically $n_{I}\approx10^{6}$ for $N\leq200$, $n_{I}\approx10^{5}$
for $N\geq256$. Ref.~\citep{Zdeborova10} described the implementation
of the $\tau$-EO heuristic used to obtain the ground state
approximations for GBP. We have presented some of the most salient
results of the extrapolation to the thermodynamic limit there.

In Ref.~\citep{Zdeborova10}, we used a fit with first-order
corrections according to Eq.~(\ref{eq:Escaling}) alone, see Tab. 1 and
Fig.~2 there. Such a first-order fit to the data from the largest
system sizes only is discussed in
Tab.~\ref{tab:first-order-cost_GBP}. While the fits produce rather
consistent results, with very similar values for $\omega$, there is
clearly a distinction between even and odd $r$ in the quality of the
fits, indicative of parity effects typical of regular graphs already
noted earlier~\citep{Boettcher03a,Zdeborova10}.

\begin{table}[!ht]
\caption{\label{tab:first-order-cost_GBP}
Fit of the data in Tab.~\ref{tab:costGBP} to Eq.~(\ref{eq:Escaling})
for the cost $c$. Only data for $64<N<2048$ has been utilized.}
\begin{centering}
\begin{tabular}{r|ccc|rrc}
\hline 
$r$  & $\left\langle c_{r}\right\rangle _{\infty}$  & $A$  & $\omega$  & ndf  & $\chi^{2}$/ndf  & $Q$ \tabularnewline
\hline 
3  & 0.114765  & 1.71  & 0.88  & 6  & 0.178  & 0.98 \tabularnewline
4  & 0.265240  & 1.77  & 0.85  & 6  & 2.124  & 0.047 \tabularnewline
5  & 0.414254  & 2.35  & 0.85  & 6  & 0.410  & 0.87 \tabularnewline
6  & 0.587974  & 2.37  & 0.84  & 6  & 0.622  & 0.71 \tabularnewline
7  & 0.755752  & 3.11  & 0.86  & 6  & 0.138  & 0.99 \tabularnewline
8  & 0.940334  & 2.93  & 0.84  & 6  & 1.053  & 0.39 \tabularnewline
9  & 1.119390  & 3.74  & 0.86  & 6  & 0.599  & 0.73 \tabularnewline
10  & 1.310780  & 3.37  & 0.83  & 6  & 1.836  & 0.088 \tabularnewline
\hline
\end{tabular}
\par\end{centering}
\end{table}

\begin{table}[!ht]
\caption{\label{tab:ln23}
Fit of the data in Tab.~\ref{tab:costGBP} to Eq.~(\ref{eq:logscaling})
with fixed $\omega=\frac{2}{3}$.  Only data for $64<N<2048$ has been
utilized.}
\begin{centering}
\begin{tabular}{r|ccc|rrc}
\hline 
$r$  & $\left\langle c_{r}\right\rangle _{\infty}$  & $A$  & $B$  & ndf  & $\chi^{2}$/ndf  & $Q$ \tabularnewline
\hline 
3  & 0.116574  & -0.69  & -1.29  & 6  & 6.347  & 1.1e-06 \tabularnewline
4  & 0.266936  & -0.56  & -1.28  & 6  & 11.886  & 2.2e-13 \tabularnewline
5  & 0.416543  & -0.75  & -1.73  & 6  & 6.770  & 3.4e-07 \tabularnewline
6  & 0.590277  & -0.71  & -1.75  & 6  & 2.311  & 0.031 \tabularnewline
7  & 0.759679  & -1.21  & -2.45  & 6  & 2.585  & 0.017 \tabularnewline
8  & 0.944030  & -1.02  & -2.29  & 6  & 0.645  & 0.69 \tabularnewline
9  & 1.124500  & -1.52  & -3.01  & 6  & 1.747  & 0.11 \tabularnewline
10  & 1.315190  & -1.14  & -2.66  & 6  & 0.600  & 0.73 \tabularnewline
\hline
\end{tabular}
\par\end{centering}
\end{table}

\begin{table}[!ht]
\caption{\label{tab:Qdouble}
Fit of the data in Tab.~\ref{tab:costGBP} for all $N$
to Eq.~(\ref{eq:Cscaling}), but restricting $\omega_{1}=2\omega$.}
\begin{centering}
\begin{tabular}{r|ccrc|rrc}
\hline 
$r$  & $\left\langle c_{r}\right\rangle _{\infty}$  & $A$  & $B$  & $\omega$  & ndf  & $\chi^{2}$/ndf  & $Q$ \tabularnewline
\hline 
3  & 0.114302  & 1.26  & 2.53  & 0.82  & 9  & 86.118  & 0 \tabularnewline
4  & 0.265357  & 1.90  & -1.37  & 0.87  & 9  & 1.950  & 0.041 \tabularnewline
5  & 0.413007  & 1.52  & 3.62  & 0.77  & 9  & 101.090  & 0 \tabularnewline
6  & 0.588208  & 2.65  & -3.16  & 0.86  & 9  & 0.351  & 0.96 \tabularnewline
7  & 0.753483  & 1.94  & 4.31  & 0.76  & 9  & 118.141  & 0 \tabularnewline
8  & 0.940714  & 3.33  & -4.55  & 0.86  & 9  & 0.547  & 0.84 \tabularnewline
9  & 1.116430  & 2.30  & 4.90  & 0.76  & 9  & 137.050  & 0 \tabularnewline
10  & 1.311290  & 3.84  & -4.83  & 0.85  & 9  & 0.975  & 0.46 \tabularnewline
\hline
\end{tabular}
\par\end{centering}
\end{table}

\begin{figure*}
\vskip 8.7in 
\includegraphics{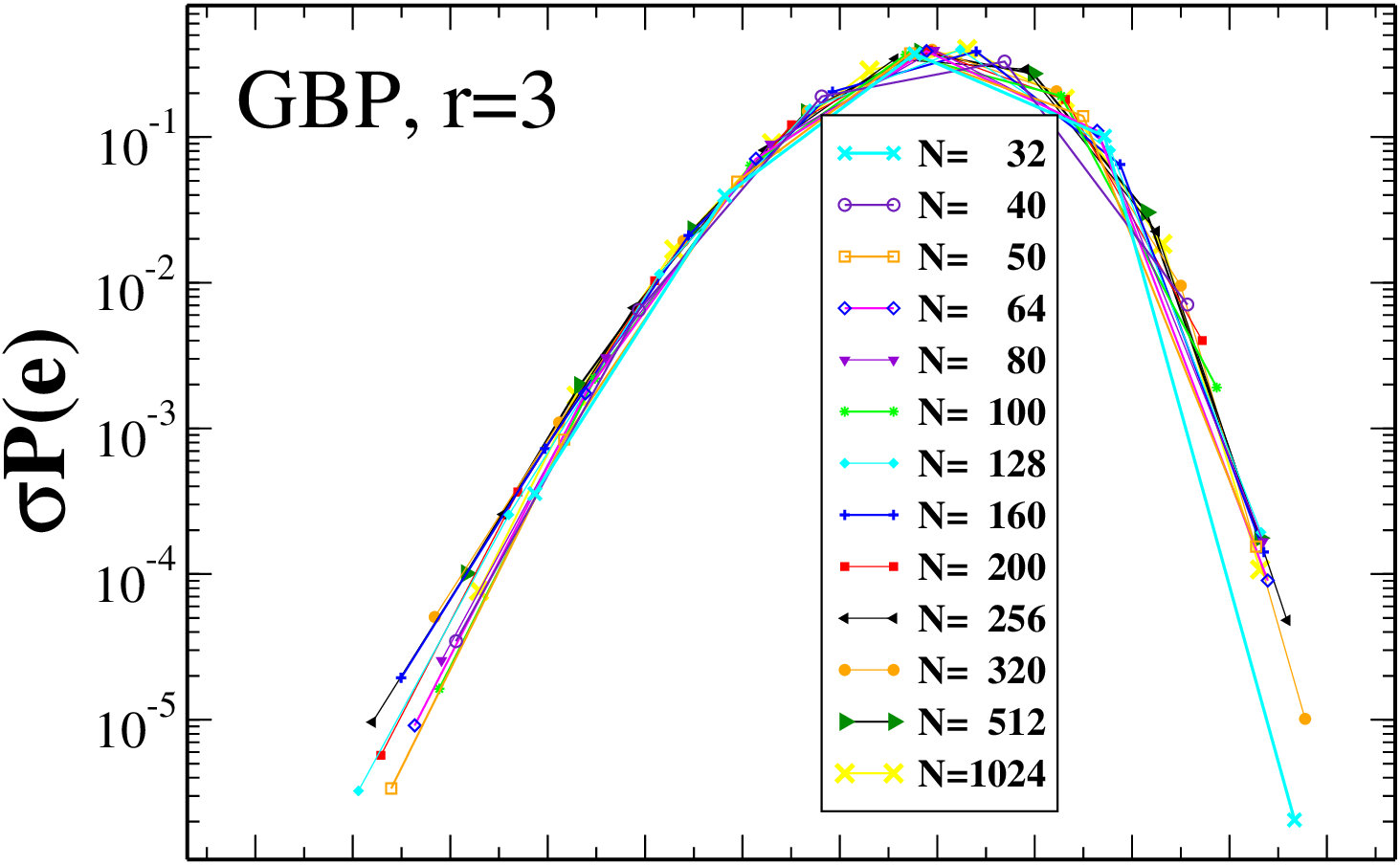} 
\includegraphics{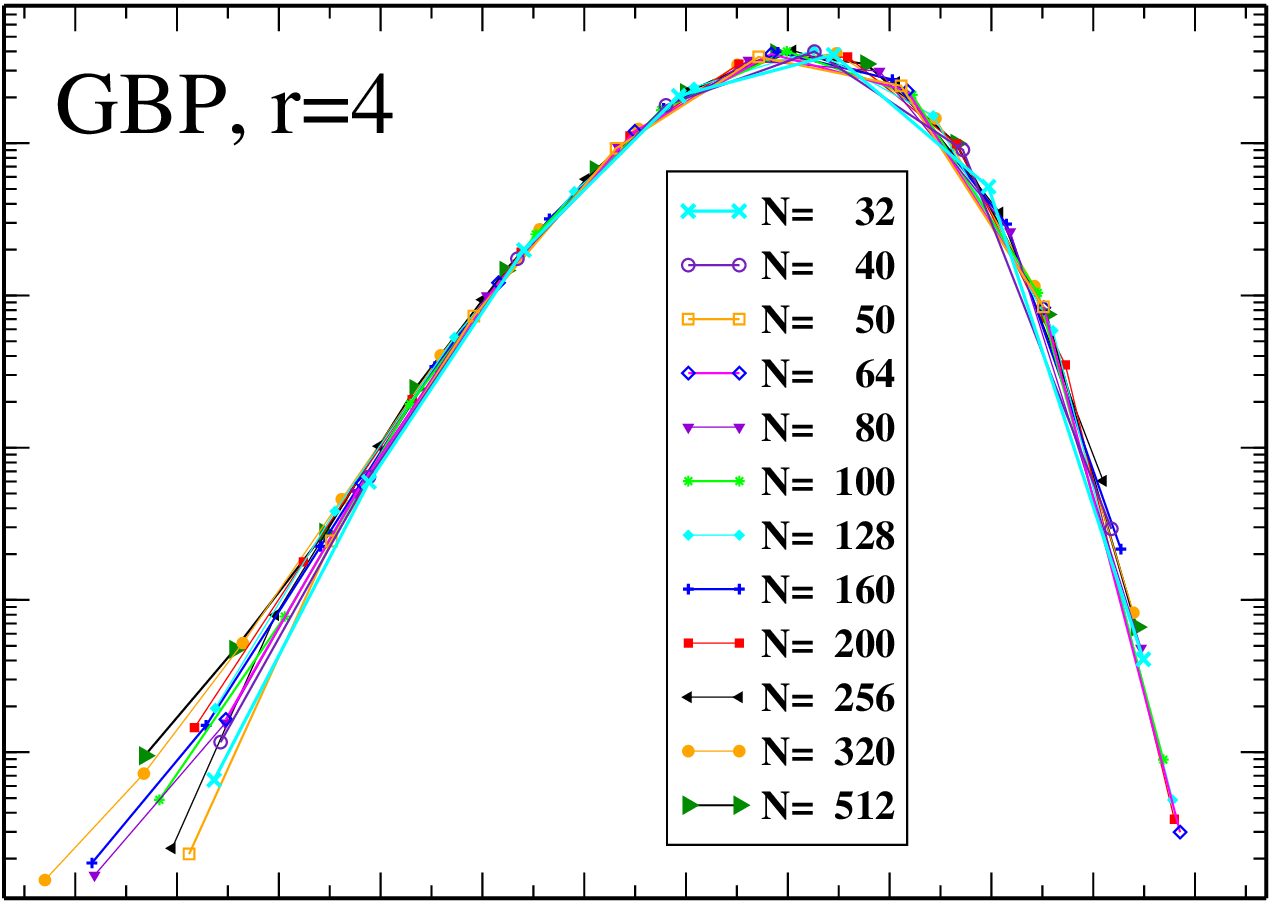} 
\includegraphics{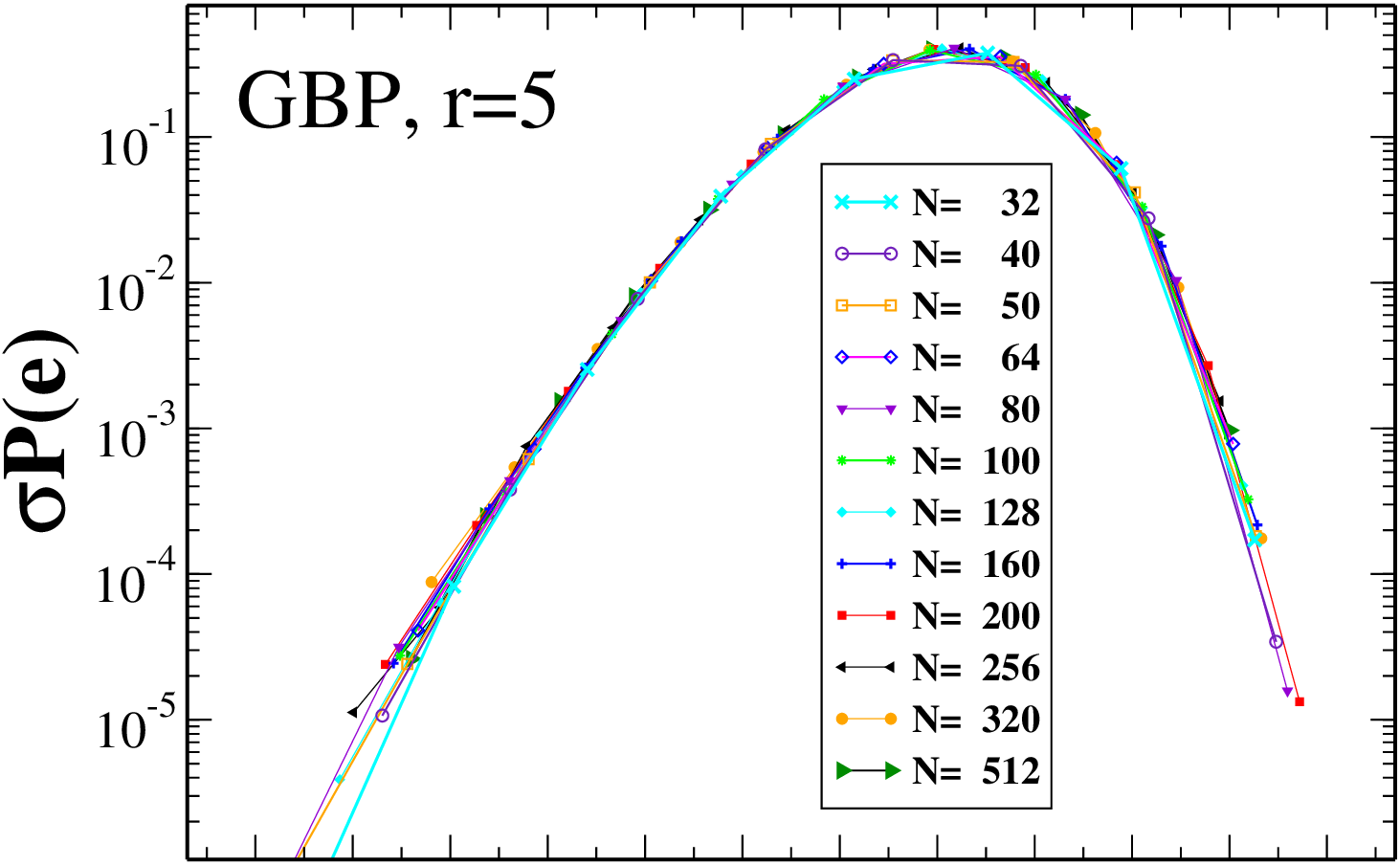} 
\includegraphics{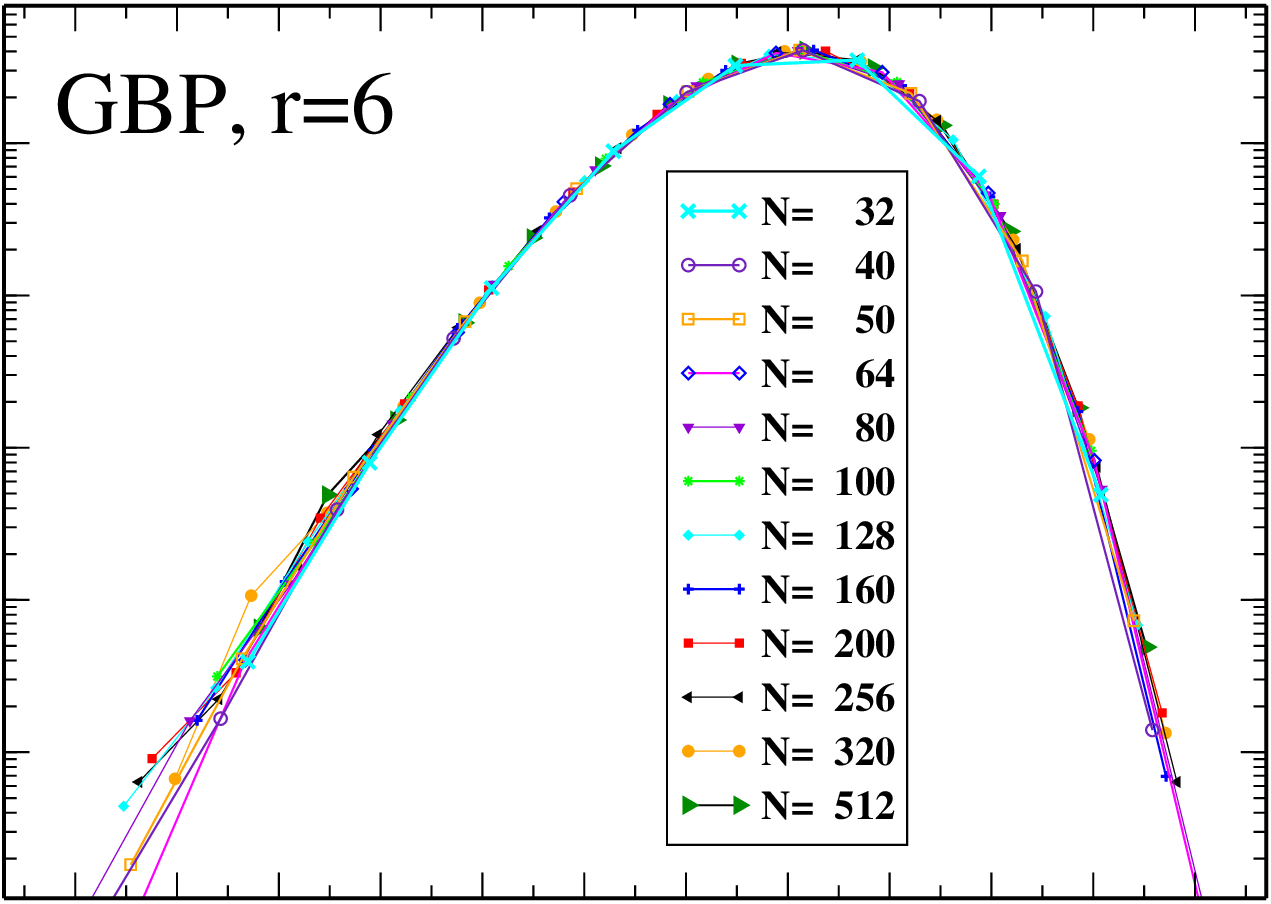} 
\includegraphics{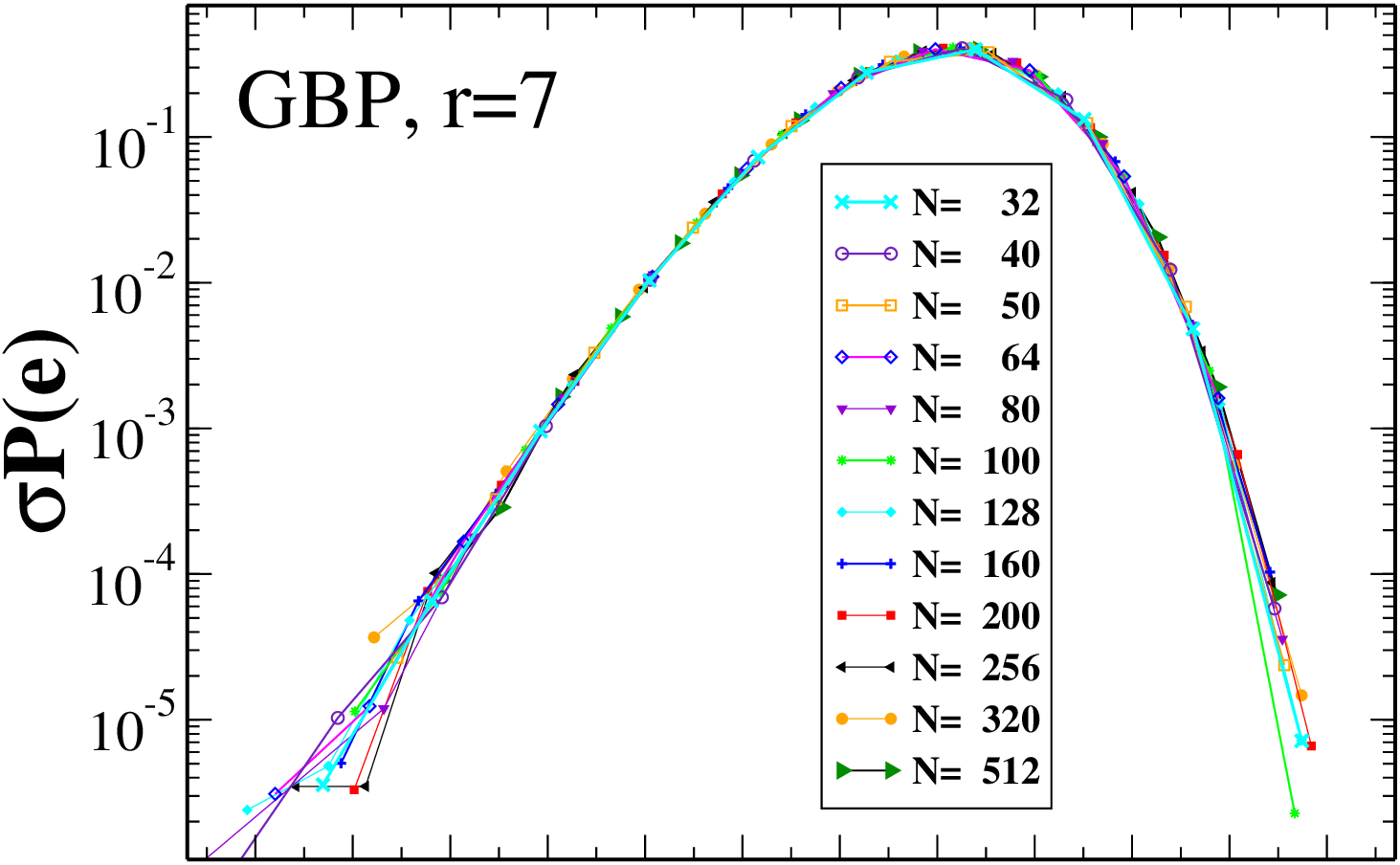} 
\includegraphics{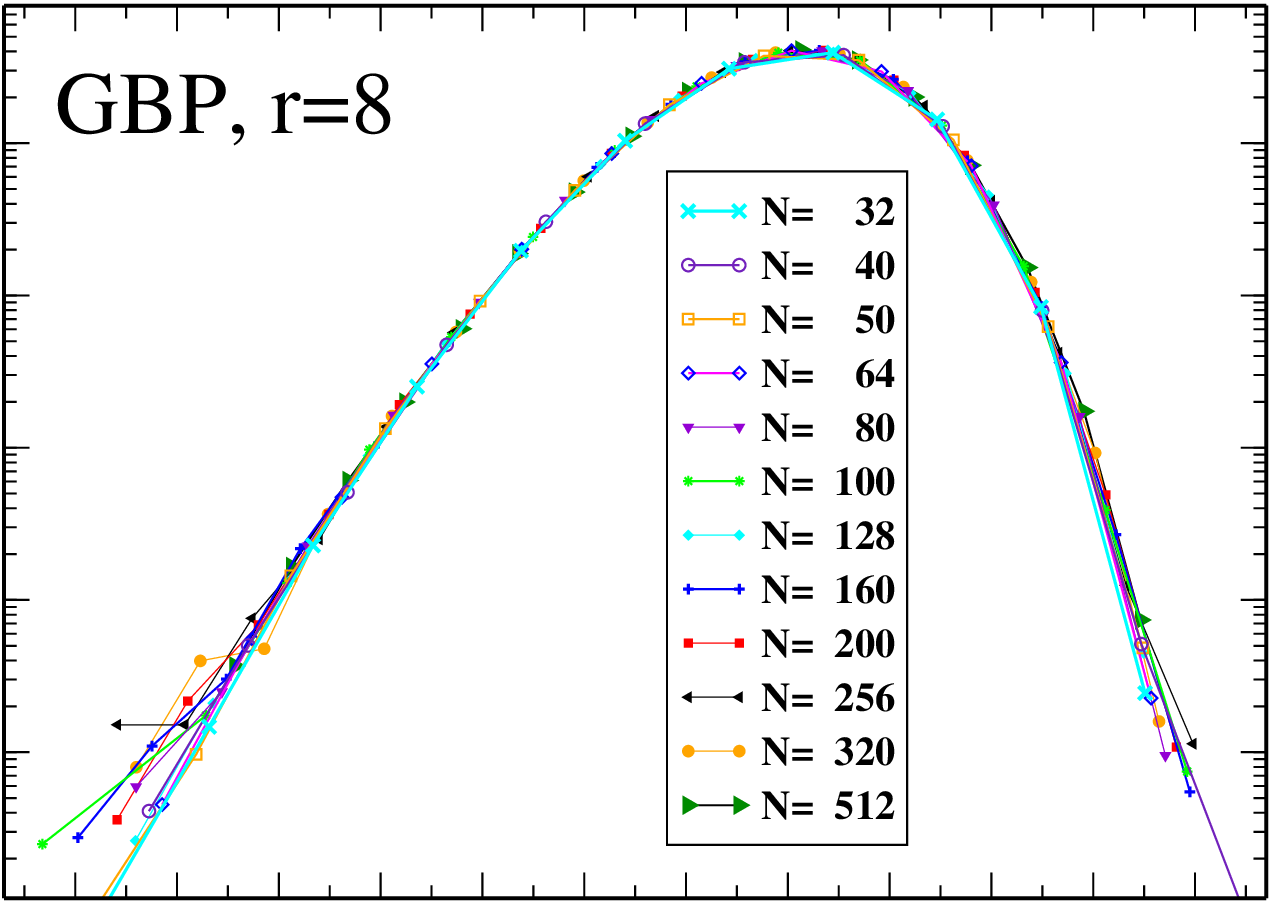} 
\includegraphics{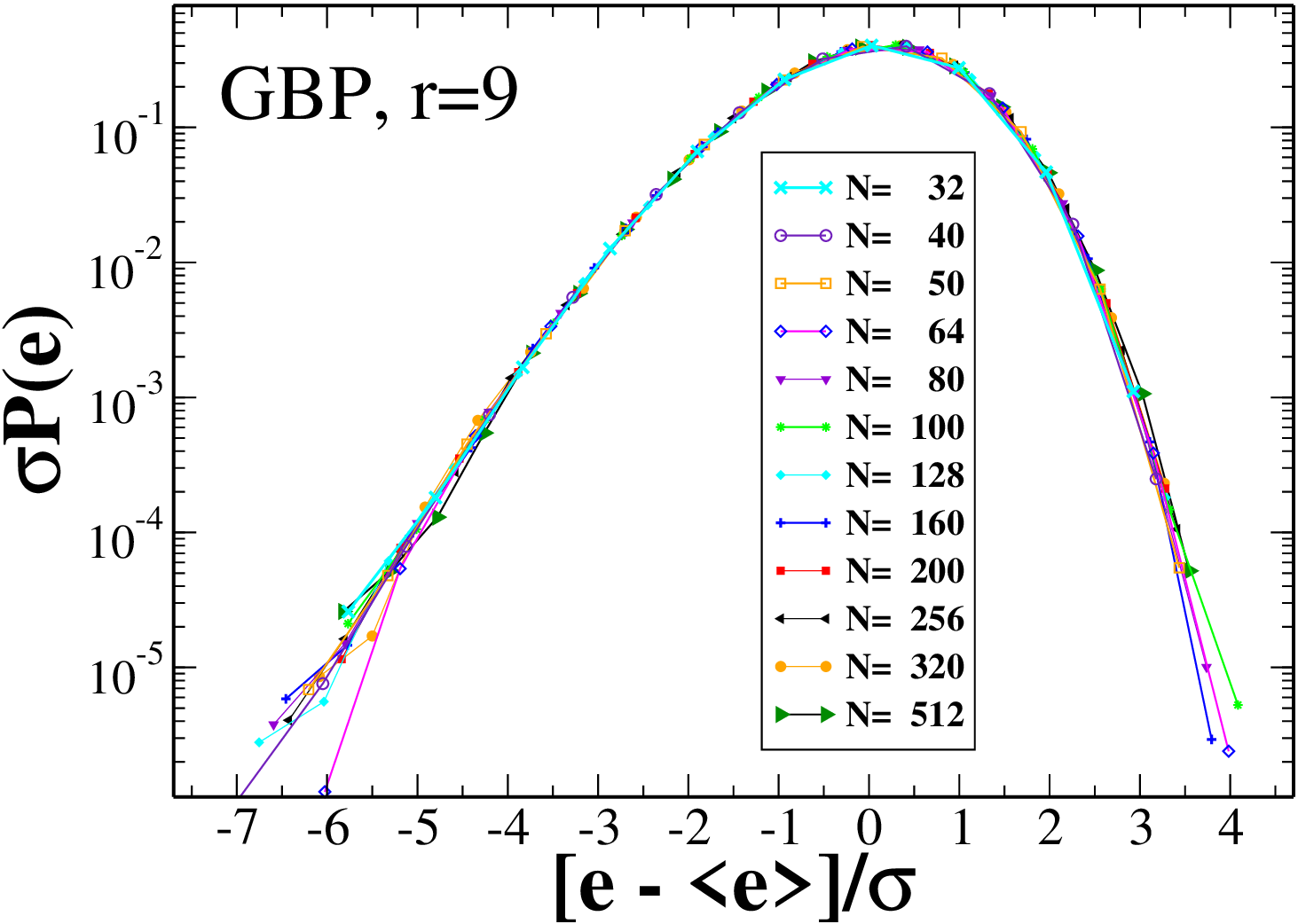} 
\includegraphics{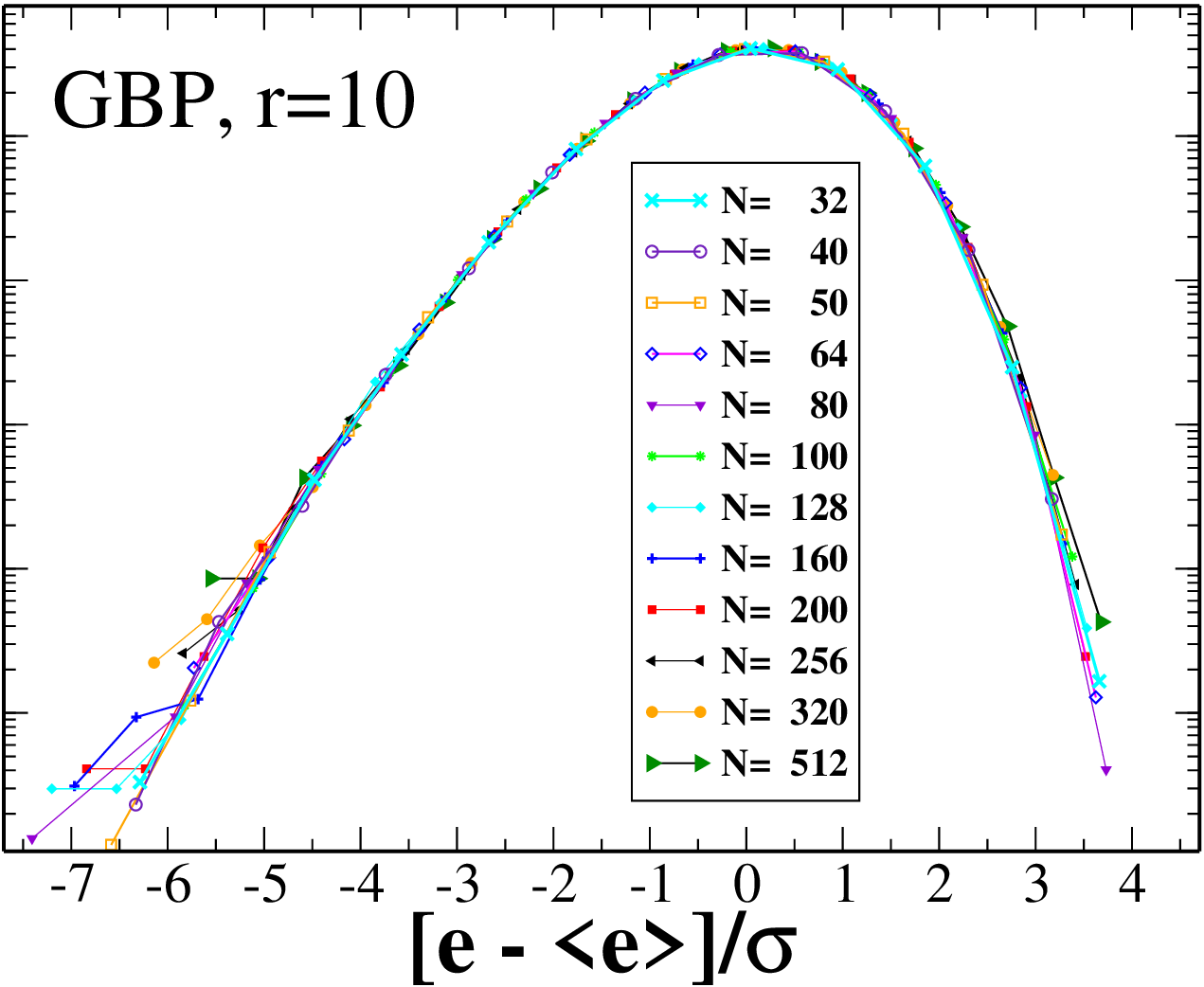} 
\caption{\label{fig:CostFluc}
Plot of the PDFs for the ensemble fluctuations of the optimal costs
per spin for GBP on Bethe lattices of degree $r=3,4,\ldots,10$. When
rescaled by their deviation $\sigma=\sigma\left(c_{r}\right)_{N}$, the
data collapses virtually for all system sizes onto a highly skewed
master-curve. There appears to be little change in skewness between
degrees $r$.}
\end{figure*}

\begin{table*}[!ht]
\caption{\label{tab:sigmaGBP} 
Standard deviation of the ground-state costs in GBP. Statistical
errors are estimated as $\sigma/\sqrt{n_{I}}$, where $n_{I}$ refers to
the number of instances considered. Appropriately rescaled, this data
is plotted also in Fig.~\ref{fig:sigma_extra}.  (Some of the data, in
particular, for larger $N$ and $r$, shows an obvious systematic bias
due to two likely sources: a bad approximation for some ground states
and an undercount of rare instances with  costs extremely
far from the average. These data points have been left out of
Fig.~\ref{fig:sigma_extra}.)}
\begin{centering}
\begin{tabular}{r|l|l|l|l|l|l|l|l}
\hline 
$N$ & $\sigma(c_{3})_{N}$  & $\sigma(c_{4})_{N}$  & $\sigma(c_{5})_{N}$  & $\sigma(c_{6})_{N}$  & $\sigma(c_{7})_{N}$  & $\sigma(c_{8})_{N}$  & $\sigma(c_{9})_{N}$  & $\sigma(c_{10})_{N}$ \tabularnewline
\hline 
32  & 0.03206(3)  & 0.04108(4)  & 0.04561(5)  & 0.05222(5)  & 0.05598(6)  & 0.06118(6)  & 0.06470(6)  & 0.06910(7) \tabularnewline
40  & 0.02664(3)  & 0.03428(3)  & 0.03814(4)  & 0.04370(4)  & 0.04680(5)  & 0.05129(5)  & 0.05419(5)  & 0.05790(6) \tabularnewline
50  & 0.02252(2)  & 0.02860(3)  & 0.03215(3)  & 0.03653(4)  & 0.03953(4)  & 0.04301(4)  & 0.04564(5)  & 0.04869(5) \tabularnewline
64  & 0.01784(2)  & 0.02333(2)  & 0.02615(3)  & 0.02998(3)  & 0.03228(3)  & 0.03539(4)  & 0.03747(4)  & 0.04009(4) \tabularnewline
80  & 0.01482(1)  & 0.01945(2)  & 0.02192(2)  & 0.02511(3)  & 0.02710(3)  & 0.02967(3)  & 0.03146(3)  & 0.03367(3) \tabularnewline
100  & 0.01253(2)  & 0.01622(2)  & 0.01837(3)  & 0.02096(3)  & 0.02280(3)  & 0.02490(4)  & 0.02639(4)  & 0.02825(4) \tabularnewline
128  & 0.01012(2)  & 0.01329(1)  & 0.01517(2)  & 0.01725(2)  & 0.01877(3)  & 0.02051(3)  & 0.02179(3)  & 0.02329(3) \tabularnewline
160  & 0.00847(1)  & 0.01113(2)  & 0.01269(2)  & 0.01445(2)  & 0.01573(2)  & 0.01716(2)  & 0.01829(3)  & 0.01948(3) \tabularnewline
200  & 0.00712(1)  & 0.00935(1)  & 0.01066(2)  & 0.01209(2)  & 0.01323(2)  & 0.01442(2)  & 0.01536(2)  & 0.01641(3) \tabularnewline
256  & 0.00582(1)  & 0.00769(1)  & 0.00878(1)  & 0.00996(2)  & 0.01092(2)  & 0.01183(2)  & 0.01271(2)  & 0.01351(2) \tabularnewline
320  & 0.00490(1)  & 0.00643(1)  & 0.00734(2)  & 0.00835(2)  & 0.00919(2)  & 0.00995(2)  & 0.01069(2)  & 0.01140(3) \tabularnewline
512  & 0.00336(1)  & 0.00441(1)  & 0.00509(1)  & 0.00580(2)  & 0.00638(2)  & 0.00697(2)  & 0.00746(2)  & 0.00798(2) \tabularnewline
1024  & 0.00195(1)  & 0.00253(1)  & 0.00295(1)  & 0.00339(2)  & 0.00386(4)  & 0.00428(4)  & 0.00463(5)  & 0.00506(6) \tabularnewline
\hline 
\end{tabular}
\par\end{centering}
\end{table*}

A surprising feature of these fits are the values for the finite-size
scaling exponent $\omega$. In our previous
studies~\citep{Boettcher03a,Boettcher03b} of bimodal spin glasses on
Bethe lattices we found throughout that this exponent was most
consistent with $\omega=\frac{2}{3}$, which has also been predicted
for SK, see Sec.~\ref{sub:Ground-State-Energy}.  While it was argued
in Ref.~\citep{Zdeborova10} that bimodal spin glass and GBP on Bethe
lattices should be equivalent in the ground state energy, this
equivalence apparently does not extend even to first-order
corrections, as the values for $\omega$ in these fits are
significantly higher. To analyze whether even higher-order corrections
could rectify this discrepancy in scaling, we attempt a fit of the
form in Eq.~(\ref{eq:Cscaling}) with $\omega=\frac{2}{3}$ fixed but
variable $B$ and $\omega_{1}$. Such a fit fails to converge, as in
each iteration $\omega_1$ moves closer to $\omega$.  Alternatively, we
prescribe the form of a plausible logarithmic higher-order correction
in Eq.~(\ref{eq:logscaling}). The results of this fit are listed in
Tab.~\ref{tab:ln23}. The quality of the fit is very poor for small
degree $r$ but gets progressively better for increasing $r$,
suggesting a possible approach to the SK result for $\omega$ for
$r\to\infty$. Yet, the results for $\left\langle c_{r}\right\rangle
_{\infty}$ are somewhat less in agreement with the conjectured
equivalence in Ref.~\citep{Zdeborova10}. Generally, if the next-order
correction contains a (possibly polynomial) logarithmic dependence of
this sort, any attempt to predict $\omega$ may be futile.

Another alternative form to fit is provided by a series expansion in
powers of $N^{-\omega}$. Thus, fitting to Eq.~(\ref{eq:Cscaling}) with
$\omega_{1}=2\omega$ we obtain the results listed in
Tab.~\ref{tab:Qdouble}.  Now, the results for $\left\langle
c_{r}\right\rangle _{\infty}$ are again in good agreement with the
conjecture but the quality of the fit is not improved over the mere
first-order fit in Tab.~\ref{tab:first-order-cost_GBP}.
Interestingly, this form is now a better fit with \emph{even} values
for $r$ but terrible for odd ones, reversing the trend from the
previous fits. Correction terms may differ between even and odd $r$
not only in their constants but in their very form.

\subsection{Ground State Fluctuations\label{sub:Ground-State-Fluctuations}}

In Fig.~\ref{fig:CostFluc}, we plot the probability density functions
(PDF) of ground-state costs over the ensembles of Bethe
lattices. Unlike for the spin glass problems below, transients due to
finite-size effects diminish quickly. This leads to a solid collapse
of the data when properly rescaled by their deviations
$\sigma\left(c_{r}\right)_{N}$ in Eq.~(\ref{eq:rho}), apparently with
little difference in the scaling function even for varying degrees
$r$.

\begin{figure}
\includegraphics[bb=0bp 20bp 710bp 540bp,clip,scale=0.33]{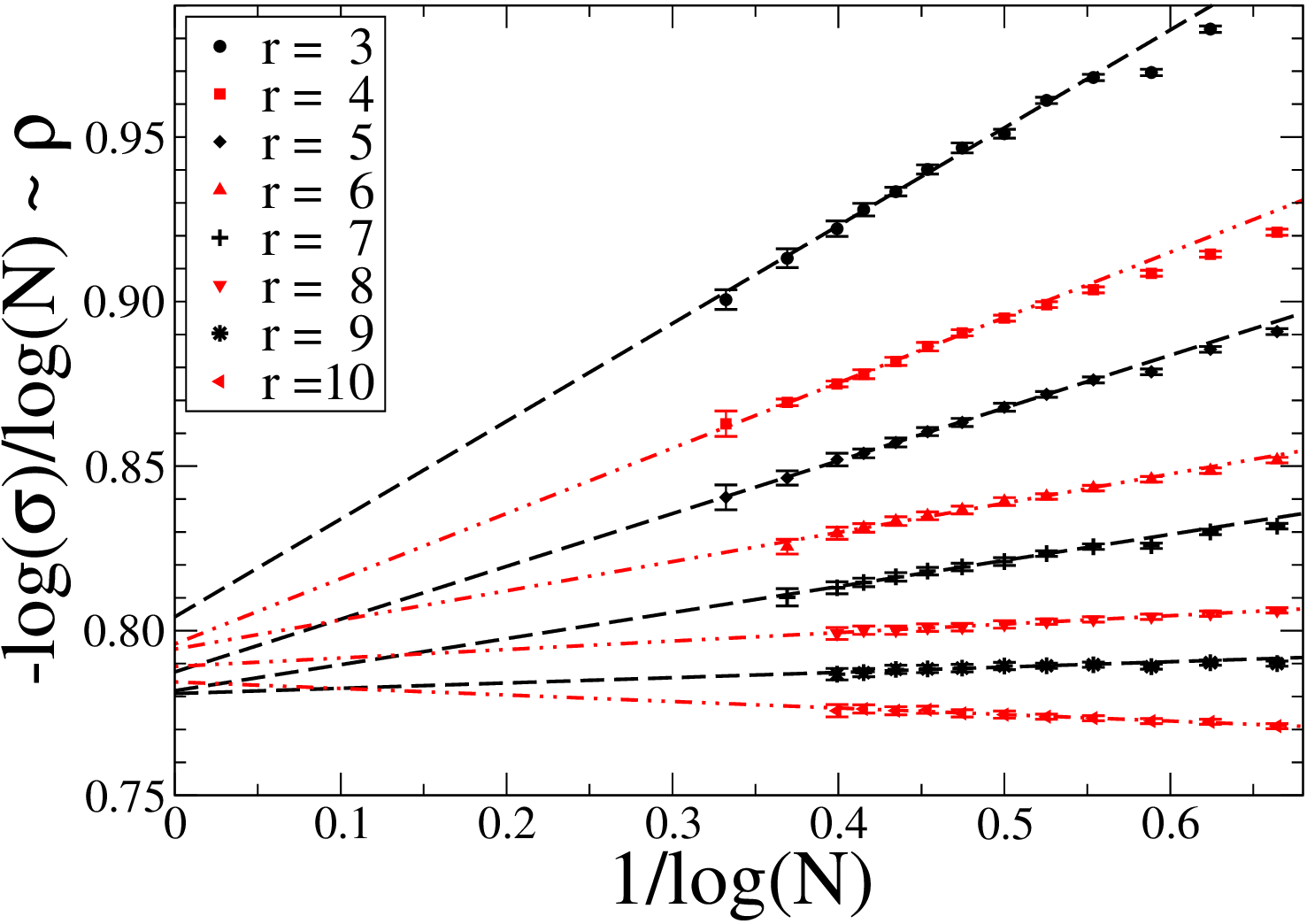}
\caption{\label{fig:sigma_extra} 
Extrapolation in the limit $N\!\to\!\infty$ for the exponent~$\rho$,
defined in Eq.~(\ref{eq:rho}), from the deviations
$\sigma\left(c_{r}\right)_{N}$ in Tab.~\ref{tab:sigmaGBP} obtained
from the distribution of GBP ground state. Appropriately rescaled via
Eq.~(\ref{eq:rhoextra}), the data for each degree $r$ extrapolates
linearly to the thermodynamic limit, with rapidly diminishing
transients. Note that for each $r$ the fitted region spans at least
one decade in system size $N$. The goodness-of-fit is $Q=1$ for all
$r$ except for $r=3$, where $Q\approx0.73$.}
\end{figure}

\begin{figure}
\includegraphics[clip,scale=0.33]{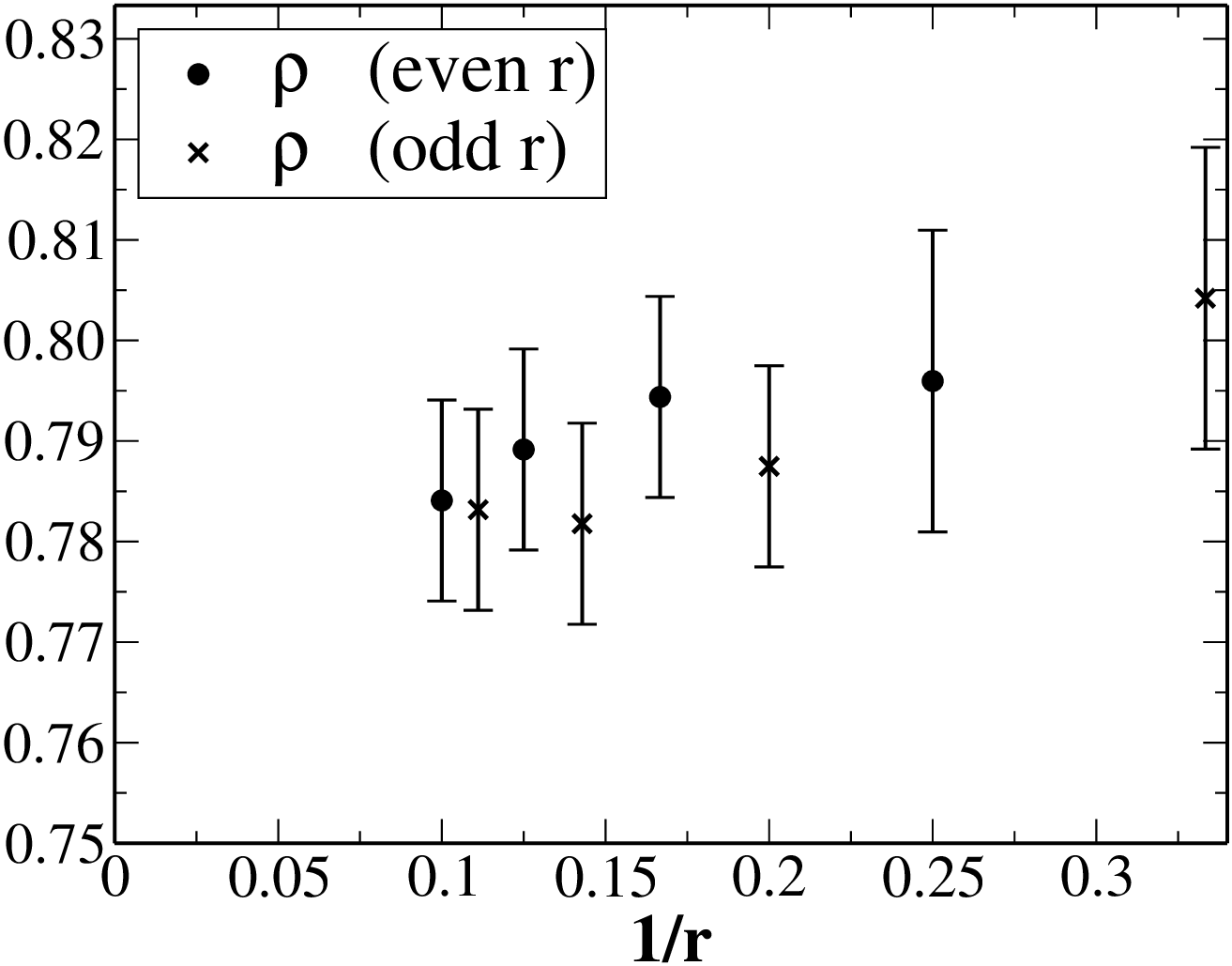}
\caption{\label{fig:rho_extra} 
Extrapolated values of $\rho$ for each degree $r$, obtained from
Fig.~\ref{fig:sigma_extra}, plotted as a function of inverse
degree. For even and odd values of $r$ separately, each sequence
extrapolates to value of $\rho_{\infty}=0.77(2)$ for
$r\to\infty$. The error bars result from the fits in
Fig.~\ref{fig:sigma_extra}; systematic errors, e.~g. due to 
higher-order corrections ignored in those fits, are potentially 
larger.}
\end{figure}

In Tab.~\ref{tab:sigmaGBP}, we list the deviations
$\sigma\left(c_{r}\right)_{N}$, as defined in Eq.~(\ref{eq:rho}), for
each degree $r$ and system size $N$. A closer look at these deviations
indeed provides a robust extrapolation for the exponent $\rho$.
Eq.~(\ref{eq:rhoextra} should yield a linear extrapolation for the
exponent $\rho$ when plotted vs. $1/\log N$, assuming negligible
corrections (i.~e. $N^{-a}\ll N^{-\rho}$). In
Fig.~\ref{fig:sigma_extra}, we show these plots for all values of
$r$. It is apparent that transient behavior decays very quickly, and a
smooth linear extrapolation is obtained. Also re-assuring is the fact
that this data shows a small even/odd effect, just as for the costs
above in Sec.~\ref{sub:Ground-State-Costs}.  The sequence of
extrapolants for $\rho$ appear to converge towards the same value of
$\rho=0.77(2)$, at least for $r\to\infty$. It is surprising to find
such an apparently non-trivial result in such a simple mean-field
model. Comparing with theory, it comes closest to the value of
$\rho=\frac{3}{4}$ previously proposed in Ref.~\citep{Bouchaud03} and
seems far from that proposed for the SK model,
$\rho=\frac{5}{6}$~\citep{Parisi08,Parisi09,Rizzo09,Rizzo09b}.  There
is, of course, no reason for it to be equal to that of SK. But the
in-between value found here for GBP is arguably close to the
corresponding extrapolations for the spin glasses below, which by
themselves remain inconclusive due to strong transients up to large
system sizes.

\section{Sherrington-Kirkpatrick Model\label{sec:Sherrington-Kirkpatrick-Model}}

We reconsider finite-size corrections in the SK model by expanding
on the simulations in Ref.~\citep{EOSK} with substantially more instances
at many more system sizes, see Tab.~\ref{Tab: AllData}. It is found
that the corrections to the average energies permit fits of reasonable
quality with systematic improvements when higher-order corrections
are incorporated, but the discussion of the energy fluctuations remains
largely inconclusive. Interestingly, this scenario is the converse
of that found in Sec.~\ref{sec:Graph-Bi-Partitioning}
for GBP.

\begin{figure*}
\includegraphics[bb=0bp 50bp 750bp 522bp,clip,scale=0.23]{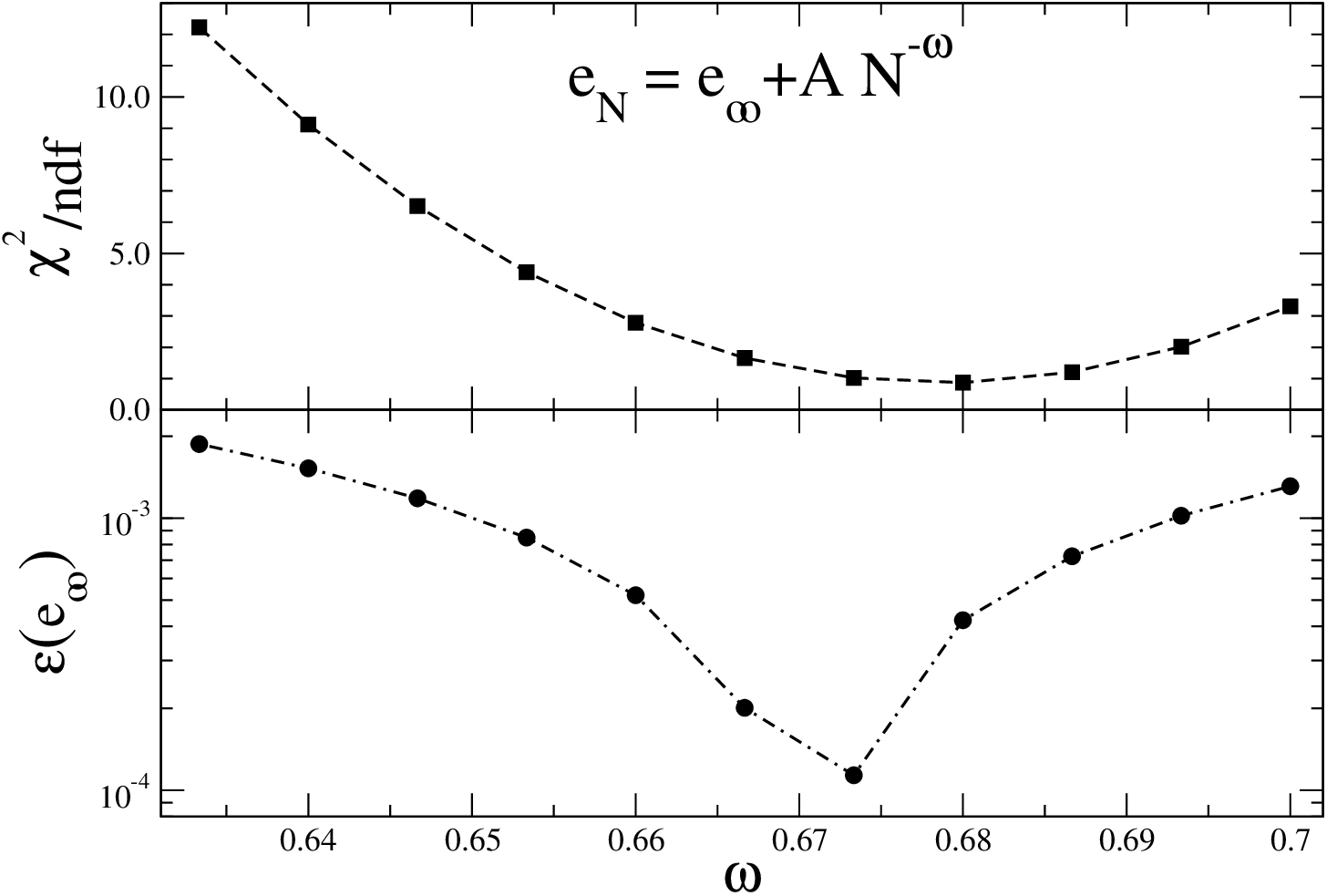}\includegraphics[bb=0bp 50bp 750bp 522bp,clip,scale=0.23]{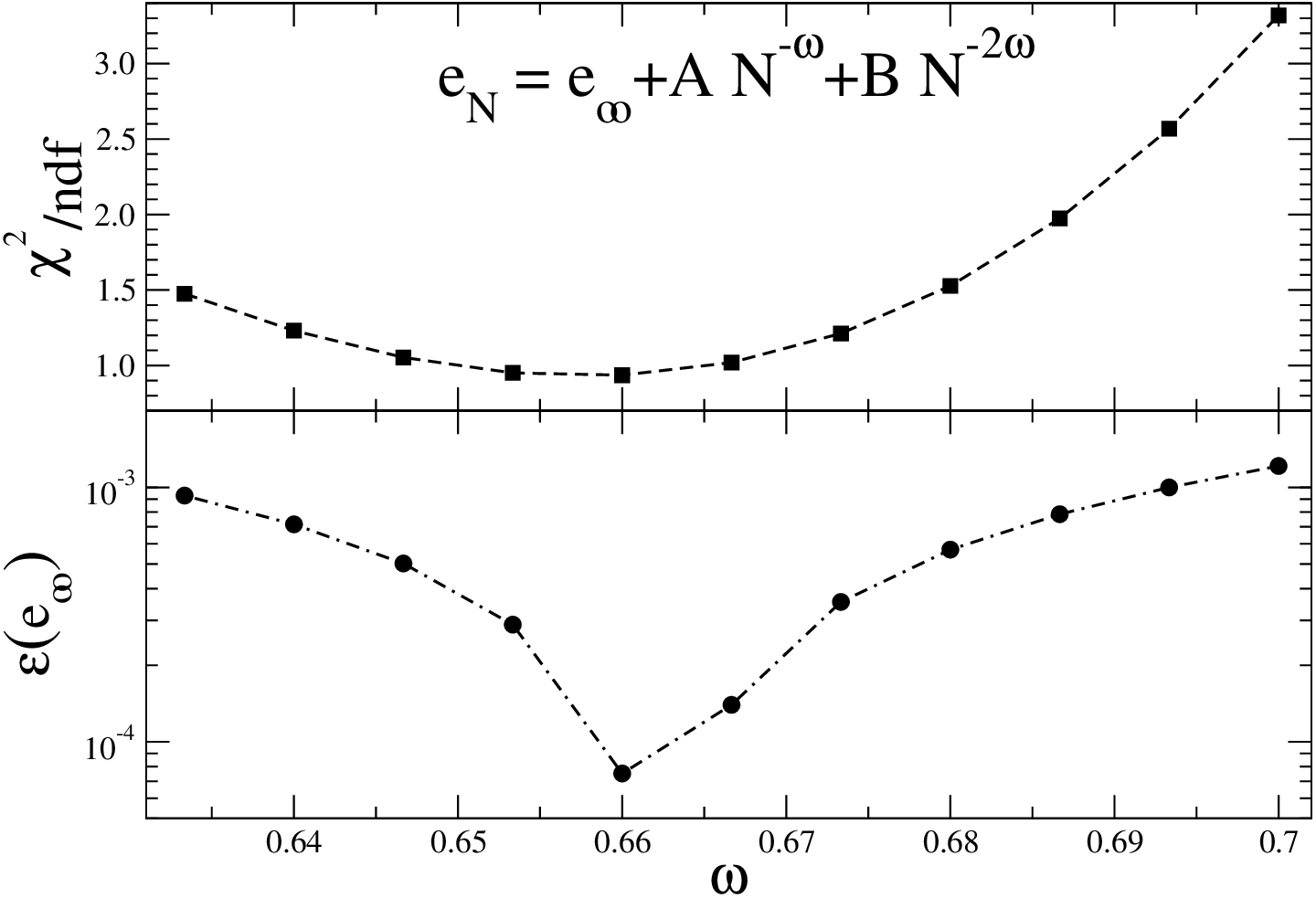}\includegraphics[bb=0bp 50bp 710bp 522bp,clip,scale=0.23]{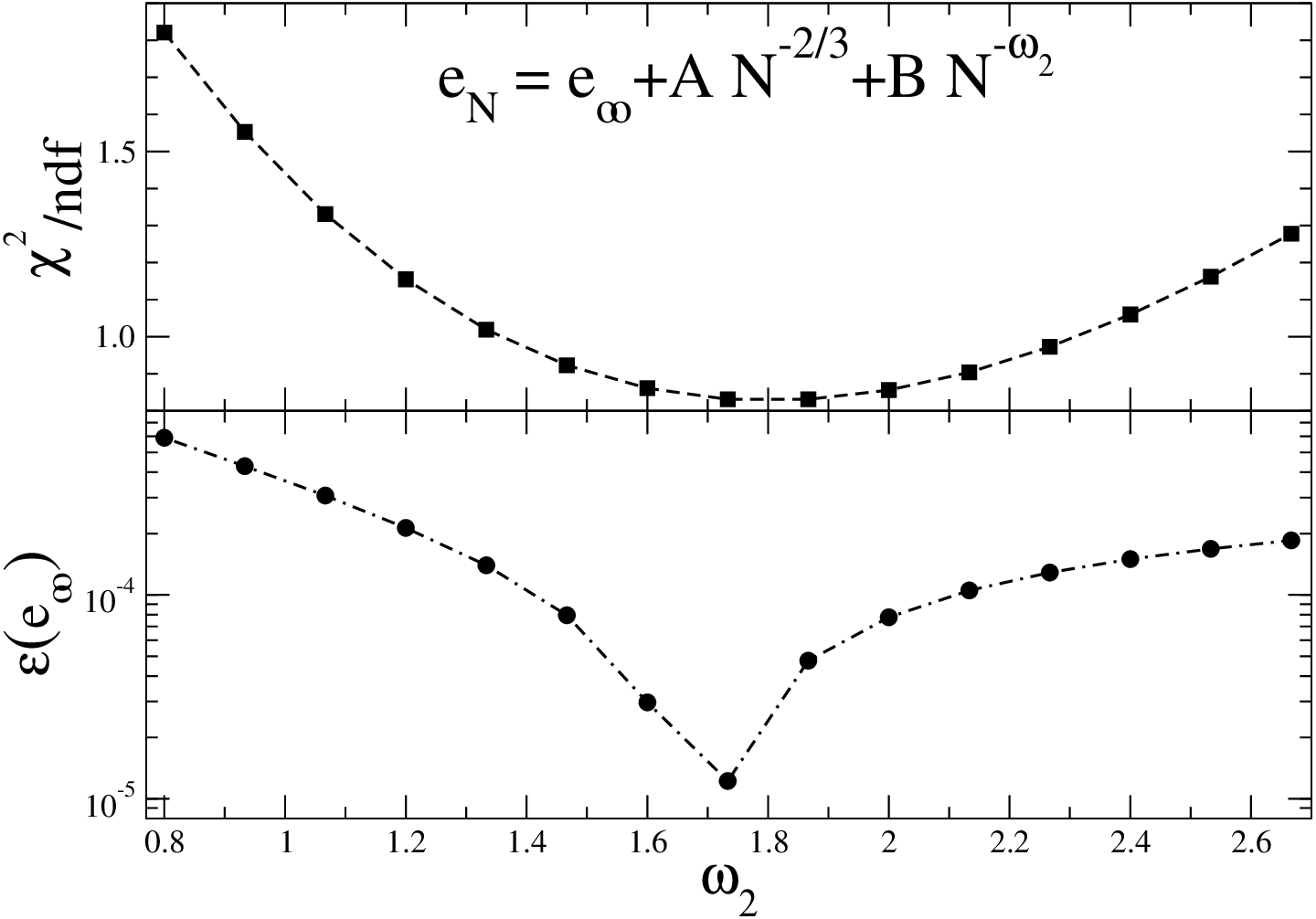}
\caption{\label{fig:Qtests}
Various fits to the average ground state energies $\left\langle
e_{0}\right\rangle _{N}$ of SK in Tab.~\ref{Tab: AllData}.  Left-most,
we consider first-order corrections only for a fit involving the
thermodynamic limit-value $e_{\infty}$ and the correction amplitude
$A$ for varying the correction exponent $\omega$. The best-quality fit
to this form, possessing minimal $\chi^{2}/ndf$, occurs just above
$\frac{2}{3}$ at $\omega\approx0.68$. Adding a higher-order correction
(with coefficient $B$, middle panel), a square of the first-order
term, bottoms out at $\omega\approx0.66$ with much lower
$\chi^{2}/ndf$ values overall. Instead, fixing the first-order
exponent to $\omega=\frac{2}{3}$ and varying the second-order exponent
$\omega_{1}$ (right panel) implicates an optimal exponent much above
$2\omega$, namely $\omega_{1}\approx1.8$. Remarkably, the optimal
choice for $\omega$ or $\omega_{1}$ about coincides in all three fits
with the lowest relative error $\epsilon\left(e_{\infty}\right)$
(lower panels) in the fitted value for $e_{\infty}$.}
\end{figure*}

\begin{figure}
\includegraphics[scale=0.33]{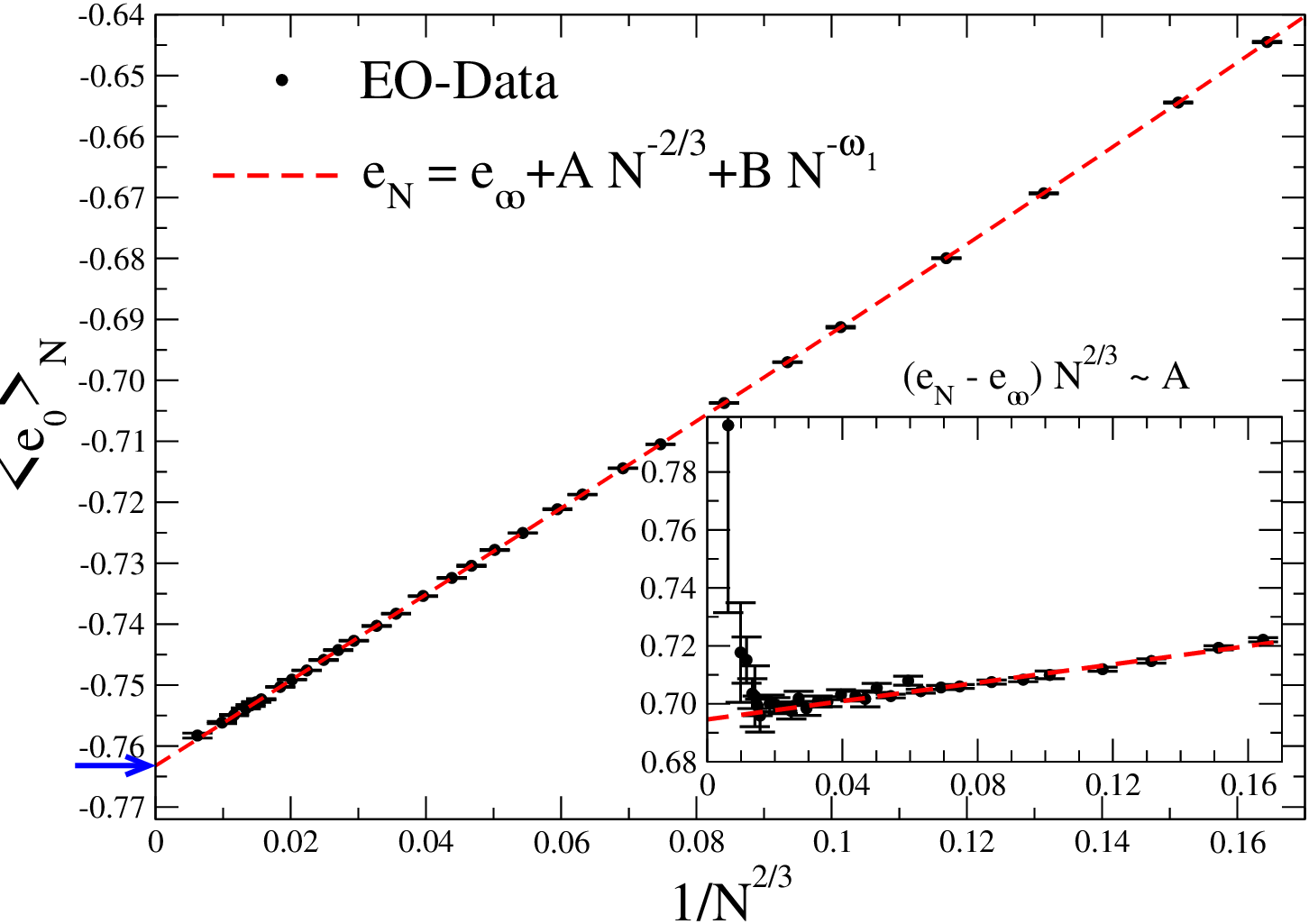}
\caption{\label{fig:ExtrapolationE}
Extrapolation plot of the average ground-state energy densities
$\left\langle e_{0}\right\rangle _{N}$ from Tab.~\ref{Tab: AllData}
as a function of the presumed finite-size corrections,
$1/N^{\frac{2}{3}}$. The statistical errors indicated are much smaller
than symbol sizes. For $N\to\infty$, the EO-data extrapolates to the
Parisi energy~\citep{Oppermann07}, $\left\langle e_{0}\right\rangle
_{\infty}=-0.7631667265(6)$, at the intercept. The indicated fit
(dashed line) predicts $e_{\infty}=-0.76323(5)$.  The slope of the
line is $A=0.70(1)$, consistent with the inset, which shows the same
data appropriately rescaled to extrapolate for $A$. It is
$B\approx0.48$, and $\omega_{1}\approx1.8$ suggests surprisingly weak
higher-order corrections, see also Fig.~\ref{fig:Qtests}.}
\end{figure}

\subsection{Average Ground State Energy\label{sub:Ground-State-Energy}}

\begin{table}
\caption{\label{Tab: AllData}List of all the data obtained with EO in sequence
of system size $N$. Given are the number of instances $n_{I}$ considered
at each $N$, and the average ground-state energy density $\left\langle e_{0}\right\rangle _{N}$
and average standard deviation $\sigma(e_{0})$ over these instances.}
\begin{tabular}{r|r|l|l||}
\hline 
$N$ & $n_{I}$ & $\left\langle e_{0}\right\rangle _{N}$ & $\sigma\left(e_{0}\right)$\tabularnewline
\hline
15 & 5000000 & -0.64449(3) & 0.0670\tabularnewline
17 & 5000000 & -0.65441(3) & 0.0614\tabularnewline
21 & 5000000 & -0.66931(2) & 0.0531\tabularnewline
25 & 5000000 & -0.67994(2) & 0.0471\tabularnewline
31 & 1480000 & -0.69127(3) & 0.0407\tabularnewline
35 & 5000000 & -0.69701(2) & 0.0374\tabularnewline
41 & 5000000 & -0.70371(2) & 0.0335\tabularnewline
49 & 5500000 & -0.71049(1) & 0.0295\tabularnewline
55 & 5000000 & -0.71442(1) & 0.0272\tabularnewline
63 & 5000000 & -0.71872(1) & 0.0246\tabularnewline
69 & 1000000 & -0.72113(2) & 0.0232\tabularnewline
79 & 5000000 & -0.72505(1) & 0.0208\tabularnewline
89 & 1000000 & -0.72783(2) & 0.0192\tabularnewline
99 & 800000 & -0.73043(2) & 0.0177\tabularnewline
109 & 816132 & -0.73242(2) & 0.0165\tabularnewline
\hline
\end{tabular}\begin{tabular}{r|r|l|l}
\hline 
$N$ & $n_{I}$ & $\left\langle e_{0}\right\rangle _{N}$ & $\sigma\left(e_{0}\right)$\tabularnewline
\hline
127 & 732463 & -0.73538(2) & 0.0147\tabularnewline
149 & 723526 & -0.73828(2) & 0.0130\tabularnewline
169 & 624094 & -0.74029(2) & 0.0119\tabularnewline
199 & 351317 & -0.74273(2) & 0.0104\tabularnewline
225 & 329043 & -0.74424(2) & 0.0095\tabularnewline
255 & 255572 & -0.74587(2) & 0.0086\tabularnewline
299 & 454555 & -0.74759(1) & 0.0077\tabularnewline
349 & 317264  &  -0.74910(1) &  0.0068\tabularnewline
399 & 204045 & -0.75030(1) & 0.0061\tabularnewline
511 & 51246 & -0.75233(2) & 0.0050\tabularnewline
549 & 144912 & -0.75278(1) & 0.0048\tabularnewline
649 & 59717 & -0.75383(2) & 0.0042\tabularnewline
799 & 25257 & -0.75491(2) & 0.0037\tabularnewline
1023 & 5338 & -0.75615(4) & 0.0031\tabularnewline
2047 & 403 & -0.7583(1) & 0.0020\tabularnewline
\hline
\end{tabular}
\end{table}

First, we want to consider the average energies, which provide an
assessment of the quality of our simulations in reference to exactly
known results about SK; a necessary step preliminary to a more
demanding analysis of the data. In Fig.~\ref{fig:ExtrapolationE}, we
plot the data for the average ground-state energy density
$\left\langle e_{0}\right\rangle _{N}$ from Tab.~\ref{Tab: AllData}
for the different $N$. It has been argued on the basis of theoretical
studies~\citep{parisi:93,parisi:93b} at or near $T_{c}$ and on previous
numerical
investigations~\citep{Boettcher03a,EOSK,Katzgraber05,Aspelmeier07,Palassini08},
that finite-size corrections to the energy behave for all $T\leq
T_{c}$ according to Eq.~(\ref{eq:Escaling}) with $\omega=\frac{2}{3}$.
We find this expectation confirmed within numerical accuracy, see
Fig.~\ref{fig:Qtests} (left). Fig.~\ref{fig:Qtests} also shows that a
higher-order correction according to Eq.~(\ref{eq:Cscaling}) even
improve on this leading behavior substantially, which is plotted as
fit to the data in Fig.~\ref{fig:ExtrapolationE}.

We can obtain a revealing insight into the quality of the EO-data
by extracting the leading behavior to explore the correction term
in more detail.  Because  the  energy in the thermodynamic
limit is well-known~\citep{Oppermann07}, $\left\langle e_{0}\right\rangle
_{\infty}=-0.7631667265(6)$, we can rewrite Eq.~(\ref{eq:Escaling}) as 
\begin{eqnarray}
A & \sim & \left[\left\langle e_{0}\right\rangle _{N}-\left\langle  e_{0}\right
\rangle _{\infty}\right]N^{\frac{2}{3}}+\ldots\qquad(N\to\infty),
\label{eq:Ecorrections}
\end{eqnarray}
and plot the EO-data in this form in the inset of
Fig.~\ref{fig:ExtrapolationE}.  Since the form of higher-order
corrections are unknown, we plot the data again as a function of
$1/N^{\frac{2}{3}}$, which provides a near-linear collapse of the data
and an extrapolated value for the amplitude $A\approx0.695(5)$, which
is consistent with the value obtained by the fit in
Fig.~\ref{fig:ExtrapolationE}. But the most important aspect of the
inset resides in the sharp crossover in the behavior of the data at
around $N\approx1000$. The consistent behavior of the data for
$N\lesssim1000$ suggest sufficient numerical accuracy in the obtained
ground state energies to this level of analysis, without any
discernible systematic bias. The data points for $N=799$ and 1024 both
exhibit a systematic error of about $\Delta A/A\approx2/70\approx3\%$
in the prediction of $A$, hence, a relative systematic error of
$\epsilon\left(e_{\infty}\right)=\Delta
e_{\infty}/e_{\infty}\sim\Delta A/N^{2/3}/e_{\infty}\approx0.03\%$
(see also Fig.~\ref{fig:Qtests}) in the prediction of typical ground
state energies overall. Unfortunately, the systematic error for the
$N=2047$ data point is about $\Delta A/A\approx1/7\approx15\%$,
leading to a relative systematic error of 0.1\% in the prediction of
putative ground states, which is sufficiently noticeable in 
Fig.~\ref{fig:ExtrapolationE} to exclude that point from the
extrapolation.

It is worthwhile to compare the inset of Fig.~\ref{fig:ExtrapolationE}
with the corresponding plot, the inset of Fig.~1, in
Ref.~\citep{Palassini08}.  While the data there is also falling for
increasing $N$ {[}arguably to the same asymptotic value of
  $A\approx0.7$, see Eq.~(\ref{eq:Ecorrections}){]}, the variation of the
data there is far more rapid. Point for point, the data there
represents a systematically higher value in the average ground-state
energy than is obtained here. This could potentially indicate a bias
in the heuristic methods used, which may fail to find true ground
states across the board. (Notably, the data there does not show
drastic degradation in the quality of the results for increasing $N$
as is found here.) Alternatively, such disagreement could be
attributed to the difference in the bond distribution used: Gaussian
there and bimodal here. Although the leading thermodynamic properties
should be universal, higher-order corrections can be sensitive to
microscopic details.

\begin{figure}
\includegraphics[bb=4bp 23bp 706bp 523bp,clip,scale=0.33]{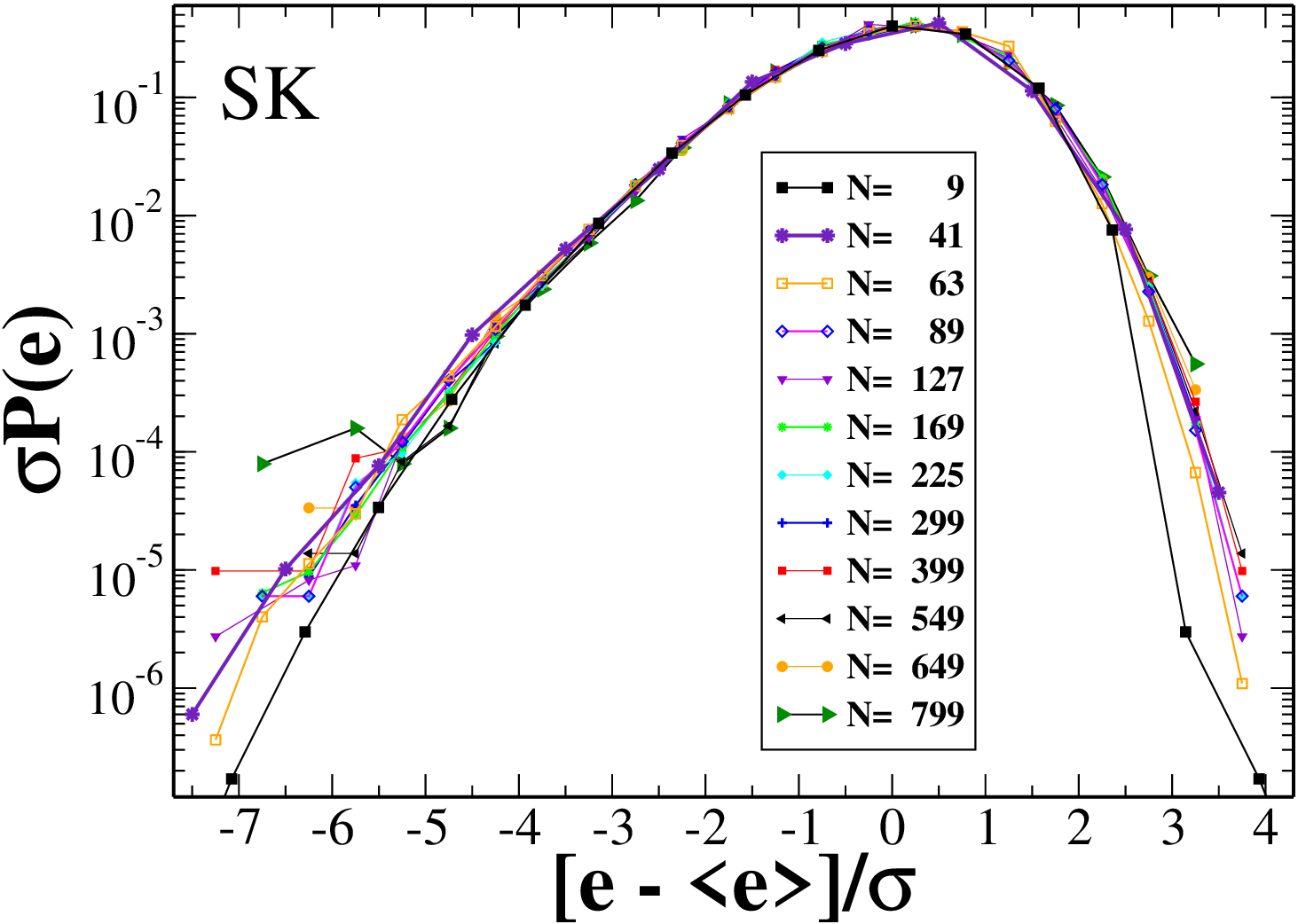}
\caption{\label{fig:PDF_SK}
Plot of the probability density functions (PDF) of obtained
ground-state energy densities $e_{0}$ for SK in units of the standard
deviation $\sigma$. For reference, the \emph{exact} probabilities for
$N=9$ are re-plotted from Ref.~\citep{Boettcher05e}.  Unlike for the PDFs
for GBP in Fig.~\ref{fig:CostFluc}, there is a significant finite-size
effect noticeable especially in the right tail of the distribution.}
\end{figure}
\begin{figure}
\includegraphics[clip,scale=0.33]{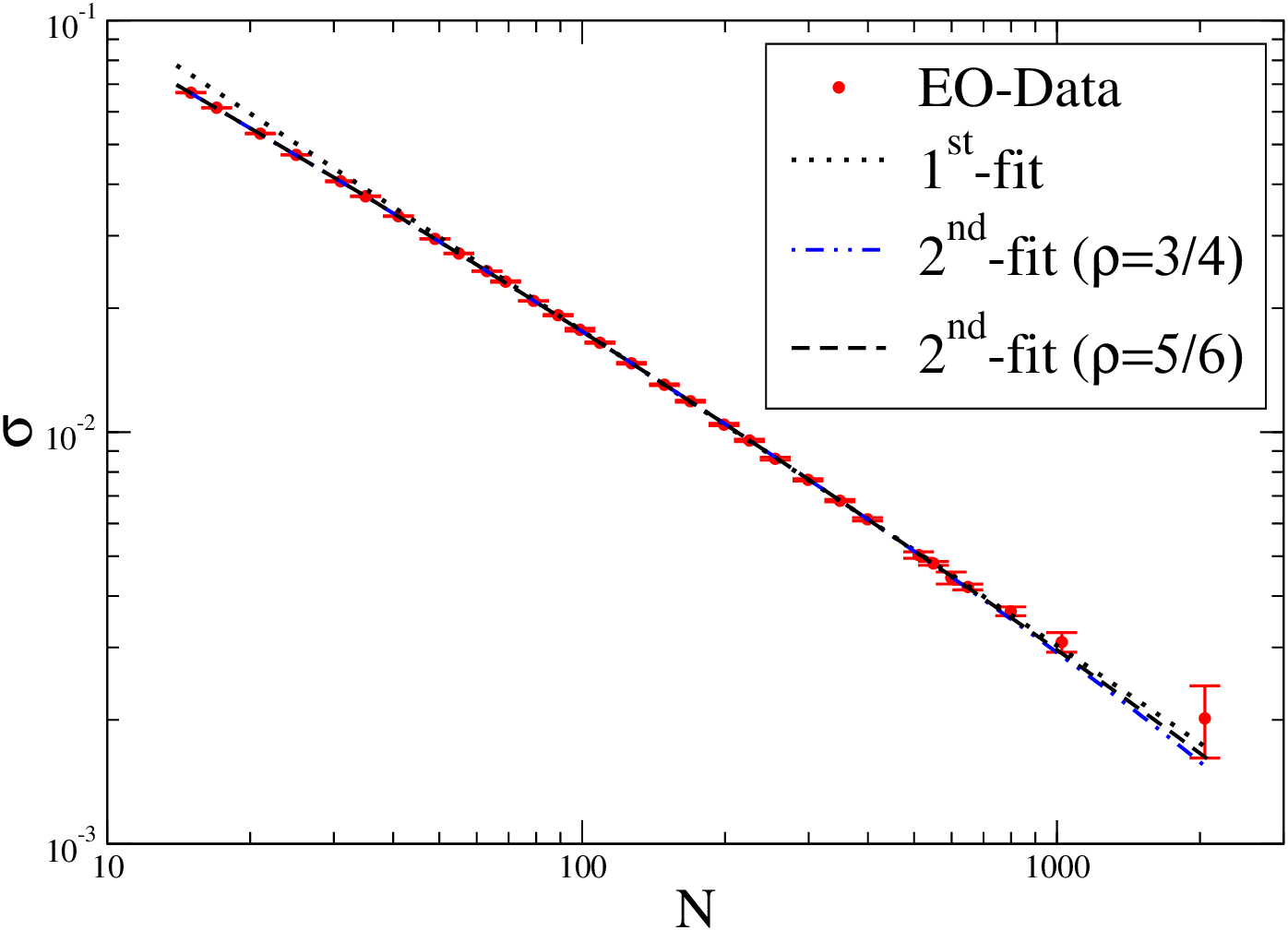}
\caption{\label{fig:rho_fit}
Double-logarithmic plot of the data from Tab.~\ref{Tab: AllData} for
$\sigma_{N}\left(e_{0}\right)$ as a function of $N$. The data is
fitted by a first-order fit with $a+bN^{-\rho}$ restricted to
$100<N<800$ (dotted line), giving $\rho\approx0.76$ with a
$\chi^{2}/ndf\approx0.24$ for $ndf=13$. Second-order fits with
$a+bN^{-\rho}+cN^{-a}$ allowing all $N<800$ and fixed
$\rho=\frac{3}{4}$ (dash-dotted line) or $\rho=\frac{5}{6}$ (dashed
line) both give essentially indistinguishable results with a
$\chi^{2}/ndf\approx0.7$ for $ndf=25$ in either case. For
$\rho=\frac{3}{4}$ we find $a\approx1.3$, and for $\rho=\frac{5}{6}$
it is quite consistent with unity, $a\approx1$. }
\end{figure}
\begin{figure}
\includegraphics[clip,scale=0.33]{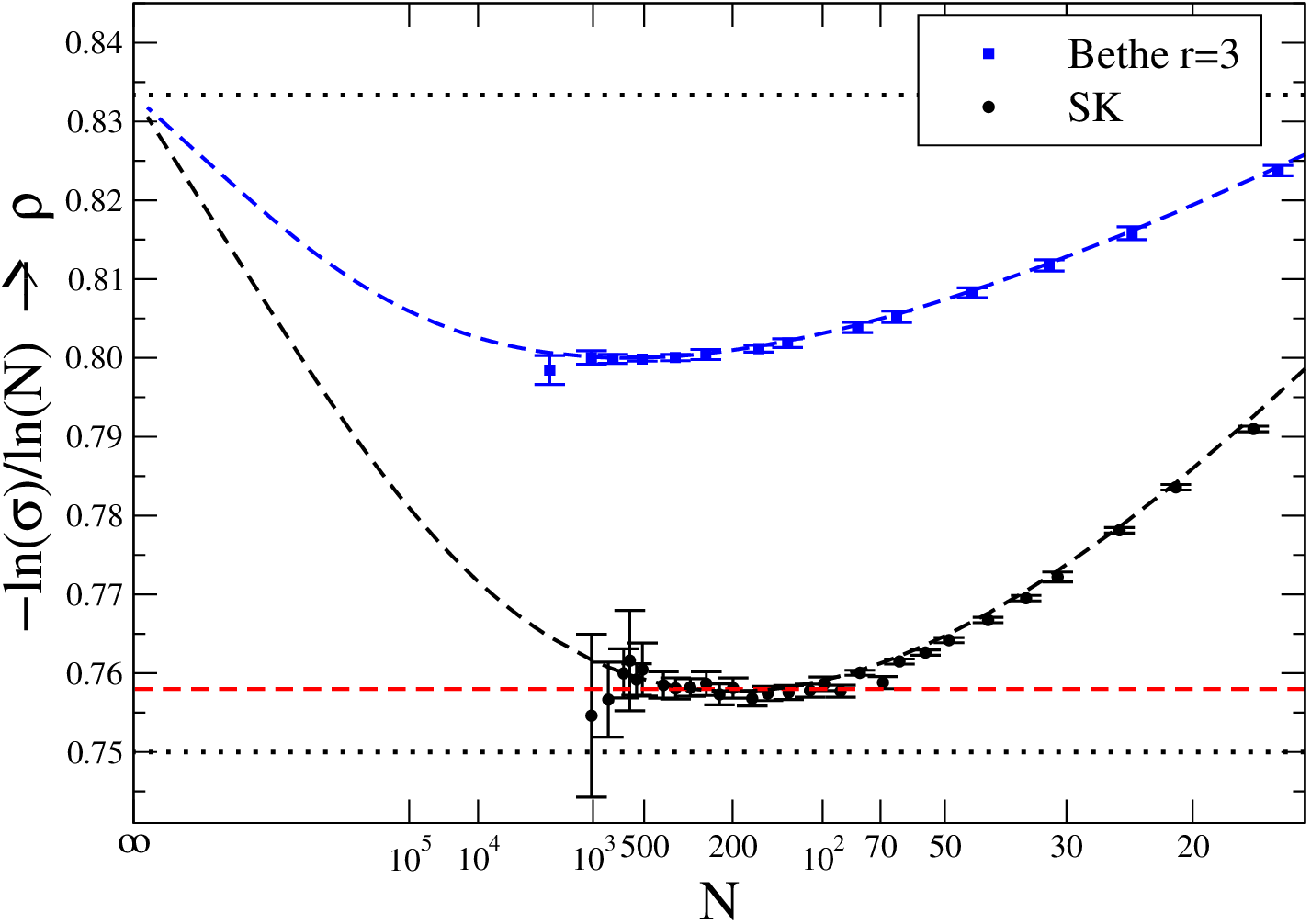}
\caption{\label{fig:rho_extrapolation}
Extrapolation plot according to Eq.~(\ref{eq:rhoextra}) towards a
prediction of $\rho$ at the ordinate-intercept for SK (black circles)
and spin glasses on Bethe lattices (blue squares).  The abscissa
denotes the system sizes $N$ on a scale of $1/\log N$.  A linear fit
(red-dashed line) of the SK data in the apparent scaling regime for
$N\geq80$ predicts a value of $\rho\approx0.76$. The inclusion of the
alternative value $\rho=\frac{5}{6}$ (dotted line) here shows the
dramatic change  required for the data to attain such a
value. Yet, a fit (black-dashed line) according to
Eq.~(\ref{eq:rhoextra}) involving non-linear corrections down to  the smallest $N$ is possible, if $\rho=\frac{5}{6}$ is assumed. The corresponding
situation for Bethe lattices makes such an
extrapolation further plausible.}
\end{figure}

In Fig.~\ref{fig:Qtests}, we present an alternative procedure to
explore corrections that also allows a probe of higher-order terms.
In this procedure, we select (the most important) one of the
parameters to be fitted as fixed and evaluate the quality of the fit
for the remaining parameters over a range of values for the selected
one.  As a measure of quality, we utilize $\chi^{2}$ per numbers of
degree of freedom (ndf), which should be minimized. The first panel
displays this procedure for just the first-order correction, again
confirming the expectation of Eq.~(\ref{eq:Escaling}). In the
remaining two panels we test possible higher-order corrections. In the
first of these, we test a fit to a regular Taylor series in powers of
$N^{-\omega}$ to second order. Incorporation of such a second-order
term improves the quality of the fit noticeably over the first-order
term alone.  Furthermore, the optimal choice for $\omega$ again proves
consistent with $\frac{2}{3}$, despite the extra degree of freedom
provided, which attests to its robustness. Then, taking
$\omega=\frac{2}{3}$ as a given, we explore an independent
second-order correction with scaling exponent $\omega_{1}$. This
yields the highest-quality fit thus far, also used in
Fig.~\ref{fig:ExtrapolationE}, but predicts that $\omega_{1}$ would be
much larger than simply $2\omega$, suggesting that such corrections
would be even weaker.

We have also tried to fit higher-order corrections of the form $1/N$
or $\ln N/N$ in addition to $1/N^{\frac{2}{3}}$ corrections, which are
plausible by analogy with the results obtained for finite-size
corrections near the critical
temperature~\citep{parisi:93,parisi:93b}.  A fit to
Eq.~(\ref{eq:logscaling}) does not produce acceptable results compared
to those found in Fig.~\ref{fig:Qtests} for any value of
$\omega$. Instead, a fit to Eq.~(\ref{eq:Cscaling}) with $\omega_1=1$
fixed produces a very narrow window of reasonable results near
$\omega=\frac{2}{3}$ but at best of the quality of what is seen
correspondingly at $\omega_{1}=1$ in the last panel of
Fig.~\ref{fig:Qtests}. Further higher-order corrections may improve on
this alternative. But if the ratio between a previous and its next
higher-order correction is a weakly falling function,
i.~e. $\left(1/N\right)/\left(1/N^{\frac{2}{3}}\right)=N^{-\frac{1}{3}}\approx0.1$
at least for $N\approx1000$ here, resolving the impact of such
corrections with the available data becomes near impossible and they
can never be fully excluded.

\begin{figure}
\includegraphics[clip,scale=0.33]{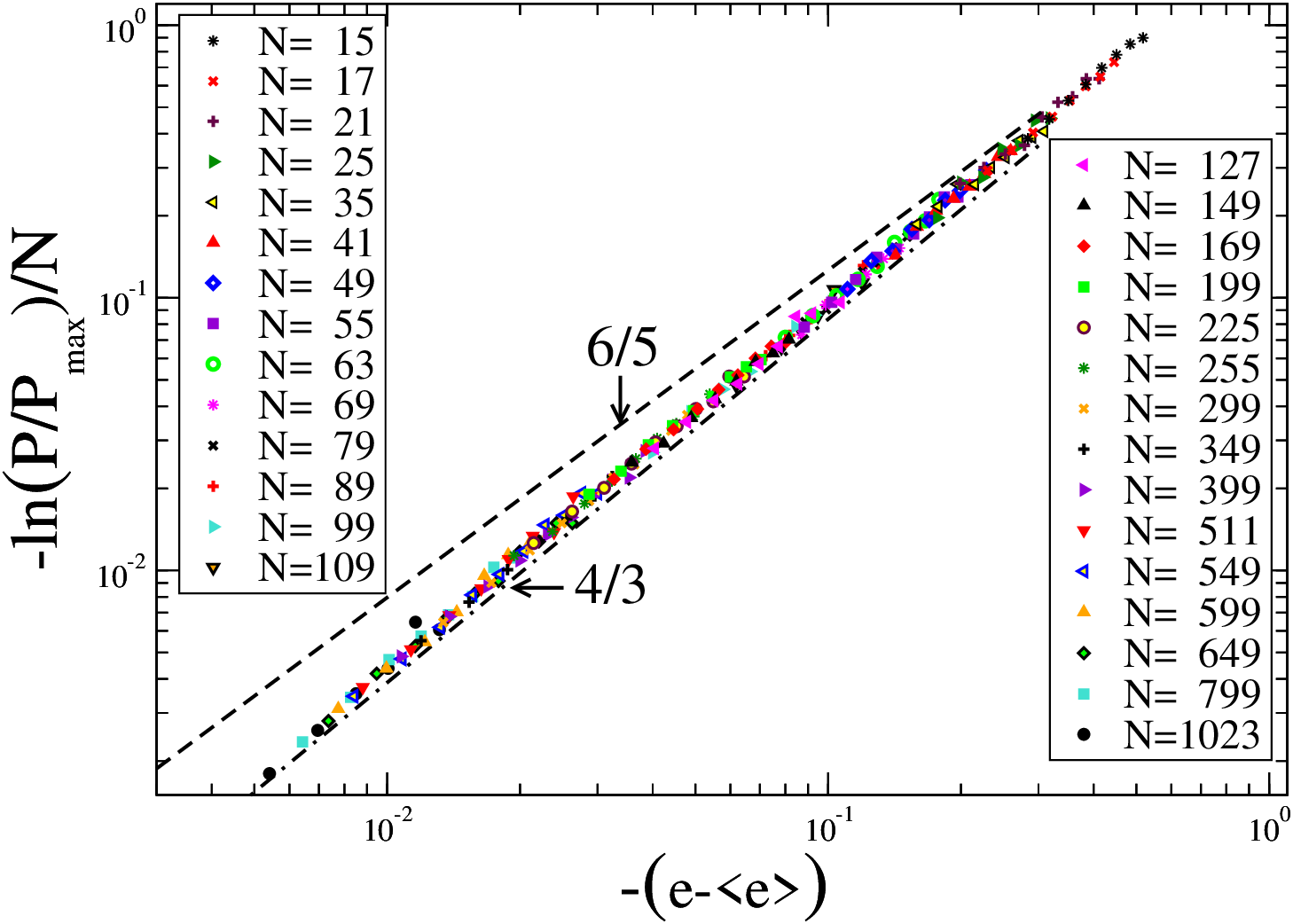}

\includegraphics[clip,scale=0.33]{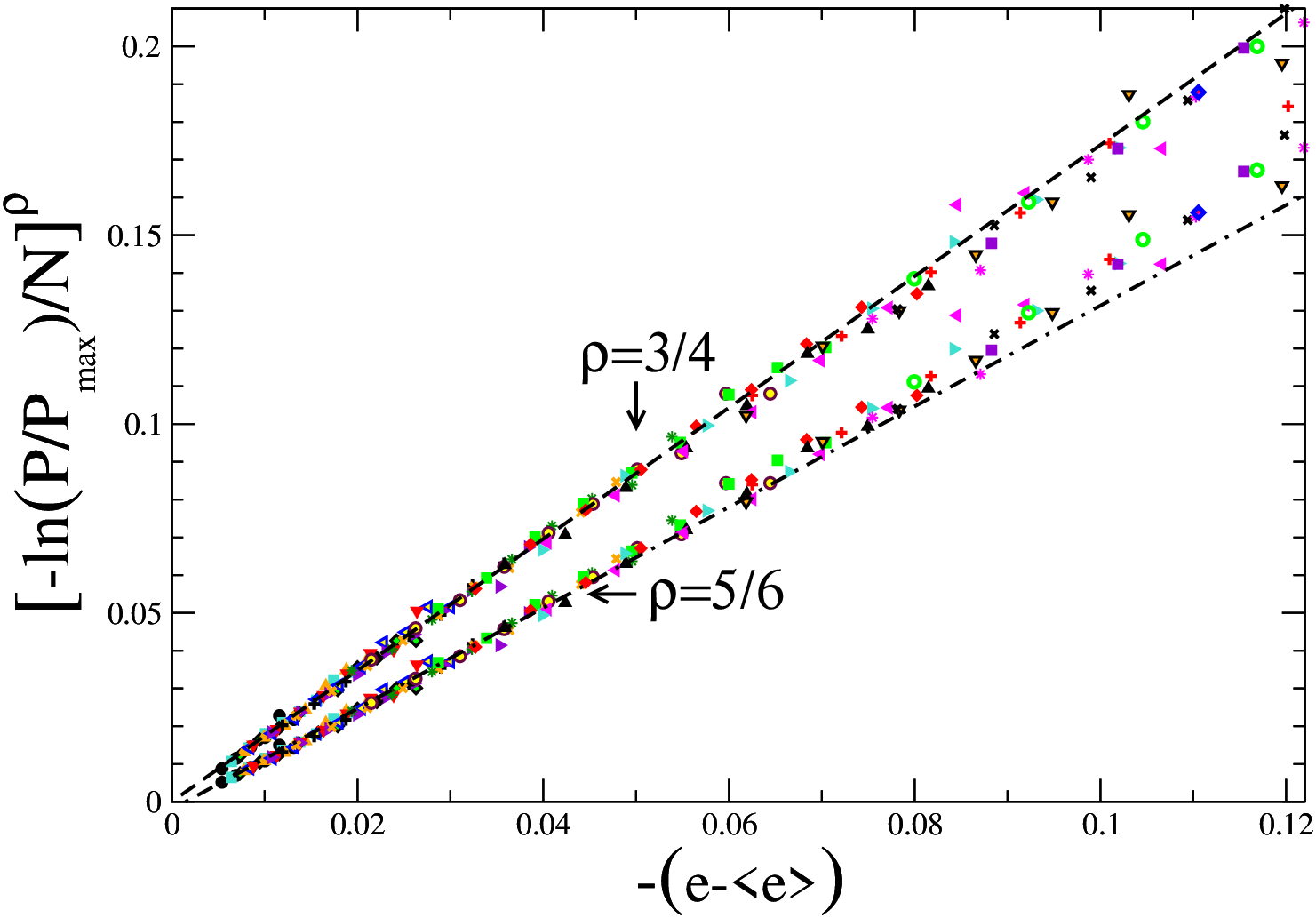}
\caption{\label{fig:ExtremeFlucNegative}
Data collapse on logarithmic scale of all SK data for energies
\emph{smaller} than the average from Fig.~\ref{fig:PDF_SK}. On top,
rescaled by an appropriately chosen $P_{{\rm max}}$, the data
collapses onto a power-law curve, which exhibits scaling over more
than a decade. (Transient data points with $P$ too close to $P_{{\rm
    max}}$ have been removed for clarity.) The collapse is most
consistent (see dash-dotted line) with $\rho^{-1}=\frac{4}{3}$ and
differs, although only slightly, from $\rho^{-1}=\frac{6}{5}$ (dashed
line). On the bottom, the same data (for $N>100$) is plotted on a
linear scale. In this form, $\rho=\frac{3}{4}$ is clearly more
consistent with linear scaling (dashed straight lines guide the eye).}
\end{figure}

\begin{figure}
\includegraphics[clip,scale=0.33]{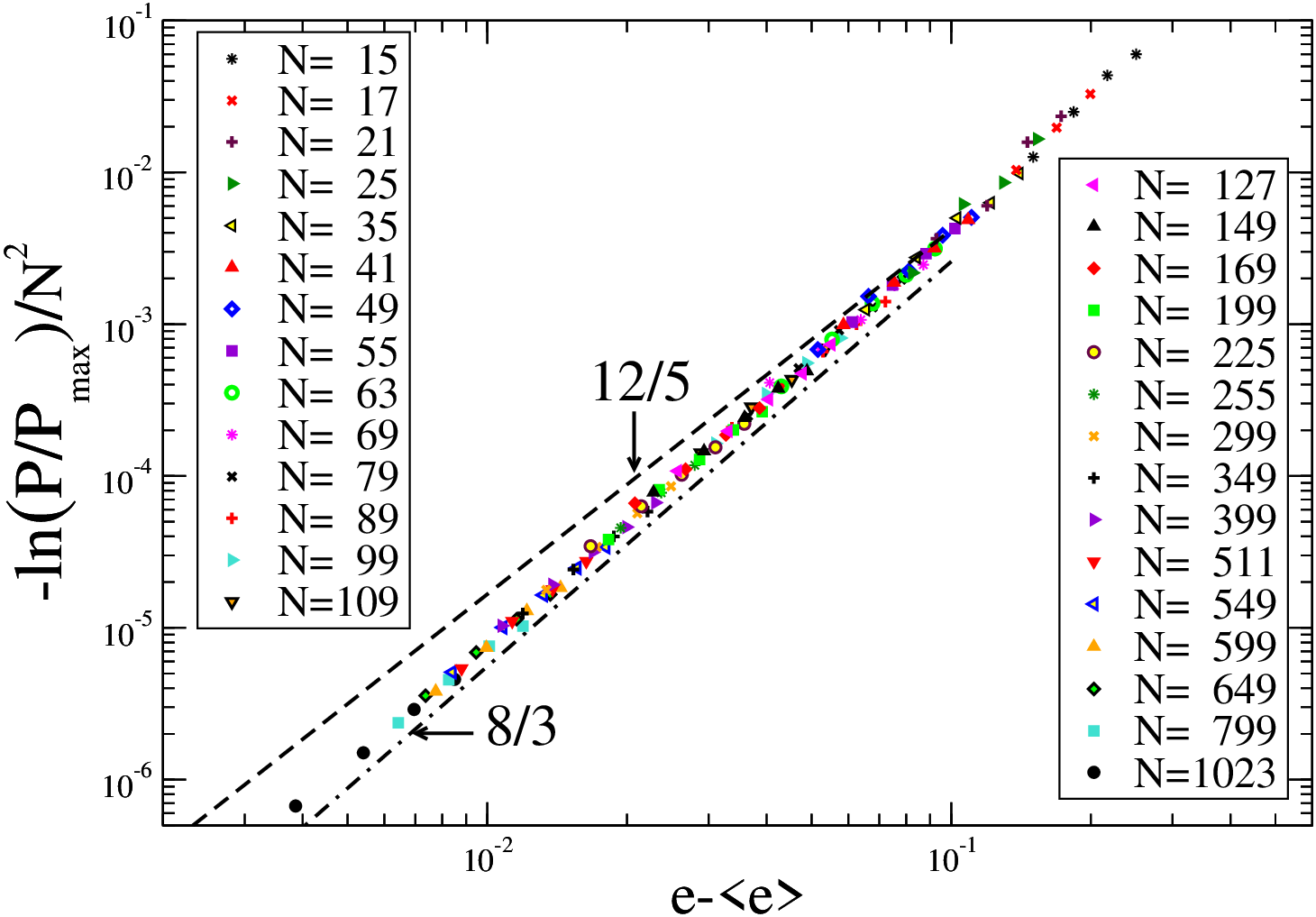}

\includegraphics[clip,scale=0.33]{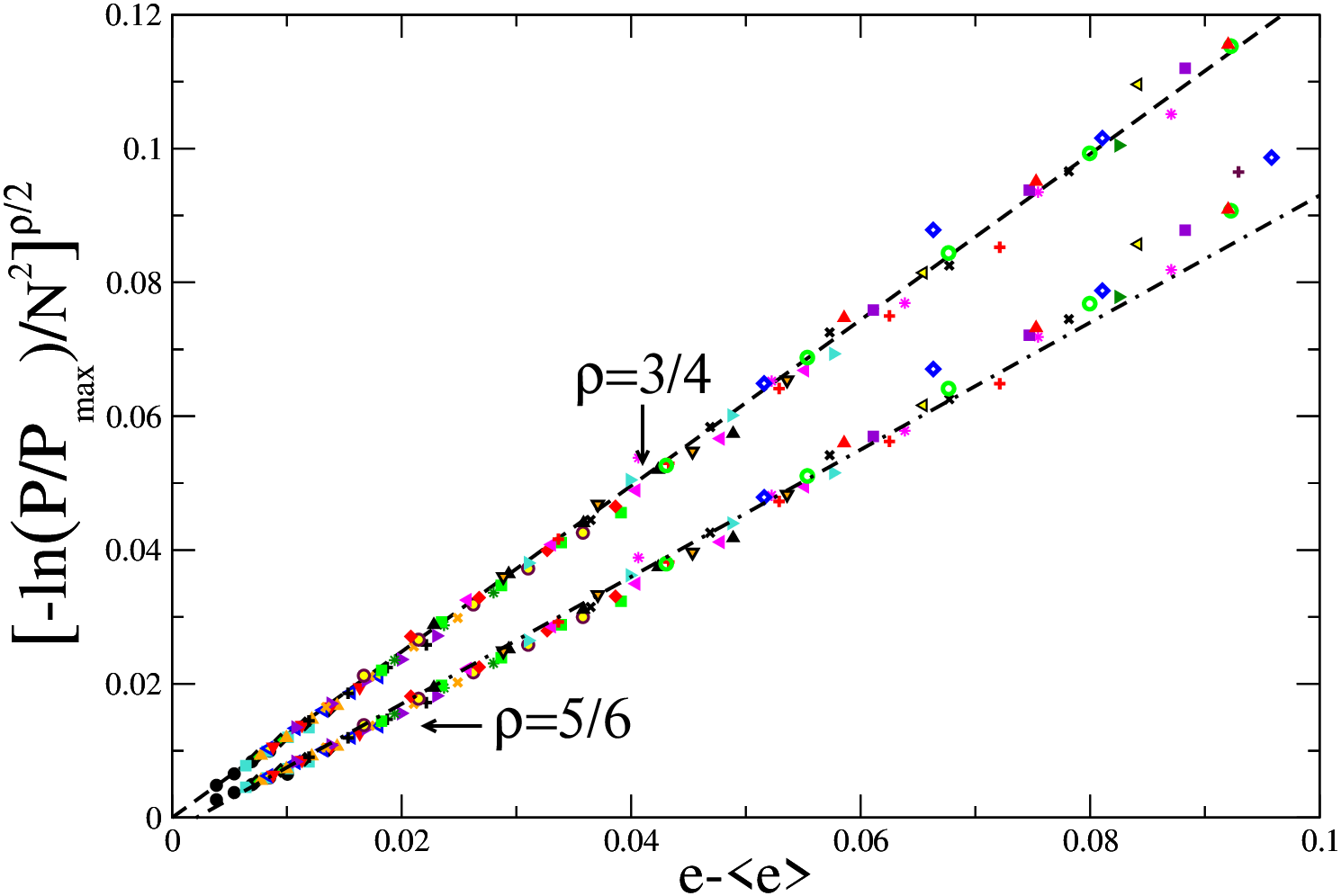}
\caption{\label{fig:ExtremeFlucPositive}
Data collapse on logarithmic scale of all SK data for energies
\emph{larger} than the average from Fig.~\ref{fig:PDF_SK}. Rescaled by
an appropriately chosen $P_{{\rm max}}$, the data collapses onto a
power-law curve, which exhibits scaling over more than a
decade. (Transient data points with $P$ too close to $P_{{\rm max}}$
have been removed for clarity.) The collapse is most consistent (see
dash-dotted line) with $\rho^{-2}=\frac{8}{3}$ and differs, again only
slightly, from $\rho^{-2}=\frac{12}{5}$ (dashed line). On the right,
the same data (for $N>50$) is plotted on a linear scale. In this form,
$\rho=\frac{3}{4}$ is clearly more consistent with linear scaling
(dashed straight lines guide the eye).}
\end{figure}

\subsection{Ground State Energy Fluctuations\label{sub:Ground-State-Energy-Fluctuations}}

Next, we consider the distribution of ground state energies around
their averages. In units of their standard deviation, the probability
density function (PDF) for each value of $N$ exhibits clearly the
asymmetric shape that is skewed towards a broader (exponential) tail
of instances with lower than average ground state energy and a cut-off
that is much sharper than exponential for those with higher energy.
This shape is largely unchanged across the sizes and can be shown, by
\emph{exhaustive} enumeration~\citep{Boettcher05e} of the
\emph{entire} ensemble for $N\leq9$, to arise already for very small
$N$. In Fig.~\ref{fig:PDF_SK} we show the PDF for all system sizes,
which demonstrates the skewness and the small variation of the shape
with $N$.  Yet, finite size effects larger than  for GBP in
Fig.~\ref{fig:CostFluc}  emerge deep in the tails of these PDFs.

Unlike the overall shape of the PDF for energy fluctuations, their
actual width, measured in terms of the standard deviation
$\sigma_{N}\left(e_{0}\right)$ in Eq.~(\ref{eq:rho}), varies in a
characteristic way with $N$.  A plot of the data for
$\sigma_{N}\left(e_{0}\right)$ in Tab.~\ref{Tab: AllData} on a
double-logarithmic scale in Fig.~\ref{fig:rho_fit} suggests a
power-law decay with $N$, but the data does not exhibit purely linear
behavior on this scale, as a simple fit reveals. Only when we restrict
to $N>100$, a fit of the data according to Eq.~(\ref{eq:rho}) with
just the leading ($\rho$-dependent) term provides satisfactory
results, with a value of $\rho$ just above $\frac{3}{4}$. A more
consistent fit of all the data is provided when higher-order
corrections are considered. This can only succeed for a reasonable,
fixed value of $\rho$, we indeed accomplish almost identical fits of
this sort for either $\rho=\frac{3}{4}$ or $\rho=\frac{5}{6}$ (and
probably any nearby value), see Fig.~\ref{fig:rho_fit}. In this
regard, the fit for fixed $\rho=\frac{5}{6}$ has the added benefit
that the higher-order term appears to scale with $a\approx1$, a likely
candidate for a next-order correction. But if leading and next-order
correction are that close, for instance $N^{-\frac{5}{6}}/N^{-1}\sim
N^{\frac{1}{6}}$, to obtain the asymptotic scaling of $\rho$ separated
by a decade from any transient behavior would require results for
$N\gtrsim10^{6}$.  Thus, with the present data, a conspiracy between
such terms leading to the observed behavior could not be excluded.

To illustrate the difficulty more clearly, we follow the procedure for
GBP in Sec.~\ref{fig:sigma_extra} and extrapolate for $\rho$ according
to Eq.~(\ref{eq:rhoextra}). The variables $y=-\log\sigma/\log N$
plotted vs $x=1/\log N\to0$ should provide an asymptotically linear
extrapolation (with exponentially small corrections $\sim
xe^{-(a-\rho)/x}$, if $a>\rho$) towards the exponent $\rho$ at the
$y$-intercept for $N\to\infty$. Plotting the SK data up to
$N\approx1000$ in this fashion in Fig.~\ref{fig:rho_extrapolation}
again indicates a value just above $\rho=\frac{3}{4}$ and apparently
far below $\rho=\frac{5}{6}$.  But unlike the GBP data in
Fig.~\ref{fig:sigma_extra}, the data for SK has still transient
features even for such large values of $N$.  Only a non-linear fit
according to all three terms in Eq.~(\ref{eq:rhoextra}), but taking an
already \emph{fixed} $\rho=\frac{5}{6}$ as given, makes such a high
value for $\rho$ plausible. Further support for such a higher value of
$\rho$ is provided by the following study of spin glasses on Bethe
lattices. On the other hand, the GBP example above and the result for
the $m$-vector model of $\rho=\frac{4}{5}$~\citep{Aspelmeier10} would
suggest that an altogether different value of $\rho$ between these two
rational values is conceivable. In fact, a purely linear extrapolation
in Fig.~\ref{fig:rho_extrapolation} and the fit in
Fig.~\ref{fig:rho_fit} for $N\gtrapprox100$ would lead to an
asymptotic value for $\rho$ very close to that of GBP in
Sec.~\ref{sec:Graph-Bi-Partitioning} above.

\subsection{Extreme Fluctuations\label{sub:Extreme-Fluctuations}}

Considering the large amount of data we have obtained for SK, we can
inspect further details of the energy fluctuations. In particular, we
can look deeper into the tails of the PDFs displayed in
Fig.~\ref{fig:PDF_SK}, where they are rescaled by their respective
$\sigma_{N}$. Here, we treat these PDFs unrescaled, according to the
form proposed in Ref.~\citep{Rizzo09}, suggested by the spherical
spin glass~\citep{Dean06}.  In Ref.~\citep{Rizzo09}, it was argued
that for ground state energies lower than the average, the
corresponding branch far in the negative tail of each PDF falls
exponentially with an argument proportional in $N$, while
configurations with larger energies are particularly rare for larger
system sizes such that the positive tail is suppressed by a factor
$N^{2}$. This system-size dependence is largely lost when each PDF is
rescaled by its width $\sigma_{N}$ in Fig.~\ref{fig:PDF_SK}.

In Figs.~\ref{fig:ExtremeFlucNegative}
and~\ref{fig:ExtremeFlucPositive}, we extract the argument of the
exponential tails and plot the data for each tail reduced by the
indicated power of $N$. In the process, the PDF-data $P(e)$ for each
system size has to be gauged by an arbitrary reference point $P_{{\rm
    max}}$ and transient behavior too close to the average or
statistically deficient data too deep in the tails has to be
discarded. The resulting collapse of all the intermediate data onto a
power-law function is presented in the upper panel of each figure on a
logarithmic scale. According to Ref.~\citep{Rizzo09}, the power-law
exponent can be interpreted as $\rho^{-1}$ for $e\ll\langle e\rangle$
and as $\rho^{-2}$ for $e\gg\langle e\rangle$. To provide a reference,
we have include lines corresponding to $\rho=\frac{3}{4}$ and
$\rho=\frac{5}{6}$ in these plots. While the differences are again
slim (and it could be argued that true asymptotic behavior has not
been reached), the consistent scaling collapse of this vast amount of
data seems to favor a value closer to $\rho=\frac{3}{4}$
again. Furthermore, both tails independently exhibit similar scaling behavior.

When viewed on a linear scale, by taking the respective power, only a
value closer to $\rho=\frac{3}{4}$ provides consistent linear behavior
for the extant data (see the lower panel of
Figs.~\ref{fig:ExtremeFlucNegative}
and~\ref{fig:ExtremeFlucPositive}). In Ref.~\citep{Rizzo09}, a similar
linear plot was provided (for the positive branch only) in which the
largest system size considered there, $N=150$, was judged consistent
with $\rho=\frac{5}{6}$. It is clear from our direct comparison here,
that even with the vast amount of additional data an ultimately
conclusive decision on the true value of $\rho$ is elusive here. The
comparison also shows that an analysis of these tails on a logarithmic
scale is favorable over a linear scale which squashes the most
interesting data points for larger system sizes. Even on a logarithmic
scale, though, it is not easy to extract the relevant asymptotic
information as ever deeper in the tails, only ever smaller-sized
systems contribute.  But overall, in this data, small-sized and
large-sized systems seem to follow similar scaling and project a
self-consistent picture.

\section{Spin Glasses on Bethe Lattices\label{sec:Spin-Glasses-on-Bethe}}

To provide a new perspective on the ground-state energy fluctuations
in SK, we revisit spin glasses with $\pm J$-bonds on Bethe lattices
(SG), in particular, on those of degree $r=3$. A similar study has
been undertaken in Ref.~\citep{Boettcher03a,Boettcher03b}, which
concerned thermodynamic averages of ground state energies and
entropies.  Here, we extend the sampling of ground state energies to
measure the PDF of ground state energy fluctuations and the scaling of
their width.  To that end, in Tab.~\ref{Tab: BL3Data} we have added a
large number of instances at each system size, up to $N=4096$. Unlike
for Bethe lattices of higher degree, at degree three we can utilize
exact methods~\citep{Boettcher04b,Boettcher04c,Boettcher08a} to reduce
the number of variables in the optimization problem by about 42\%,
hence, making larger system sizes accessible at sufficient statistics.

\begin{table}
\caption{\label{Tab: BL3Data}
List of all the data obtained with EO for Bethe lattices of degree
$r=3$ for various system sizes $N$. Given are the number of instances
$n_{I}$ considered at each $N$, and the average ground-state energy
density $\left\langle e_{0}\right\rangle _{N}$ and  standard
deviation $\sigma(e_{0})$ over these instances.}
\begin{tabular}{r|r|l|l}
\hline 
$N$  & $n_{I}$  & $\langle e_{0}\rangle_{N}$  & $\sigma\left(e_{0}\right)$\tabularnewline
\hline 
16  & 1000000  & -1.1467(1)  & 0.10177\tabularnewline
32  & 1000000  & -1.19375(6)  & 0.06003 \tabularnewline
44  & 1680320  & -1.20895(4)  & 0.04695\tabularnewline
64  & 1000000  & -1.22314(4)  & 0.03515 \tabularnewline
80  & 1486688  & -1.23005(2)  & 0.02952 \tabularnewline
128  & 2000000  & -1.24153(1)  & 0.02043 \tabularnewline
160  & 3273585  & -1.24583(1)  & 0.01714\tabularnewline
256  & 1498807  & -1.25296(1)  & 0.01181 \tabularnewline
350  & 4015909  & -1.25660(1)  & 0.00921 \tabularnewline
512  & 7638942  & -1.26007(1)  & 0.00680 \tabularnewline
750  & 1718511  & -1.26274(1)  & 0.00501\tabularnewline
1024  & 743404  & -1.26444(1)  & 0.00390 \tabularnewline
2048  & 113389  & -1.26710(1)  & 0.00226 \tabularnewline
\hline
\end{tabular}
\end{table}

\subsection{Average Ground State Energy\label{sec:Ground-State-EnergyBL3}}

\begin{figure}
\includegraphics[clip,scale=0.33]{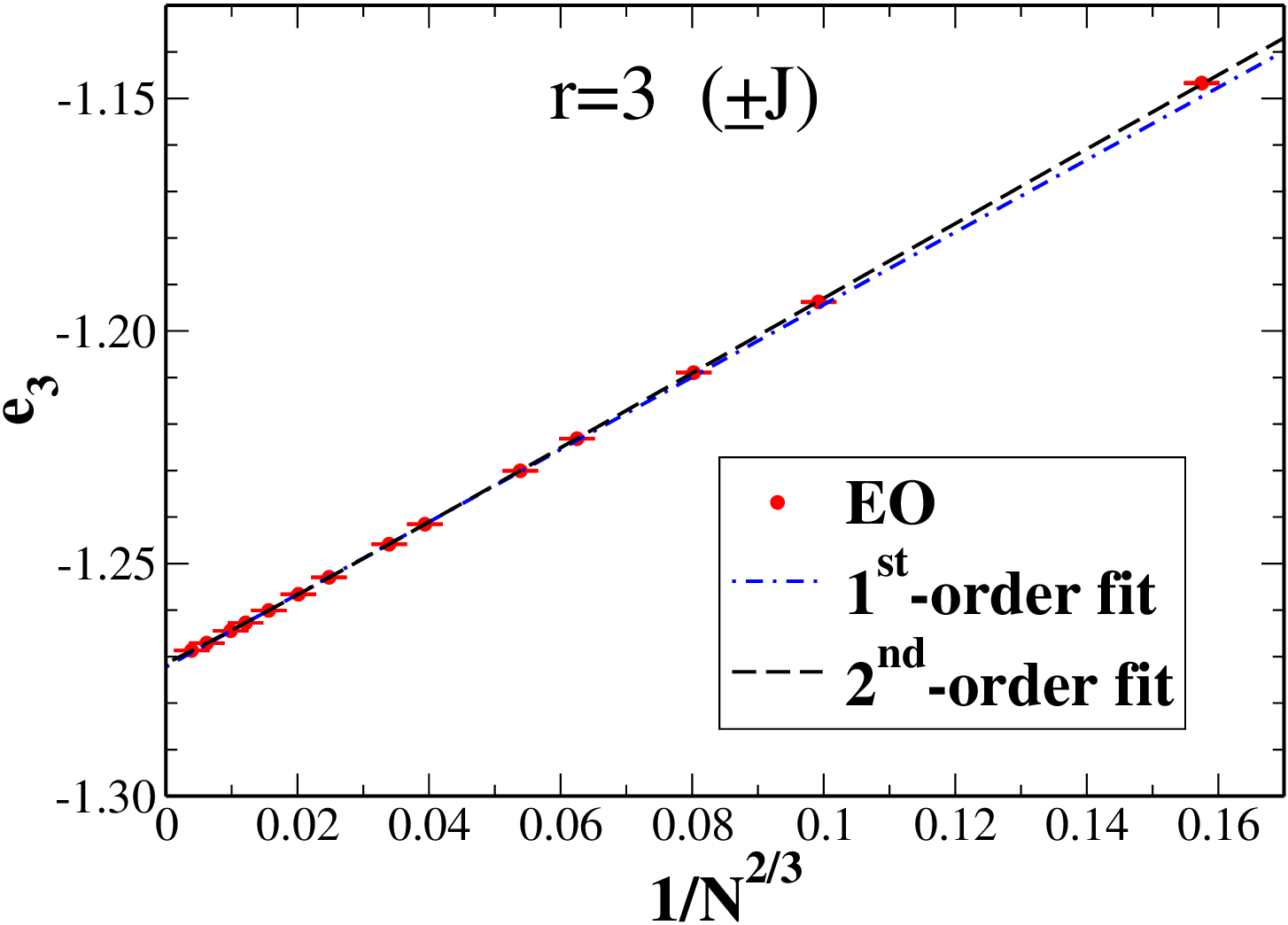}
\caption{\label{fig:ExtrapolationBethe3}
Extrapolation plot of the average ground-state energy densities
$\left\langle e_{0}\right\rangle _{N}$ for Bethe lattices of degree
$r=3$ as a function of the presumed finite-size corrections,
$1/N^{\frac{2}{3}}$. As in Fig.~\ref{fig:ExtrapolationE}, the
statistical errors indicated are much smaller than symbol sizes.  For
$N\to\infty$, irrespective of the order of the fit, the data
extrapolates to $\left\langle e_{0}\right\rangle _{\infty}=-1.2715(1)$
at the intercept. }
\end{figure}

\begin{figure}
\includegraphics[clip,scale=0.33]{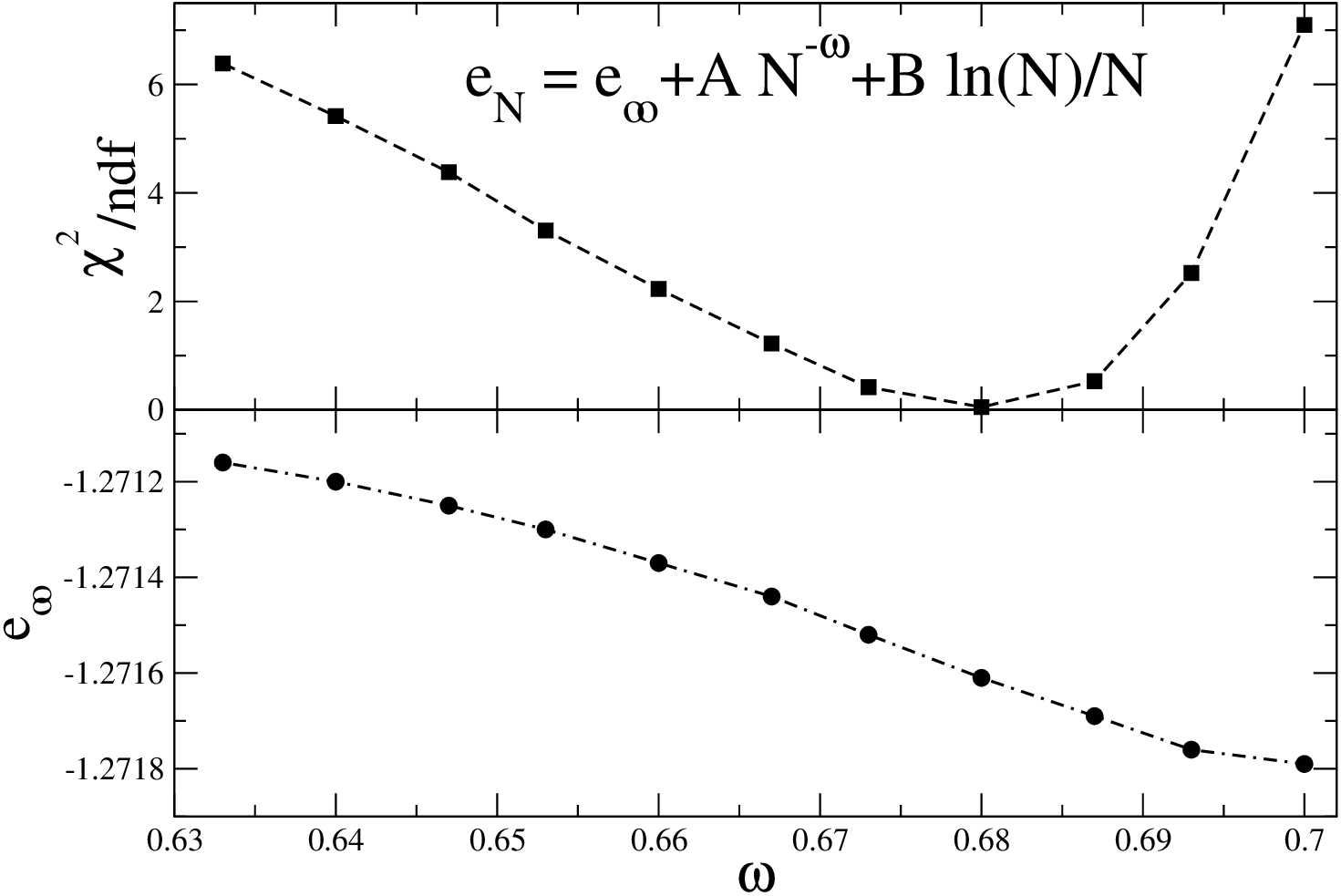}
\caption{\label{fig:Qtests-BL3}
Second-order fit over a range of fixed $\omega$, with logarithmic
corrections, to the average ground state energies $\left\langle
e_{0}\right\rangle _{N}$ for spin glasses on Bethe lattices of degree
$r=3$ in Tab.~\ref{Tab: BL3Data}. }
\end{figure}

In Fig.~\ref{fig:ExtrapolationBethe3}, we present the average energy
densities obtained with EO at the system sizes simulated. As in
Ref.~\citep{Boettcher03a}, the extrapolation of the data is virtually
linear when plotted as function of $N^{-\frac{2}{3}}$. Such a linear
extrapolation yields $\left\langle e_{3}\right\rangle
_{\infty}=-1.2715(1)$ for the thermodynamic energy density, consistent
with the value determined in Ref.~\citep{Boettcher03a} and consistent
with the one-step replica-symmetry breaking result reported in
Refs.~\citep{Mezard03,Mezard06}. Remarkably, an attempt at adding a
higher-order correction term contrasts with the same discussion for
SK. Neither of the two types of higher-order fits presented in
Fig.~\ref{fig:Qtests} provide reasonable results here.  In turn, a fit
to Eq.~(\ref{eq:logscaling}), which failed for SK, does converge on
this data, see Fig.~\ref{fig:Qtests-BL3}.  Across the plotted regime,
the thermodynamic value for $e_{\infty}$ remains quite robust. The
optimum is rather close to $\omega=\frac{2}{3}$; such a
2\emph{nd}-order fit including $\omega$ also shown in
Fig.~\ref{fig:ExtrapolationBethe3} converges to $\omega\approx0.677$.

\begin{figure}
\includegraphics[clip,scale=0.33]{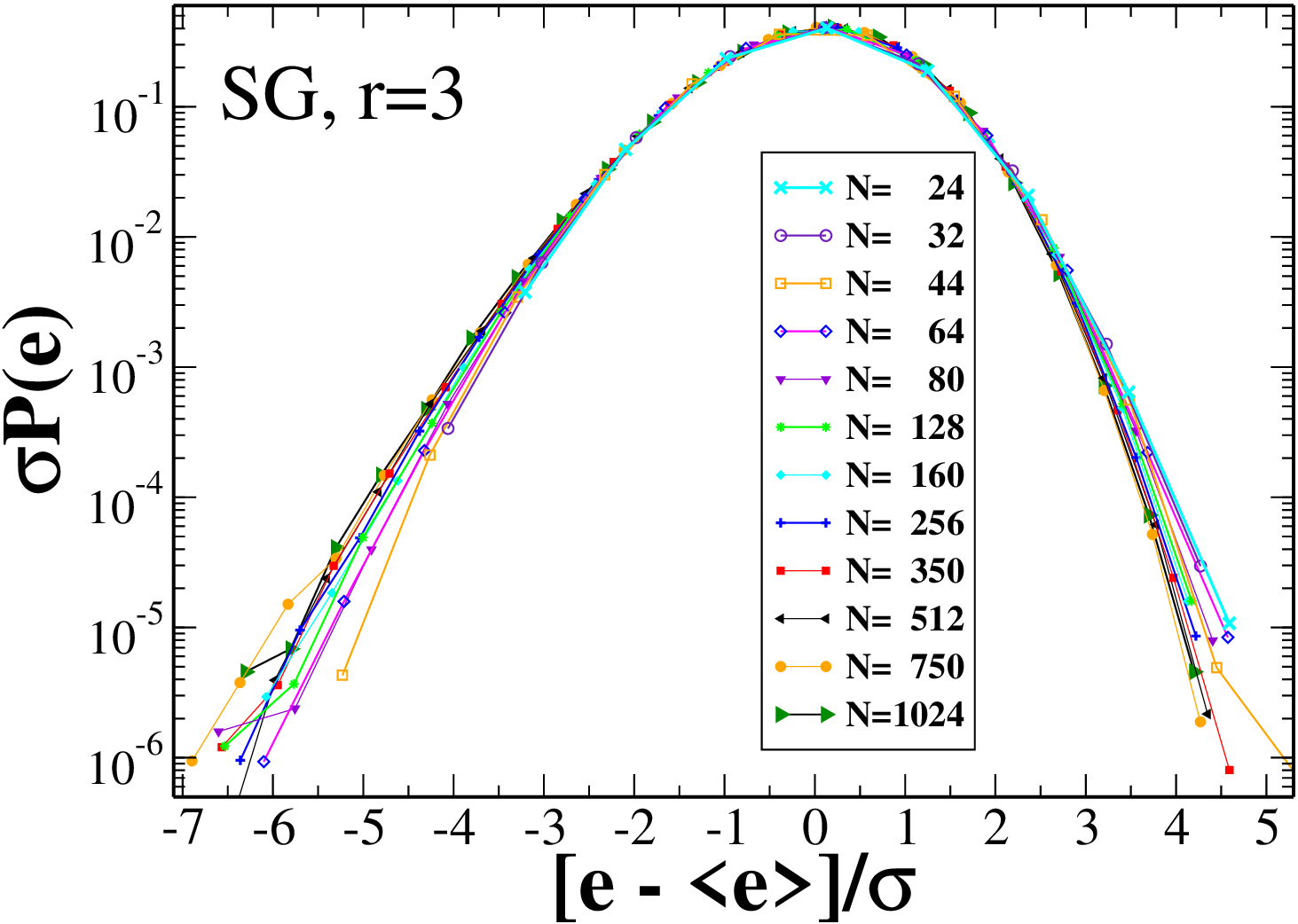}
\caption{\label{fig:PDF_Bethe3}
Plot of the probability density functions (PDF) of obtained
ground-state energy densities $e_{0}$ for the spin glass on a Bethe
lattice of degree $r=3$ in units of the standard deviation
$\sigma$. Here, in comparison with Figs.~\ref{fig:CostFluc} and
~\ref{fig:PDF_SK}, significant finite-size effects are detectable in
both tails. }
\end{figure}

\subsection{Ground State Energy Fluctuations\label{sub:Ground-State-Energy-FluctuationsBL3}}

In Fig.~\ref{fig:PDF_Bethe3}, we show the probability density
functions (PDF) of the energy densities around those
averages. Overall, those PDFs are a bit more symmetrical than for SK
in Fig.~\ref{fig:PDF_SK}, but exhibit even more finite-size effects in
both tails, especially in comparison with the corresponding PDFs of
GBP in Fig.~\ref{fig:CostFluc}.  Despite their more symmetrical
appearance, the scaling  with $N$ of the deviations $\sigma$ listed in
Tab.~\ref{Tab: BL3Data} seems to indicate an even higher
value of $\rho$, as Fig.~\ref{fig:rho_extrapolation} suggests. There,
those $\sigma$ for the Bethe lattice are displayed in an extrapolation
plot together with that of SK, to highlight their similarity. This
data is somewhat smoother than for SK, but just as much beset with
transients. A family of extrapolants for each degree $r$ seems
conceivable, reaching all the way to the SK-limit at $r=\infty$.  In
parallel with the discussion for SK in
Sec.~\ref{sub:Ground-State-Energy-Fluctuations}, we can at best argue
that $\rho=\frac{5}{6}$ is consistent with the trend of the
extrapolation. In this plot, a value of $\rho=\frac{3}{4}$ or even
that from GBP seems to be ruled out by that trend.

\subsection{Extreme Fluctuations\label{sub:Extreme-FluctuationsBL3}}

\begin{figure}
\includegraphics[clip,scale=0.33]{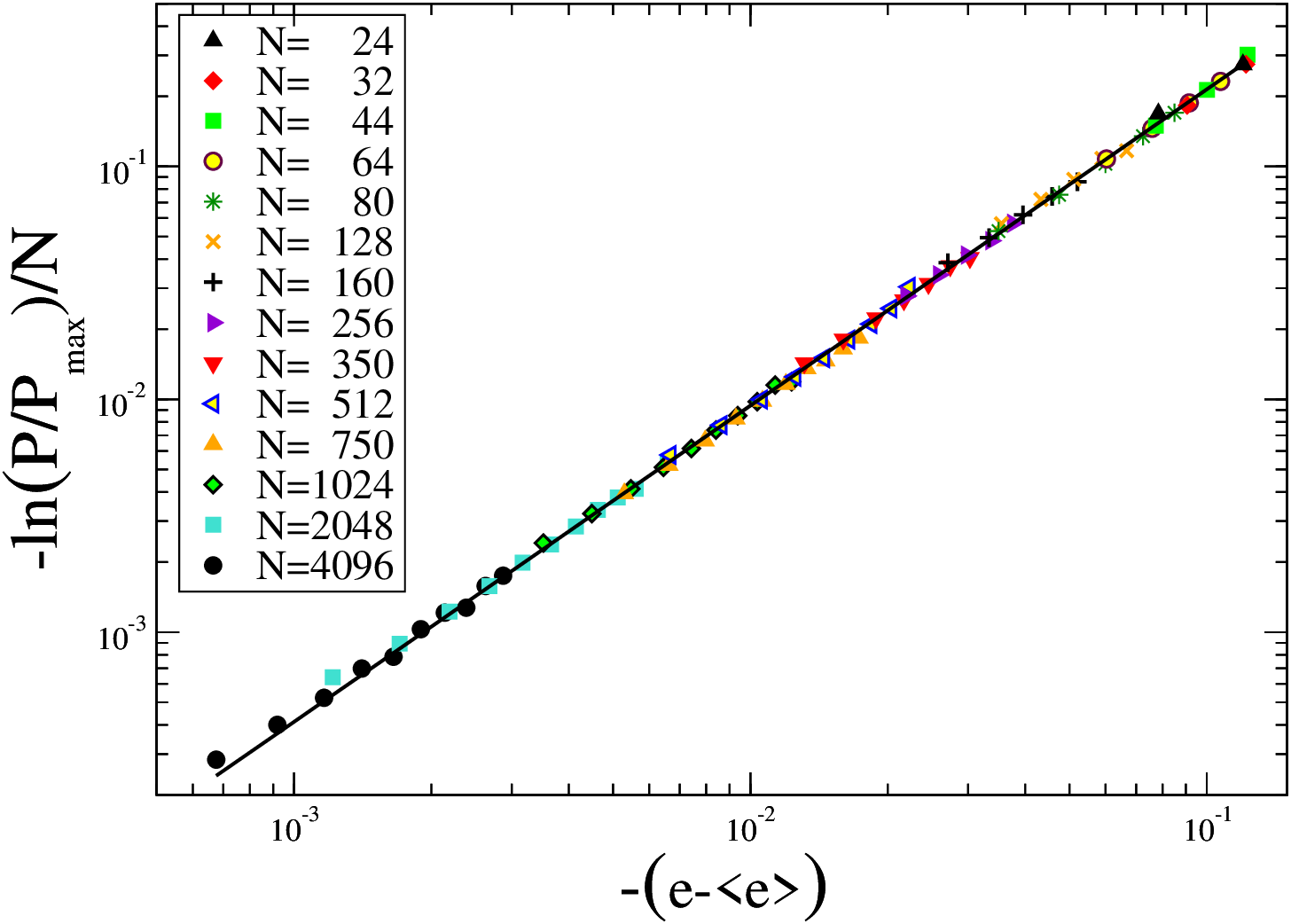}

\includegraphics[clip,scale=0.33]{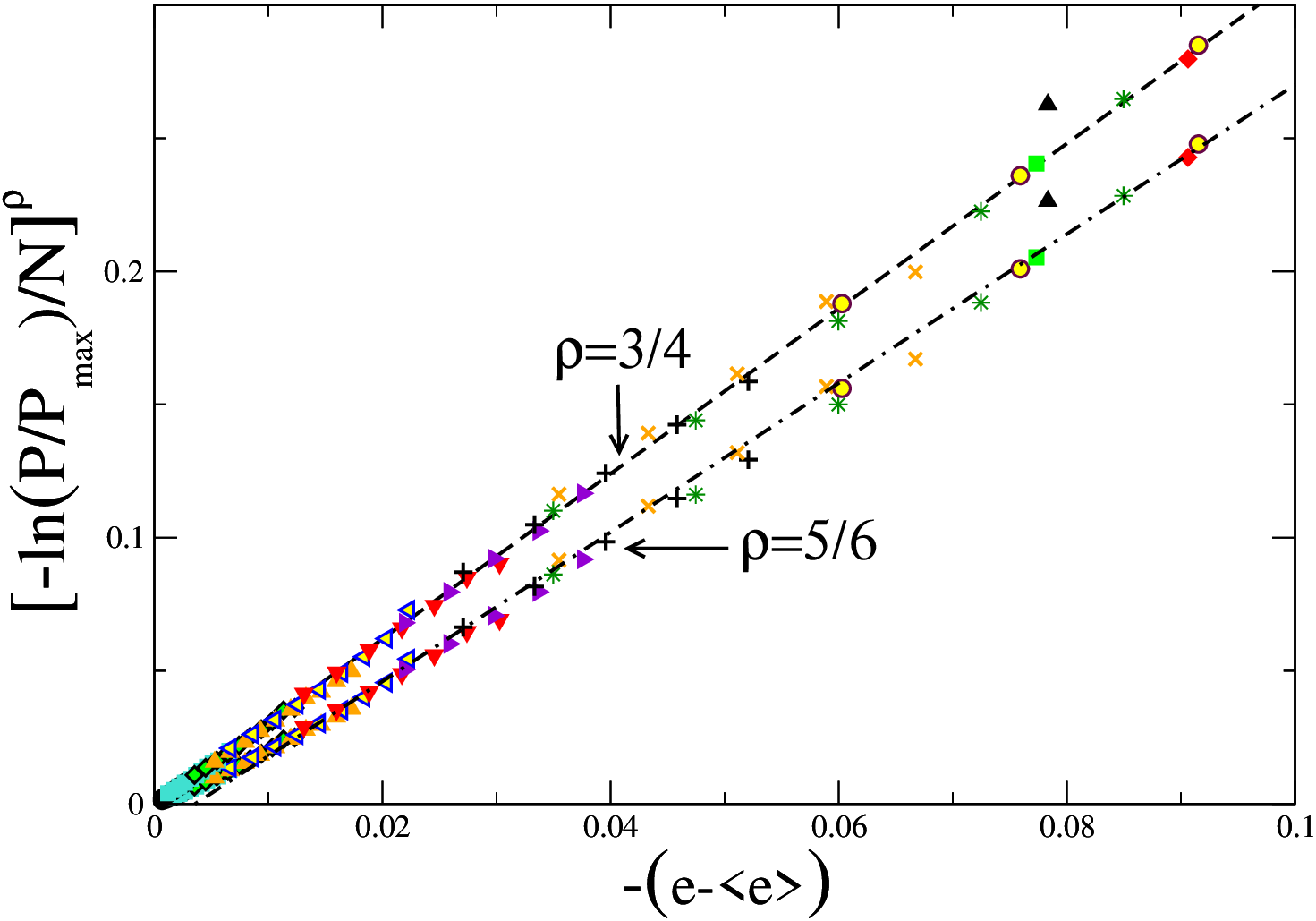}
\caption{\label{fig:ExtremeFlucNegativeBL3}
Data collapse on logarithmic scale of all Bethe lattice data for
energies \emph{smaller} than the average from
Fig.~\ref{fig:PDF_Bethe3}. Top, rescaled by an appropriately
chosen $P_{{\rm max}}$, the data collapses onto a power-law curve,
which exhibits scaling over more than a decade. The fit (straight
line) gives an exponent of $\rho^{-1}\approx1.36$. Bottom, the
same data is plotted on a linear scale. In this form,
$\rho=\frac{3}{4}$ is more consistent with linear scaling (dashed
straight lines guide the eye).}
\end{figure}

\begin{figure}
\includegraphics[clip,scale=0.33]{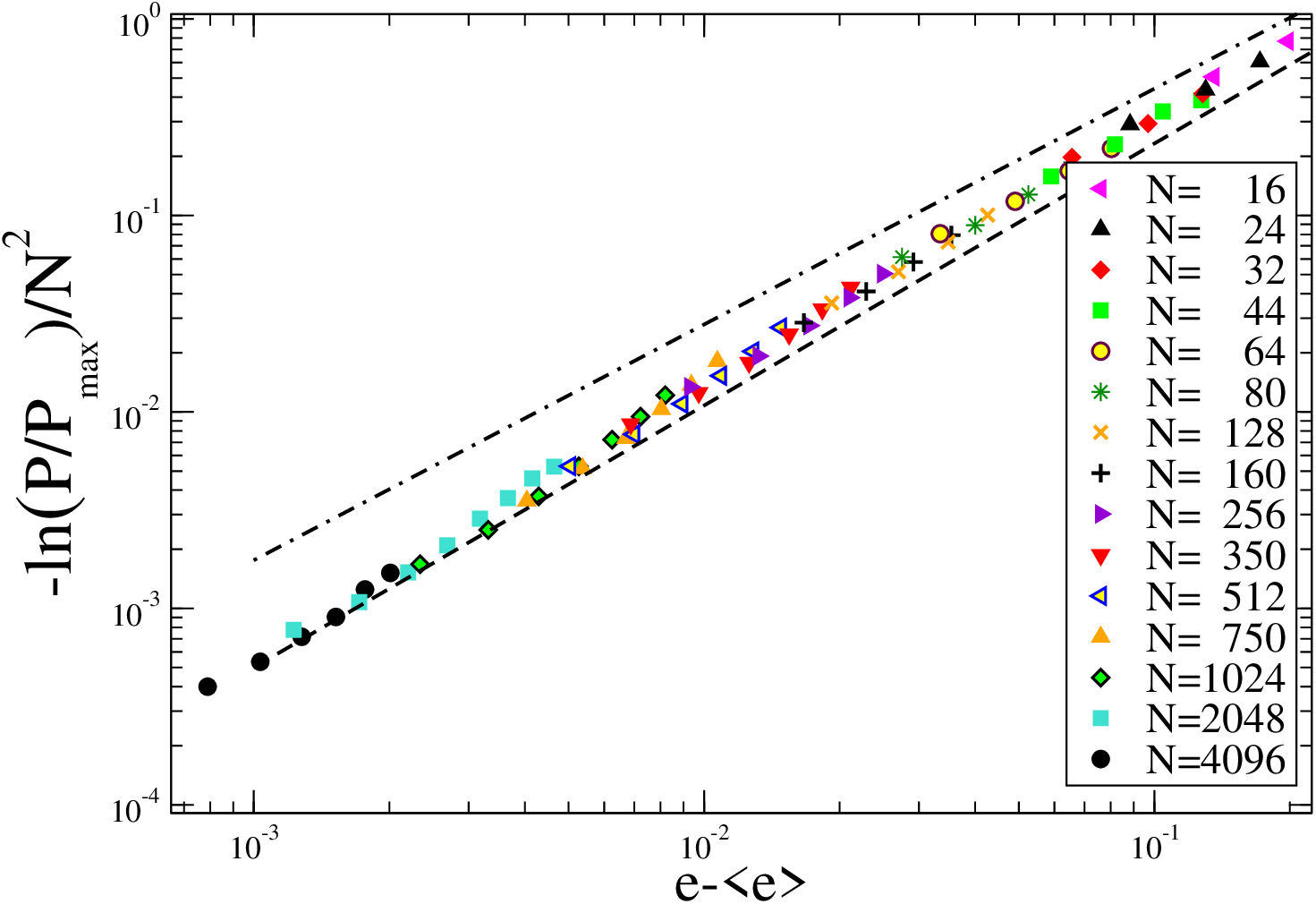}

\includegraphics[clip,scale=0.33]{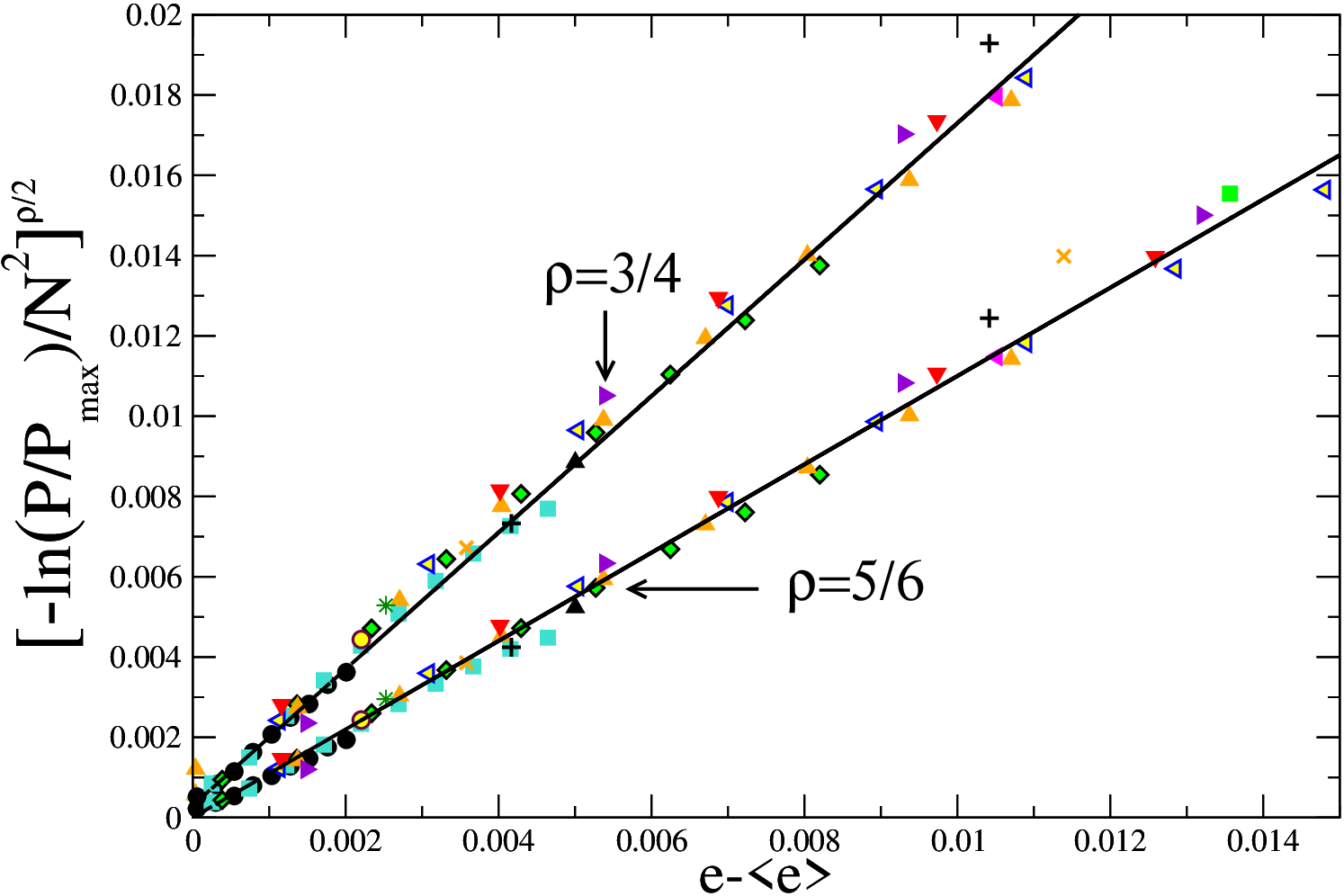}
\caption{\label{fig:ExtremeFlucPositiveBL3}
Data collapse on a logarithmic scale (top) of all Bethe lattice data
for energies \emph{larger} than the average from
Fig.~\ref{fig:PDF_Bethe3}. Rescaled by an appropriately chosen
$P_{{\rm max}}$, the data collapses onto a power-law curve, which
exhibits scaling over a decades. Bottom, the same data is plotted on a
linear scale. Here, both set of data are somewhat consistent with
linear scaling, but the situation is clear than in
Fig.~\ref{fig:ExtremeFlucNegativeBL3} (dashed straight lines guide the
eye).}
\end{figure}

We have attempted a detailed analysis of the tails of the fluctuations
for Bethe lattices with the identical approach as conducted in
Sec.~\ref{sub:Extreme-Fluctuations} for SK. The results shown in
Figs.~\ref{fig:ExtremeFlucNegativeBL3}-\ref{fig:ExtremeFlucPositiveBL3}
are indistinguishable from those for SK above, and would also suggest a
value closer to $\rho=\frac{3}{4}$ for the Bethe lattice, which seems
to contradict the indication provided by the extrapolation of $\sigma$
in Fig.~\ref{fig:rho_extrapolation}.

\section{Spin Glasses on Random Graphs\label{sec:Spin-Glasses-RG}}

As a useful reference point to the previous studies, we also include a
comparison with a spin glass on sparse, ordinary random graphs of mean
degree $C=2$. It provides an example where fluctuations in the ground
states, whether the energy or the cost, appear to converge to a normal
distribution. As argued in Sec.~\ref{data}, it is essential for the
case of a fluctuating geometry to focus on the actual cost, i.~e. the
total absolute weight of the violated bonds, of the ground state.
Still, even the PDF for these ground state costs seems asymptotically
Gaussian, as has been predicted recently in Ref.~\citep{Rizzo09b}.

\subsection{Average Ground State Costs\label{sub:Average-Ground-State Cost}}

In Tab.~\ref{tab:RG2Data}, we have listed the average ground state
costs and their deviations for a number of system sizes up to
$N=4096$.  A large number of instances has been averaged over, even at
the largest sizes, since the exact graph reduction
methods~\citep{Boettcher04b,Boettcher04c,Boettcher08a} used for the
Bethe lattice above are even more effective here: Even at the largest
size, those reductions result in graphs of at most $15\%$ of the
original size that need to be optimized with the EO heuristic.

\begin{figure}
\includegraphics[bb=8bp 22bp 706bp 523bp,clip,scale=0.33]{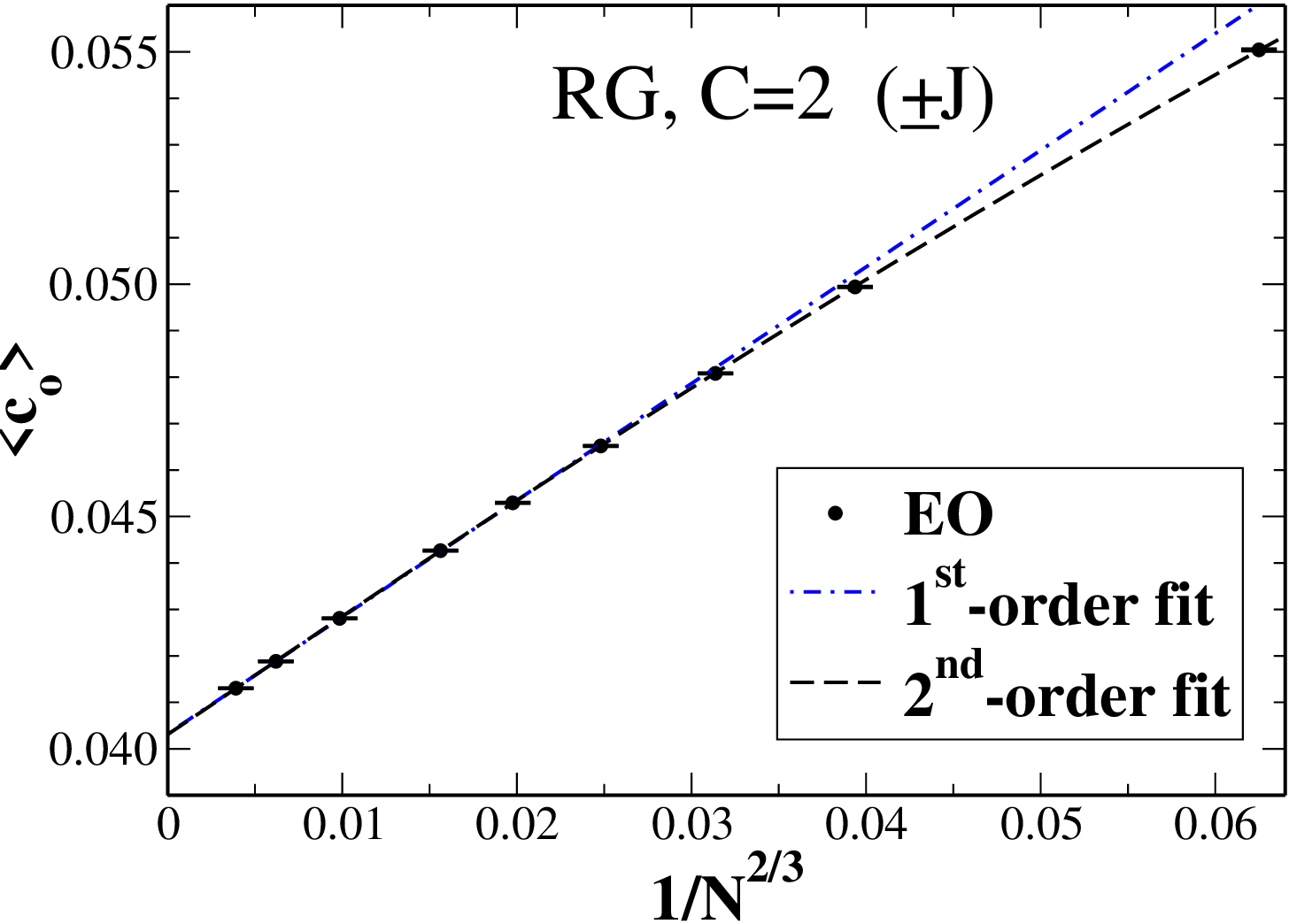}
\caption{\label{fig:ExtrapolationRG2}
Extrapolation plot of the average ground-state cost densities
$\left\langle c_{0}\right\rangle _{N}$ for ordinary random graphs of
average degree $C=2$ as a function of the presumed finite-size
corrections, $1/N^{\frac{2}{3}}$. As in Fig.~\ref{fig:ExtrapolationE},
the statistical errors indicated are much smaller than symbol sizes.
Shown are also a first-order (blue dash-dotted line) and a
second-order fit (black dashed line) in powers of
$1/N^{\frac{2}{3}}$. For $N\to\infty$, the data extrapolates to
$\left\langle c_{0}\right\rangle _{\infty}=0.04030(5)$ at the
intercept. }
\end{figure}

\begin{figure}
\includegraphics[clip,scale=0.33]{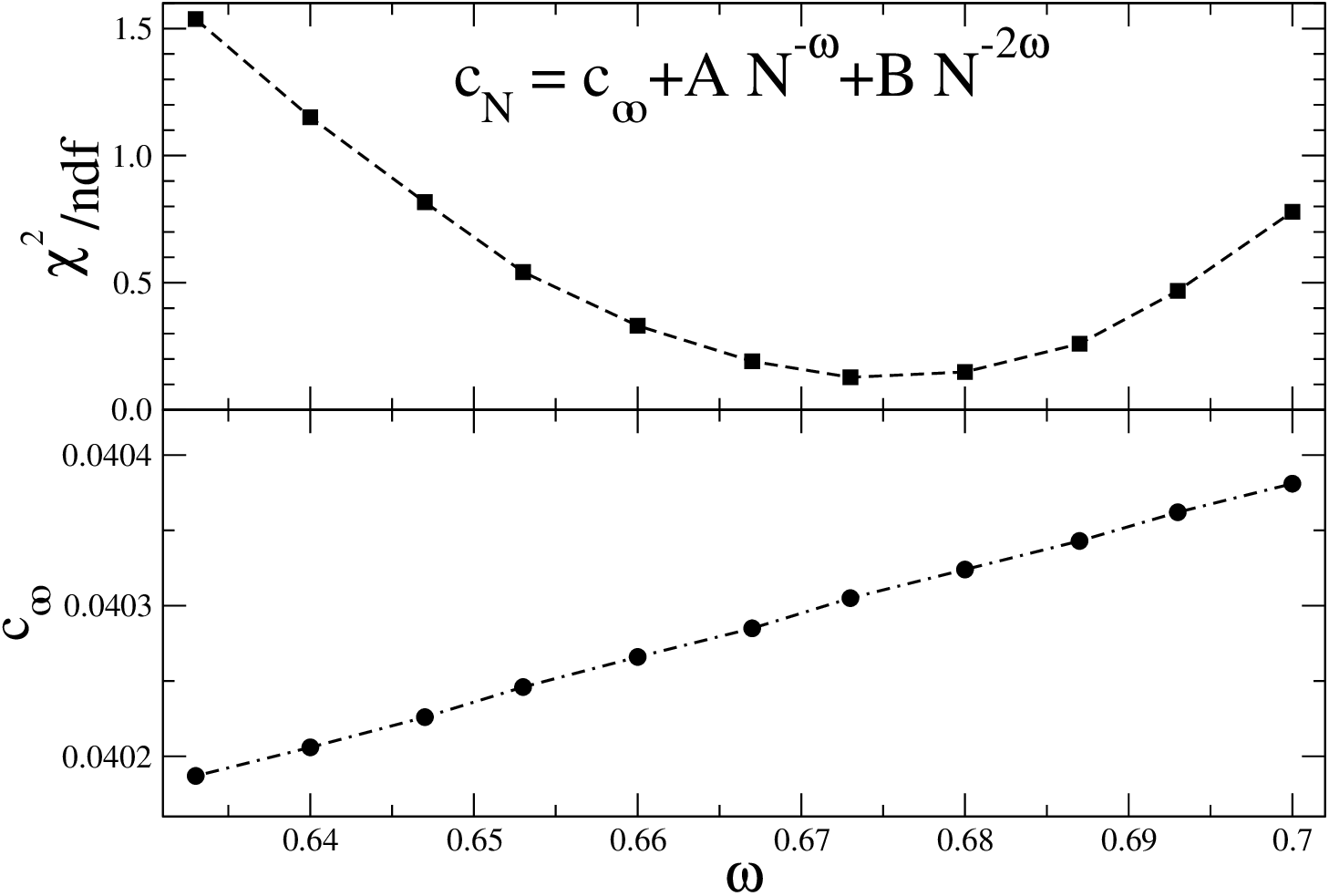}
\caption{\label{fig:Qtests-RG2}
Second-order fit in powers of $N^{-\omega}$ over a range of fixed
$\omega$, to the average ground state costs $\left\langle
c_{0}\right\rangle _{N}$ for spin glasses on random graphs of degree 2
in Tab.~\ref{tab:RG2Data}. The minimum in $\chi^{2}/ndf$ in the upper
plot strongly suggests a pure power series with $\omega=\frac{2}{3}$;
such a fit is included in Fig.~\ref{fig:ExtrapolationRG2}. The lower
panel shows the range of extrapolated values in the thermodynamic cost
density.}
\end{figure}

In Fig.~\ref{fig:ExtrapolationRG2}, we plot the extrapolation of those
average ground state costs to the thermodynamic limit. Again, the
extrapolation proves most consistent with $N^{-\frac{2}{3}}$
corrections at finite size, although stronger transients are apparent
here. In fact, Fig.~\ref{fig:Qtests-RG2} indicates that finite-size
corrections may be a pure power series in $N^{-\frac{2}{3}}$, the
first two orders of which are also shown as asymptotic fits in
Fig.~\ref{fig:ExtrapolationRG2}.

\begin{table}
\caption{\label{tab:RG2Data}
List of all the data obtained with EO for spin glasses on random
graphs of average degree $C=2$ at system sizes $N$. Given are the
number of instances $n_{I}$ considered at each $N$, and the average
ground-state cost density $\left\langle c_{0}\right\rangle _{N}$ and
average standard deviation $\sigma(c_{0})$ over these instances.}
\begin{tabular}{r|r|l|l}
\hline 
$N$  & $n_{I}$  & $\langle e_{0}\rangle_{N}$  & $\sigma_{N}\left(e_{0}\right)$\tabularnewline
\hline 
64 & 1050000 & 0.05504(2) & 0.01693 \tabularnewline
128 & 1050000 & 0.04994(1) & 0.01062 \tabularnewline
180 & 1050000 & 0.04808(1) & 0.00846 \tabularnewline
256 & 1050000 & 0.04652(1) & 0.00671 \tabularnewline
360 & 1050000 & 0.045296(5) & 0.00537 \tabularnewline
512 & 1050000 & 0.044266(4) & 0.00429 \tabularnewline
1024 & 395722 & 0.042808(4) & 0.00278 \tabularnewline
2048 & 668638 & 0.041883(4) & 0.00183 \tabularnewline
4096 & 319036 & 0.041304(4) & 0.00123 \tabularnewline
\hline
\end{tabular}
\end{table}

\subsection{Ground State Cost Fluctuations\label{sub:Ground-State-Cost RG2}}

\begin{figure}
\includegraphics[bb=8bp 27bp 706bp 523bp,clip,scale=0.33]{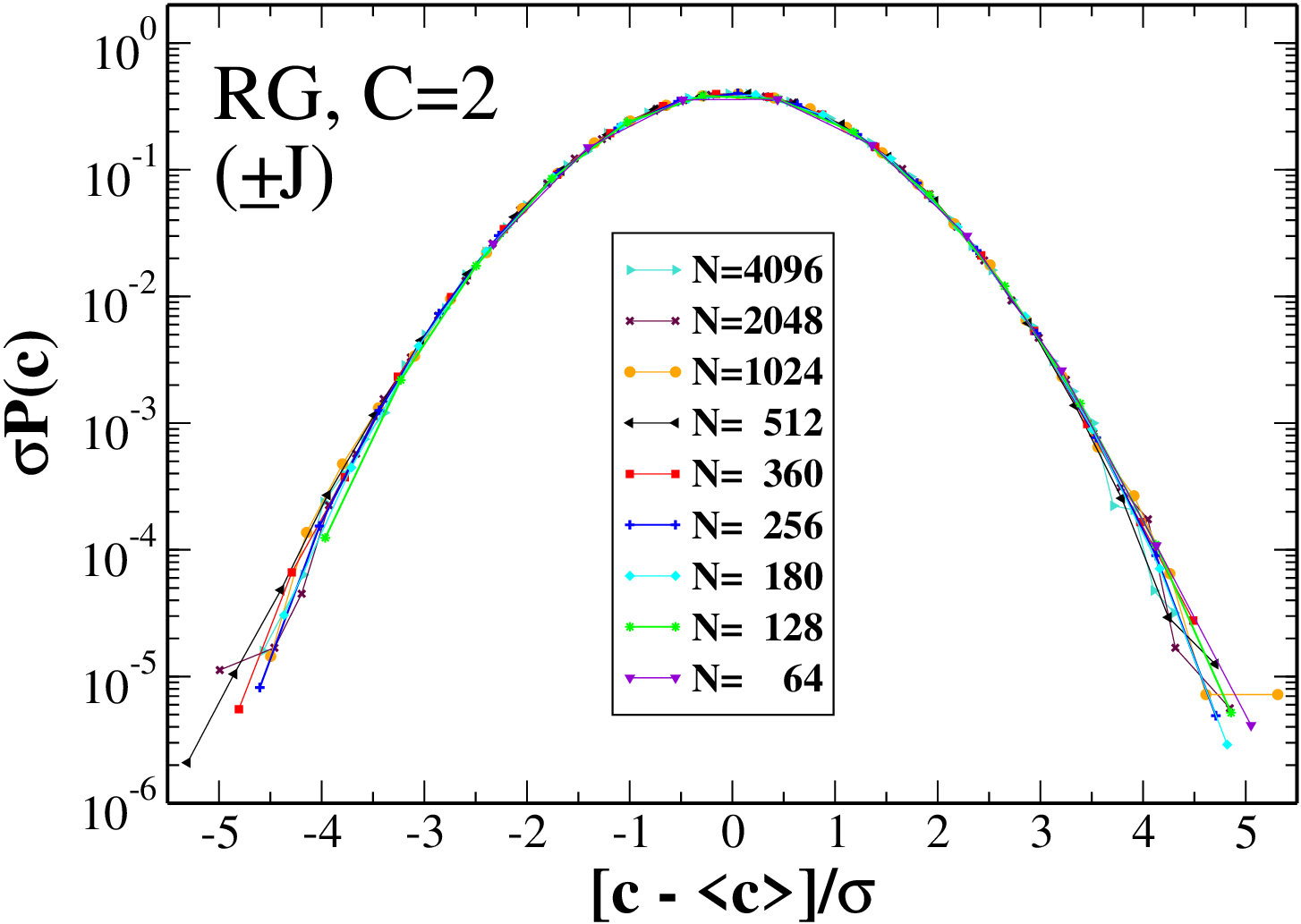}
\caption{\label{fig:PDF_RG2}
Plot of the probability density functions (PDF) of obtained
ground-state cost densities $c_{0}$ for the spin glass on an ordinary
random graph of average degree 2 in units of the standard deviation
$\sigma$. }
\end{figure}

\begin{figure}
\includegraphics[clip,scale=0.33]{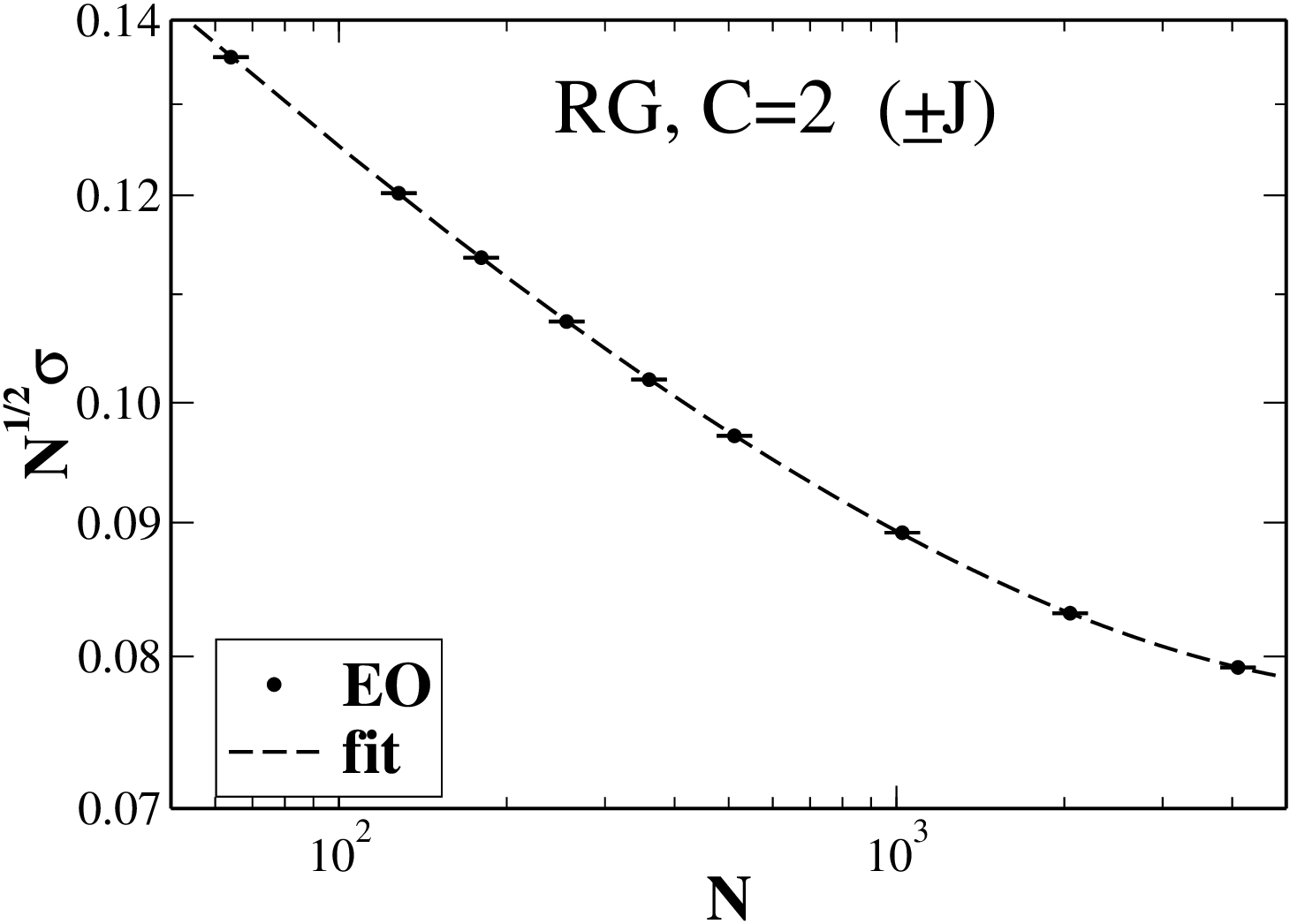}
\caption{\label{fig:sigmaRG2}
Double-logarithmic plot with system size $N$ of the deviations in the
ground state costs, here given as
$N^{\frac{1}{2}}\sigma_{N}\left(c_{0}\right)$ to highlight any
difference from Gaussian behavior, for a spin glass on an ordinary
random graphs of average degree $C=2$. The plot is far from flat or
linear, suggesting significant corrections to scaling.  Expecting
asymptotically normal scaling with $\rho=\frac{1}{2}$, we fitted the
data with an additional higher-order correction term with $\sim 
N^{-a}$ (dashed line), as in Eq.~(\ref{eq:rho}). The fit determines
$a\approx0.75$.}
\end{figure}

\begin{figure}
\includegraphics[bb=7bp 26bp 706bp 523bp,clip,scale=0.33]{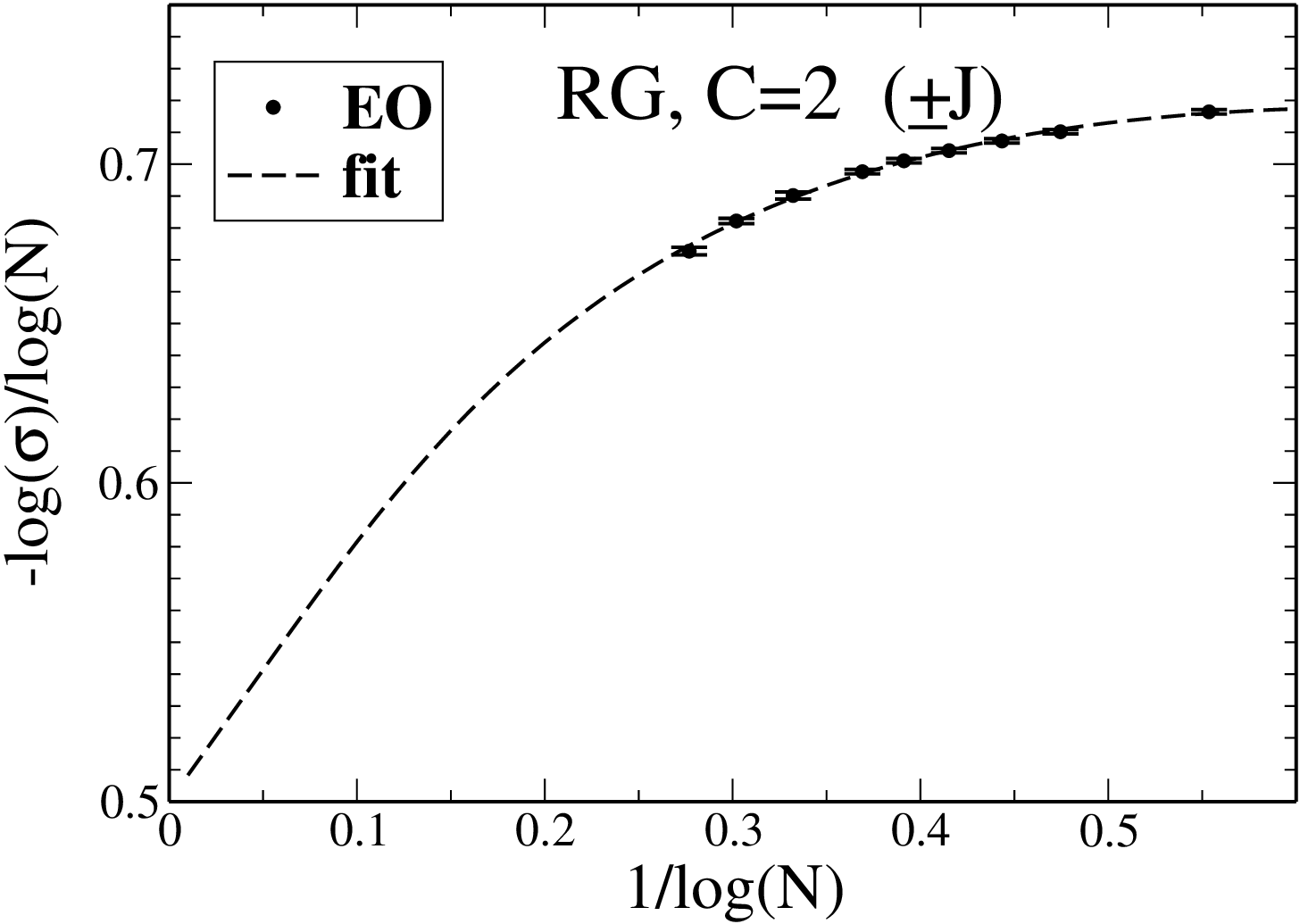}
\caption{\label{fig:rho_extrapolationRG2}
Extrapolation plot of the data for the deviation $\sigma$ in
Tab.~\ref{tab:RG2Data} according to Eq.~(\ref{eq:rhoextra}). A
fit is only possible for  a fixed $\rho=\frac{1}{2}$. }
\end{figure}

In Fig.~\ref{fig:PDF_RG2}, we have plotted the PDF for the ground
state cost fluctuations. It shows no sign of asymmetry for any size
$N$. Therefore, it is quite surprising that the finite size values
obtained for the deviation $\sigma_{N}\left(c_{0}\right)$ plotted
in Fig.~\ref{fig:sigmaRG2} and extrapolated in Fig.~\ref{fig:rho_extrapolationRG2}
exhibit a rather slow convergence to a normal width. These results
serve as a warning how, even for seemingly trivial fluctuations, the
asymptotic behavior might be reached only quite slowly.

\section{Three-Spin Interactions on Bethe Lattices\label{sec:Three-Spin-Interactions-on}}

To complement the discussion of glasses with two-spin interactions,
and to compare with related work~\citep{Nakajima09,Franz01}, we have
also considered spin glasses on Bethe lattices with three-spin
interactions, both, with discrete $(\pm J)$ and Gaussian bonds, as
listed in Tabs.~\ref{tab:BL43Data_discrete}
and~\ref{tab:BL43Data_Gauss}. This study succeeds in confirming recent
observations regarding finite-size corrections to the average ground
state energy~\citep{Nakajima09}.  But they can merely give a crude
picture of deviations within the distribution of those energies. This
inadequacy has two origins: First, it proves inherently challenging to
determine ground states for instances of this problem with any
accuracy already at moderately sized systems.  The systems are large
enough to predict average energies with reasonable errors, but
insufficient for the asymptotic analysis for the deviations.  Second,
those deviations in their own right appear to be far narrower for this
problem than for any of the two-spin models above. In fact, for
discrete bonds the grounds states seem to cover only a few states
above and below the average, with almost an invariant width, such that
the deviations in the density seem to fall with $\sim1/N$.

A somewhat wider distribution is observed for Gaussian bonds, which
provides for a smoother appearance for the PDF at all system sizes
compared to the discrete case. Any skewness can only be observed when
plotted for ground state \emph{cost} fluctuations; energy fluctuations
would always be normal, originating from the random fluctuations in
the total bond-weight themselves, as described in Sec.~\ref{data}.
Surprisingly, these cost fluctuations skew exactly in the opposite
direction from any previous studied PDF, such as those above. Although
the deviations $\sigma$ extracted from those PDFs indeed seem
inconsistent with $1/N$ scaling, the system sizes attained in this
study are rather small, $N\leq100$, and asymptotic behavior may not
have been reached in this study.

\begin{table}
\caption{\label{tab:BL43Data_discrete}
List of all the data obtained with EO for $p=3$-spin glasses on Bethe
lattices of degree $r=4$ at system sizes $N$ for discrete bonds. Given
are the number of instances $n_{I}$ considered at each $N$, and the
average ground-state energy density $\left\langle e_{0}\right\rangle
_{N}$ and  standard deviation $\sigma$ over these
instances. The largest system size has unacceptable systematic errors
and is ignored in any fit.}
\begin{tabular}{r|r|l|l||}
\hline 
$N$  & $n_{I}$  & $\langle e_{0}\rangle_{N}$  & $\sigma_{N}\left(e_{0}\right)$\tabularnewline
\hline 
15 & 10000 & -1.104(1) & 0.0995\tabularnewline
18 & 100000 & -1.1171(3) & 0.0834\tabularnewline
24 & 100000 & -1.1406(2) & 0.0584\tabularnewline
30 & 30000 & -1.154(3) & 0.0500\tabularnewline
33 & 100000 & -1.1576(2) & 0.0460\tabularnewline
36 & 100000 & -1.1619(1) & 0.0411\tabularnewline
39 & 100000 & -1.1665(1) & 0.0365\tabularnewline
45 & 30000 &   -1.173(2) & 0.0329\tabularnewline
48 & 800000 & -1.17435(3) & 0.0319\tabularnewline
\hline
\end{tabular}\begin{tabular}{r|r|l|l}
\hline 
$N$  & $n_{I}$  & $\langle e_{0}\rangle_{N}$  & $\sigma_{N}\left(e_{0}\right)$\tabularnewline
\hline 
51 & 130000 & -1.1760(1) & 0.0296\tabularnewline
54 & 800000 & -1.17843(3) & 0.0267\tabularnewline
60 & 25000 & -1.1829(2) & 0.0242\tabularnewline
75 & 400000 & -1.18892(3) & 0.0189\tabularnewline
90 & 35000 & -1.1929(1) & 0.0157\tabularnewline
120 & 11000 & -1.1973(1) & 0.0121\tabularnewline
150 & 200000 & -1.20087(2) & 0.0103\tabularnewline
180 & 80000 & -1.20264(3) & 0.0087\tabularnewline
240 & 350 & -1.1995(4) & 0.0067\tabularnewline
\hline
\end{tabular}
\end{table}

\begin{table}
\caption{\label{tab:BL43Data_Gauss}
List of all the data obtained with EO for $p=3$-spin glasses on Bethe
lattices of degree $r=4$ at system sizes $N$ for Gaussian bonds. Given
are the number of instances $n_{I}$ considered at each $N$, and the
average ground-state cost density $\left\langle c_{0}\right\rangle
_{N}$ and  standard deviation $\sigma$ over these
instances. The largest system size has unacceptable systematic errors
and is ignored in any fit.}
\begin{tabular}{r|r|l|l}
\hline 
$N$  & $n_{I}$  & $\langle c_{0}\rangle_{N}$  & $\sigma_{N}\left(c_{0}\right)$\tabularnewline
\hline 
33 & 100000 & 0.02366(3) & 0.00916\tabularnewline
39 & 120000 & 0.02261(2) & 0.00783\tabularnewline
51 & 120000 & 0.02122(2) & 0.00616\tabularnewline
66 & 80000 & 0.02016(3) & 0.00495\tabularnewline
99 & 25000 & 0.01901(2) & 0.00356\tabularnewline
144 & 3500 & 0.0193(1) & 0.00274\tabularnewline
\hline
\end{tabular}
\end{table}

\begin{figure}
\includegraphics[clip,scale=0.33]{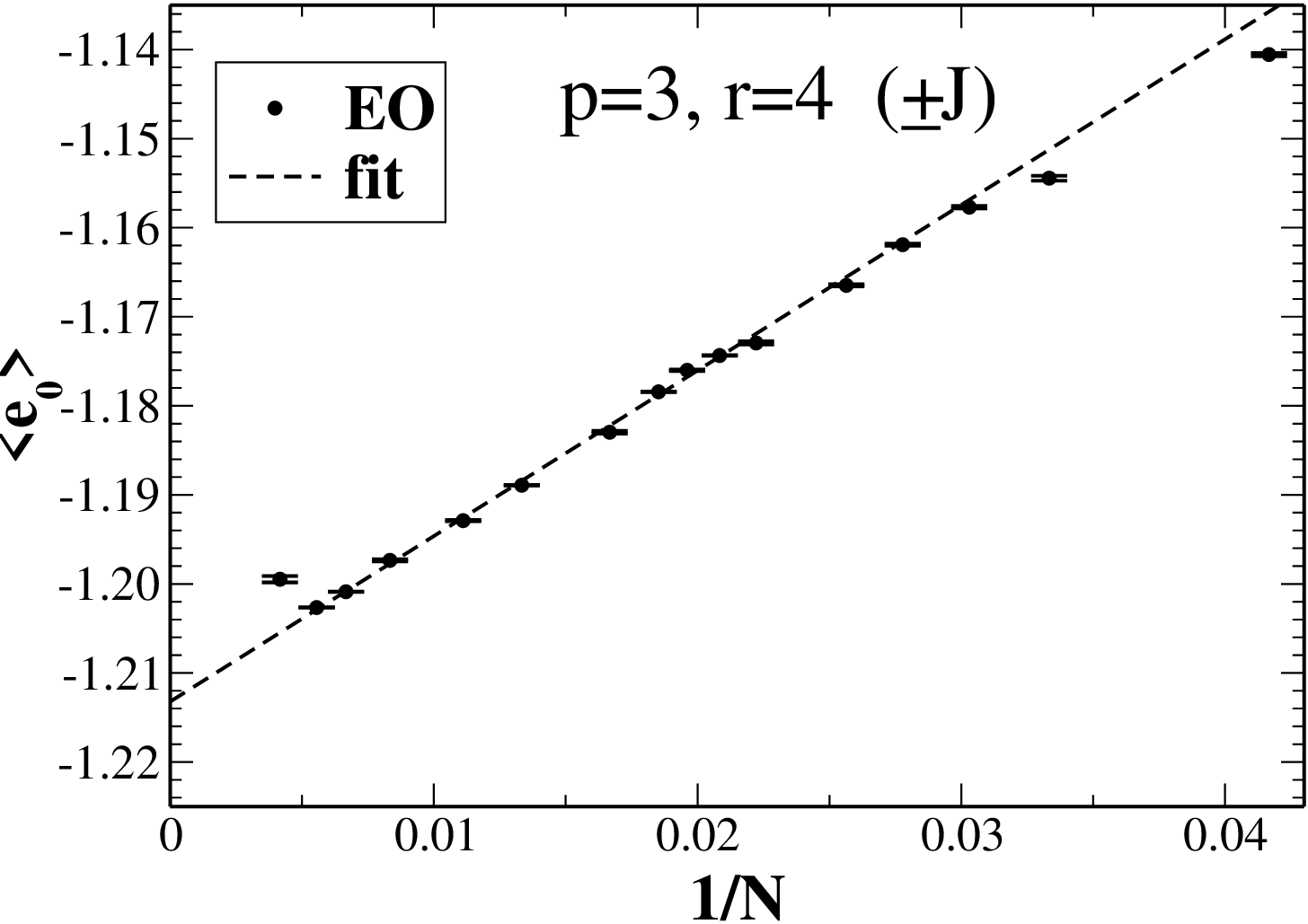}
\caption{\label{fig:ExtrapolationBethe43}
Extrapolation plot of the average ground-state energy densities
$\left\langle e_{0}\right\rangle _{N}$ for a spin glass with discrete
bonds in which $p=3$ spins mutually interact on a Bethe lattices of
degree $r=4$, as a function of the presumed finite-size corrections,
$1/N$. }
\end{figure}
\begin{figure}
\includegraphics[clip,scale=0.33]{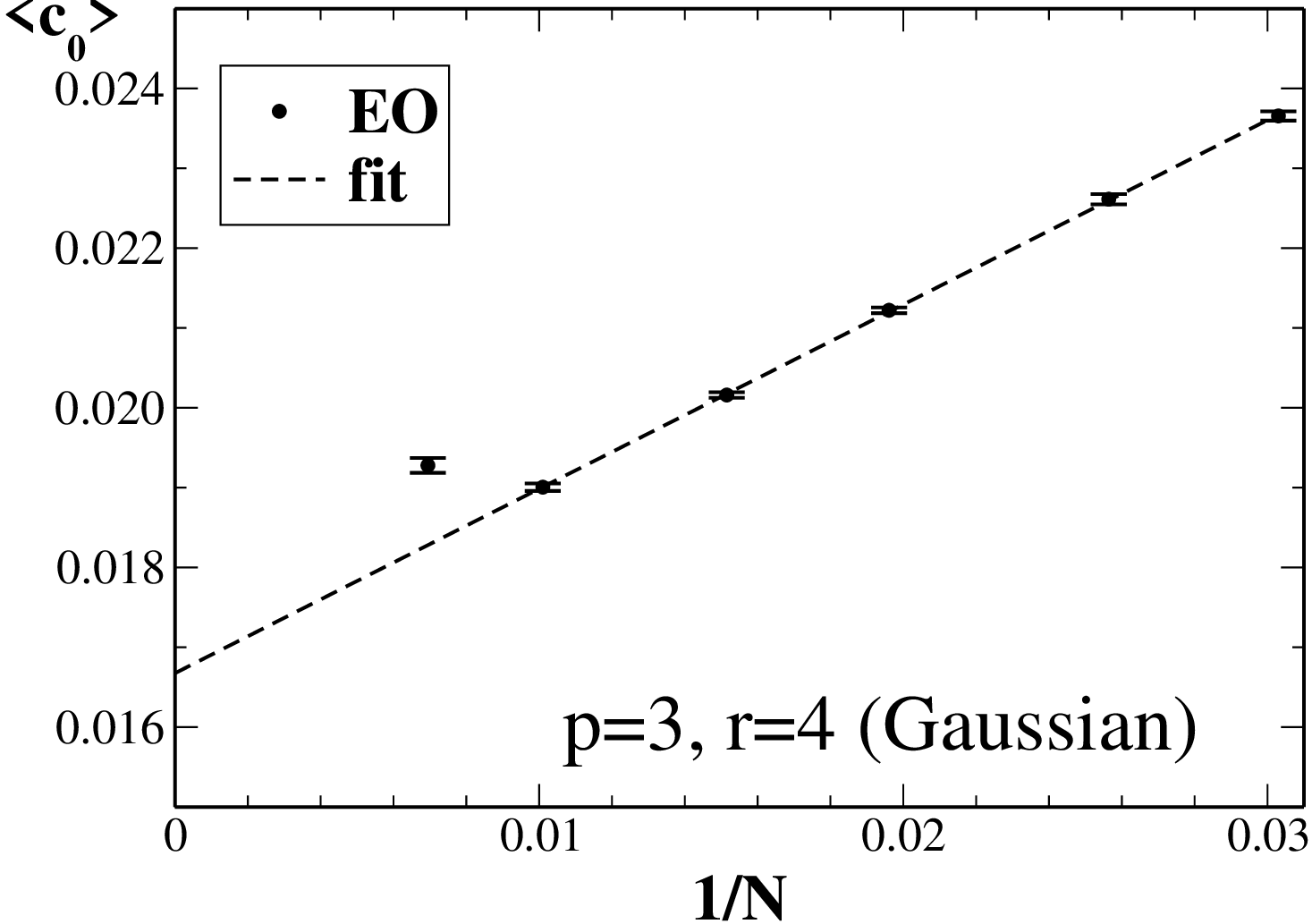}
\caption{\label{fig:ExtrapolationBethe43_Gauss}
Extrapolation plot of the average ground-state cost densities
$\left\langle e_{0}\right\rangle _{N}$ for a Gaussian spin glass with
three-spin interactions on a Bethe lattices of degree $r=4$, as a
function of the presumed finite-size corrections, $1/N$. }
\end{figure}
\begin{figure}
\includegraphics[clip,scale=0.33]{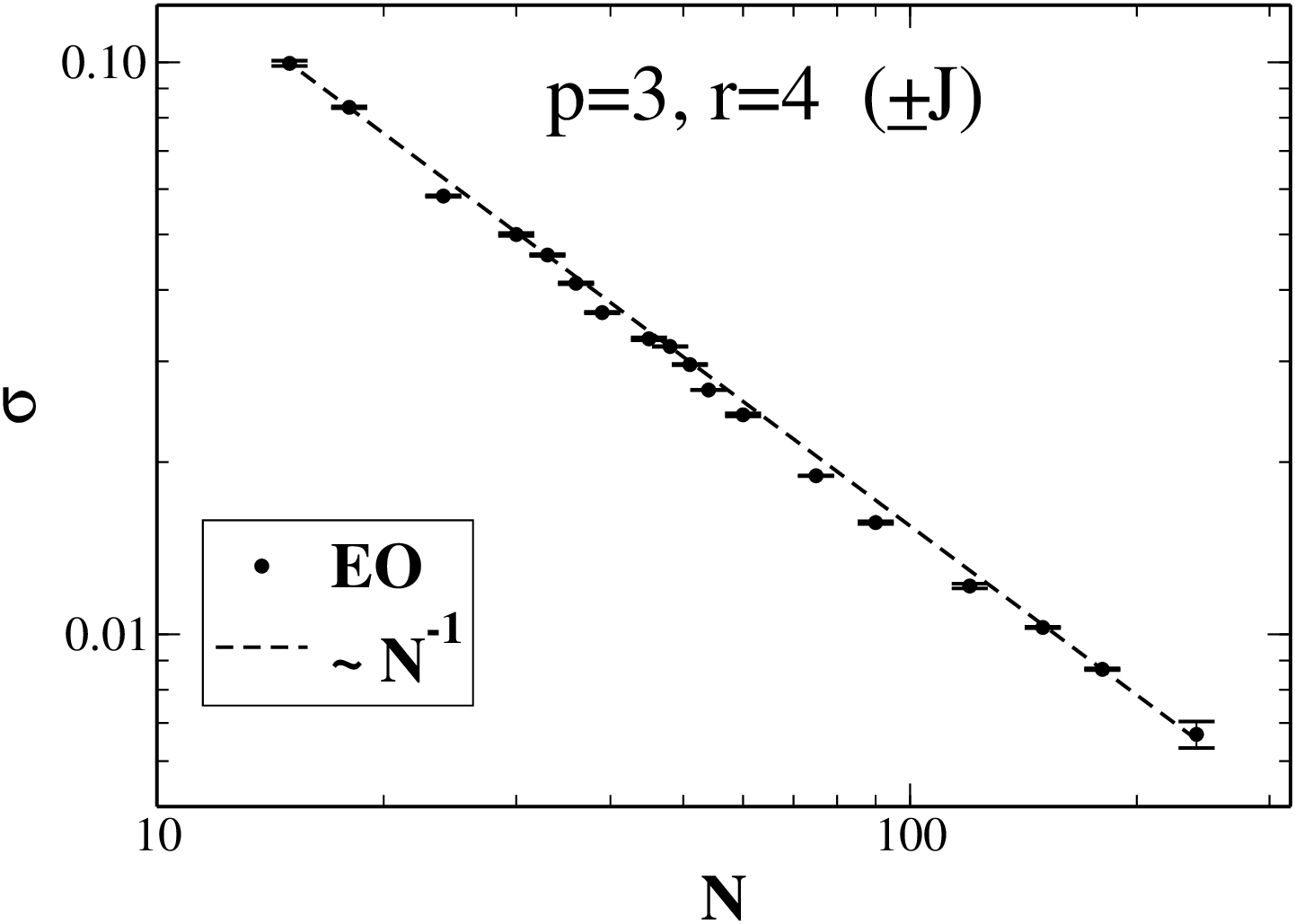}
\caption{\label{fig:sigma_Bethe34}
Log-log-plot of the deviations $\sigma$ in the ground state energy
fluctuations of a $p=3$ spin glass on Bethe lattices of degree $r=4$
as a function of system size $N$.  The data is very difficult to fit,
and we only provide the dashed line $\sim1/N$ as a guide to the eye.}
\end{figure}
\begin{figure}
\includegraphics[bb=14bp 29bp 706bp 531bp,clip,scale=0.33]{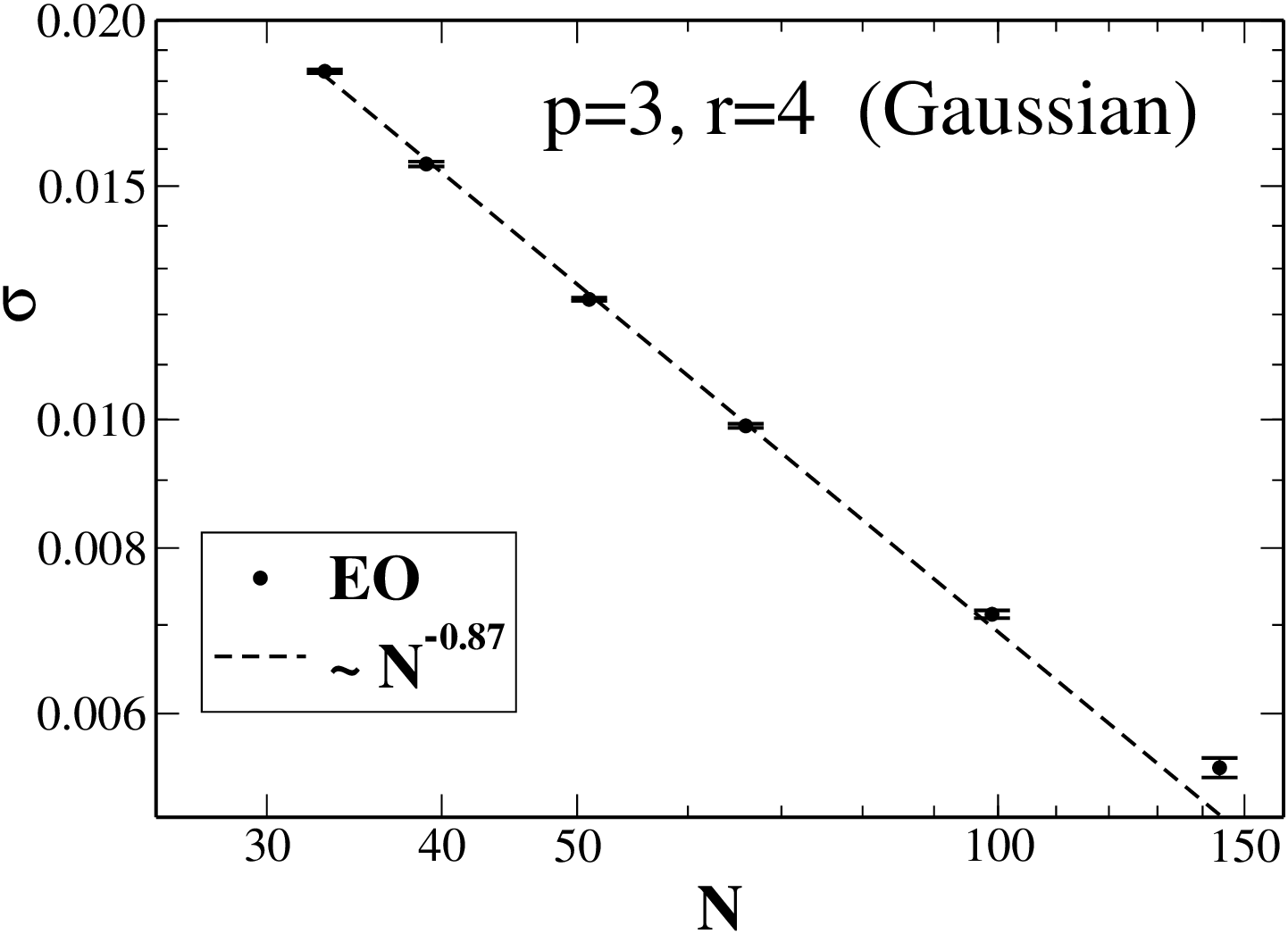}
\caption{\label{fig:sigma_Bethe34_Gauss}
Log-log-plot of the deviations $\sigma$ in the ground state energy
fluctuations for a $p=3$-spin glass on Bethe lattices of degree $r=4$
as a function of system size $N$, here for Gaussian bonds. As for the
plot for discrete bonds in Fig.~\ref{fig:sigma_Bethe34}, this data
is also difficult to fit, and there is a visible trend to even lower
values in the exponent then $0.87$ found by only fitting to sizes
$N\leq100$ (dashed line).}
\end{figure}

\subsection{Ground State Energy\label{sub:Ground-State-Energyp3}}

In
Figs.~\ref{fig:ExtrapolationBethe43}-\ref{fig:ExtrapolationBethe43_Gauss},
we extrapolate the obtained average ground state energies (for
discrete bonds) or costs (for Gaussian bonds) in
Tabs.~\ref{tab:BL43Data_discrete} and~\ref{tab:BL43Data_Gauss}.  In
both cases, finite size corrections appear to decay with volume
corrections, like $1/N$, stronger than in any of the $p=2$-spin models
above. Hence, even though the system sizes are small, quite reasonable
extrapolations are achieved. In the discrete case in
Fig.~\ref{fig:ExtrapolationBethe43} there appears to be significant
structure in the transients. By reproducing the exact ground states
for a sample of those instances up to size $N=51$ with exact
(branch-and-bound) algorithms, we have verified that these are not due
to systematic errors in the optimization heuristic, as 100\% agreement
was achieved for each instance. Rather, we expect that those effects
are due to the constraints in the formation of 4-regular hyper-graphs
at small $N$ in combination with the discrete set of bonds
available. Correspondingly, the Gaussian data is free of any such
structure, and thus extrapolates with comparable accuracy despite the
overall smaller system sizes used.

For $N\to\infty$, the discrete data in
Fig.~\ref{fig:ExtrapolationBethe43} extrapolates to $\left\langle
e_{0}\right\rangle _{\infty}=-1.213(2)$ at the intercept, slightly
above the one-step replica symmetry-breaking (RSB)
prediction~\citep{Franz01}, as would be expected for the true ground
state at full RSB. With bonds drawn from a Gaussian distribution, we
instead plot the actual cost of violated bonds. It is difficult to
obtain good minima already at $N\approx100$, such that we can only
extrapolate data for smaller $N$. For $N\to\infty$, the extrapolation
yields a cost of $\left\langle c_{0}\right\rangle _{\infty}=0.0167(4)$
at the intercept. Since average cost and energy density for Gaussian
bonds are related via Eq.~(\ref{eq:cost_energy}), we obtain with
$\left\langle \left|J\right|\right\rangle =\sqrt{\frac{2}{\pi}}$,
$r=4$, and $p=3$ that $\left\langle e_{0}\right\rangle =-1.030(1)$.

\subsection{Ground State Fluctuations\label{sub:Ground-State-Fluctuationsp3}}

In light of the limited ability to produce ground states at larger
system sizes, any prediction for the fluctuation exponent $\rho$ is
poor. Furthermore, extreme fluctuations are difficult to attain, since
the width is rather narrow. Yet, for discrete bonds the data plotted
for $\sigma$ in Fig.~\ref{fig:sigma_Bethe34} is quite consistent with
$1/N$-decay, i.~e. $\rho=1$.  But there does not seem to be a
clear trend towards asymptotic scaling in the case of Gaussian bonds
shown in Fig.~\ref{fig:sigma_Bethe34_Gauss}.  If anything, the data
appears to exhibit upward curvature, away from an $1/N$ scaling
regime, indicating that $\rho$ in this case may be even lower than the
fitted value of $\approx0.87$. Such a discrepancy between discrete and
Gaussian bonds on Bethe lattices was also noted for the $p=2$-spin
model~\citep{Boettcher10a}.

\begin{figure}
\includegraphics[clip,scale=0.33]{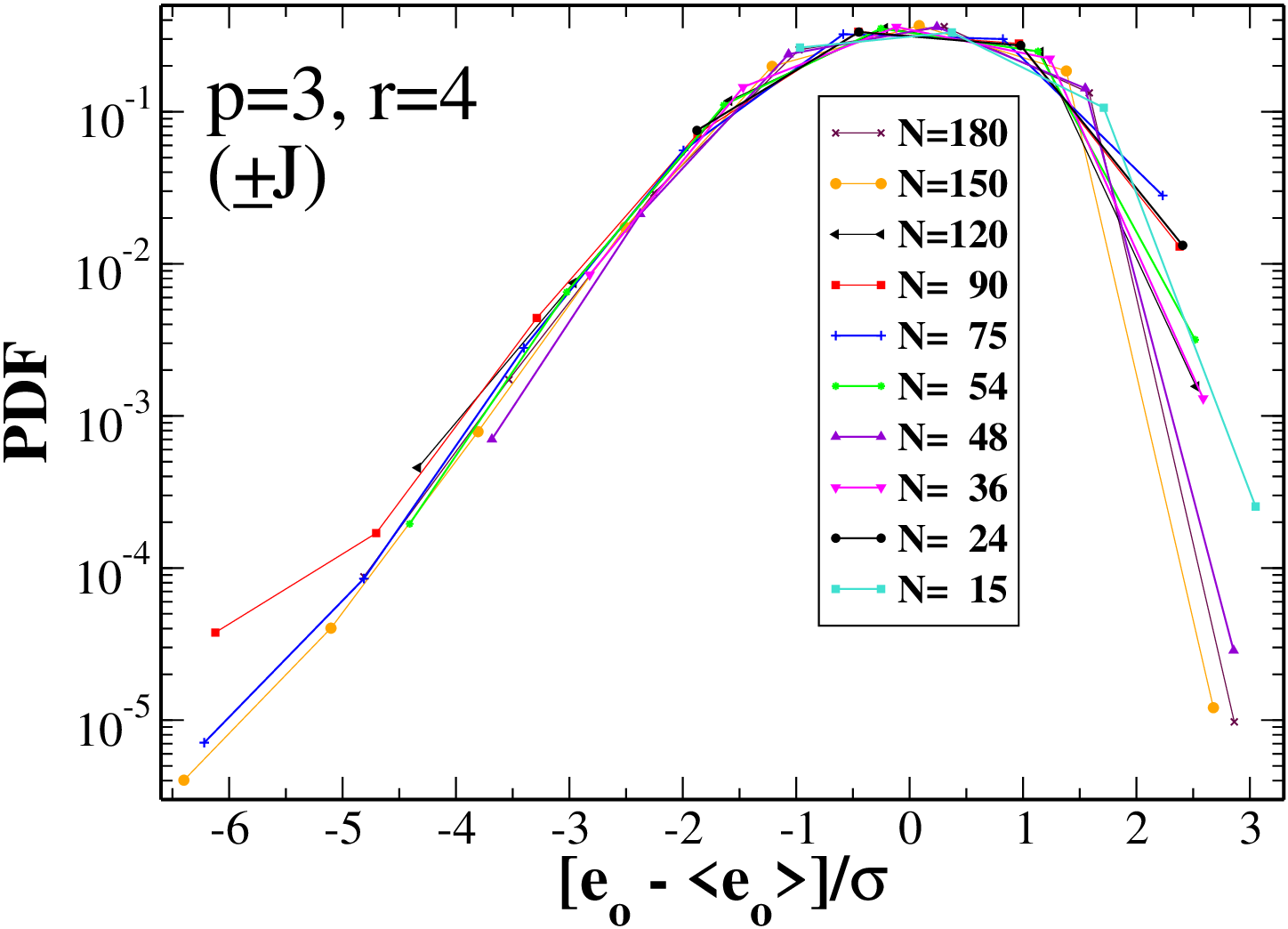}
\caption{\label{fig:PDF_p3discrete}
Plot of the probability density functions (PDF) of obtained
ground-state energy densities $e_{0}$, in units of the standard
deviation $\sigma$, for the $p=3$-spin glass with discrete bonds on an
$r=4$-regular random graph. The rugged appearance of the data is due
to the discreteness of the bonds and the fact that only a very narrow
set of energy values around the mean are taken on.}
\end{figure}
\begin{figure}
\vskip 2.3in 
\includegraphics{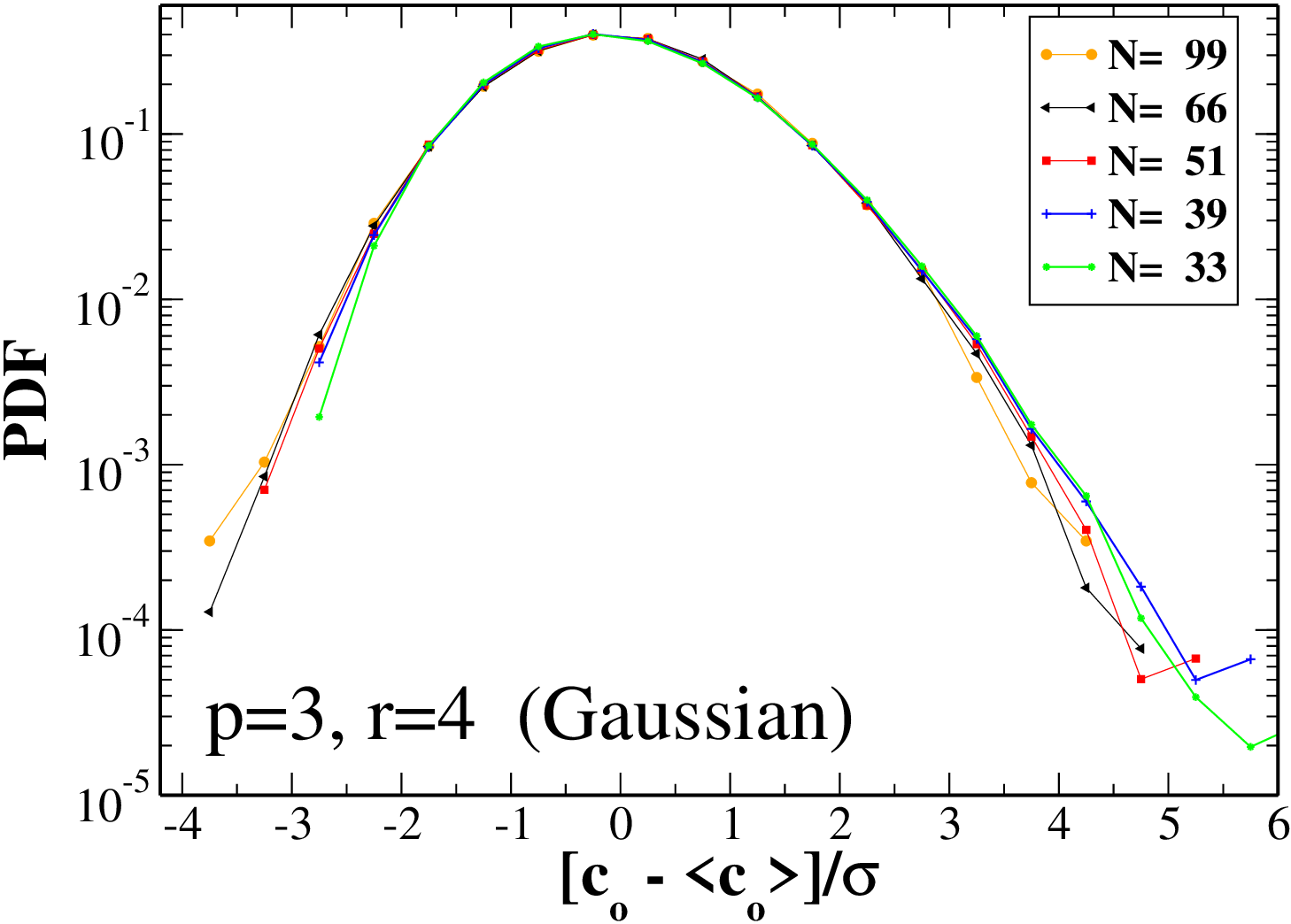} 
\includegraphics{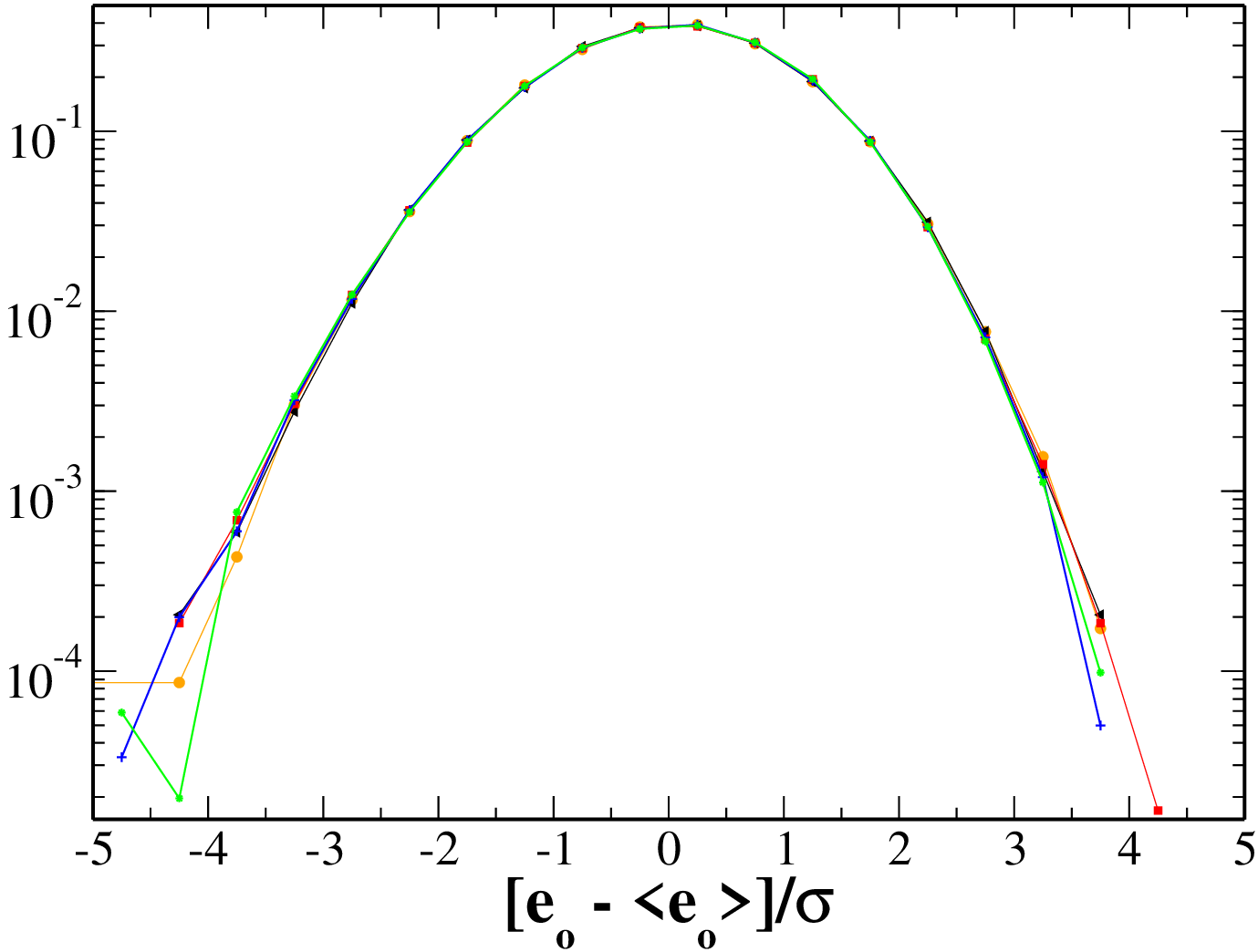} 
\caption{\label{fig:PDF_p3Gauss}
Plot of the probability density functions (PDF) of obtained
ground-state \emph{cost} densities $c_{0}$, in units of the standard
deviation $\sigma$, for the $p=3$ spin glass with Gaussian bonds on an
$r=4$-regular random graph. The data is skewed with a sharp cut-off
for costs less than the average and a long tail for larger cost,
exactly opposite to any previous PDF for energy fluctuations. The
inset shows the same data plotted as energy fluctuation PDF, which are
purely normal distributed. }
\end{figure}

\begin{figure*}
\vskip 5.0in 
\includegraphics{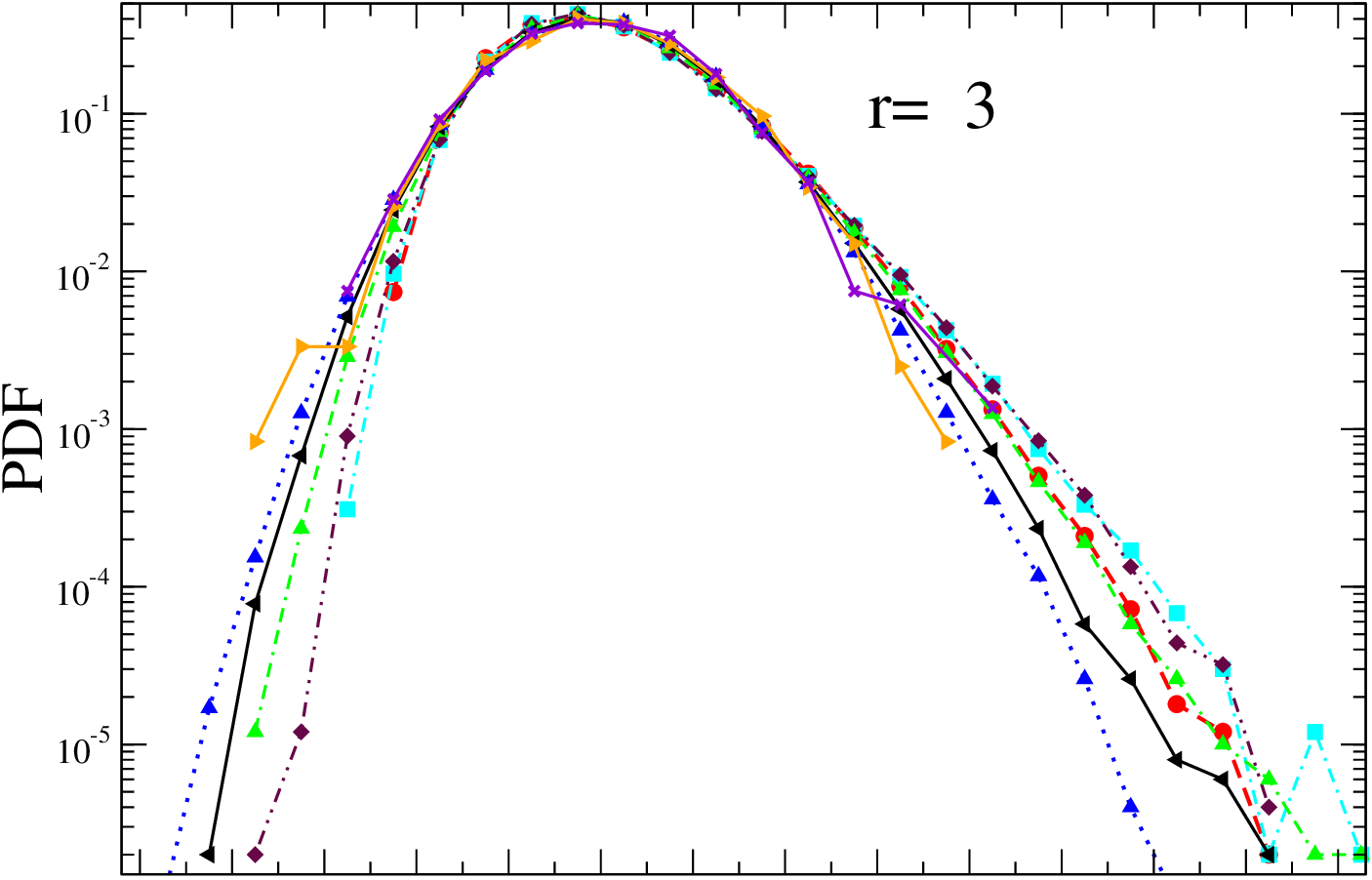} 
\includegraphics{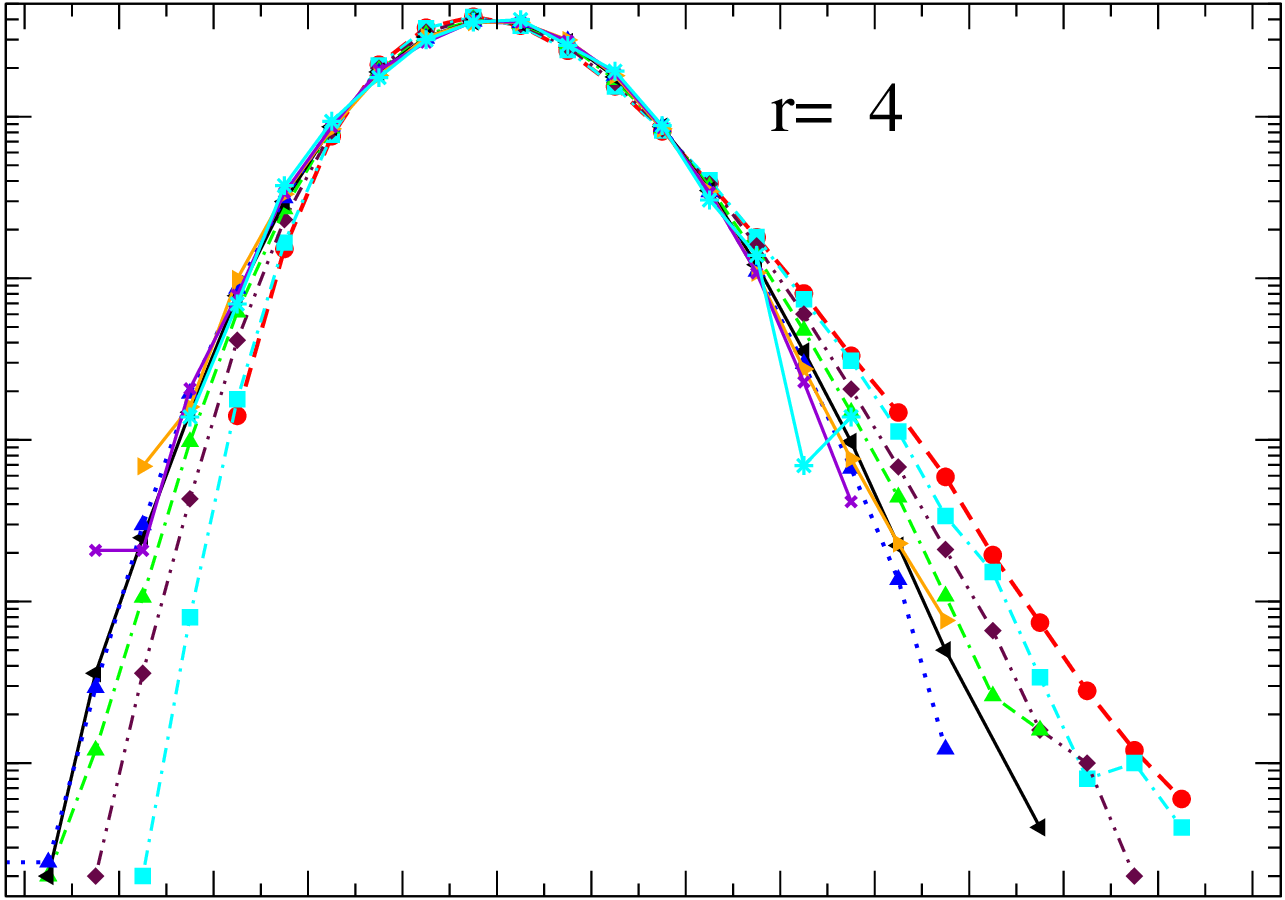} 
\includegraphics{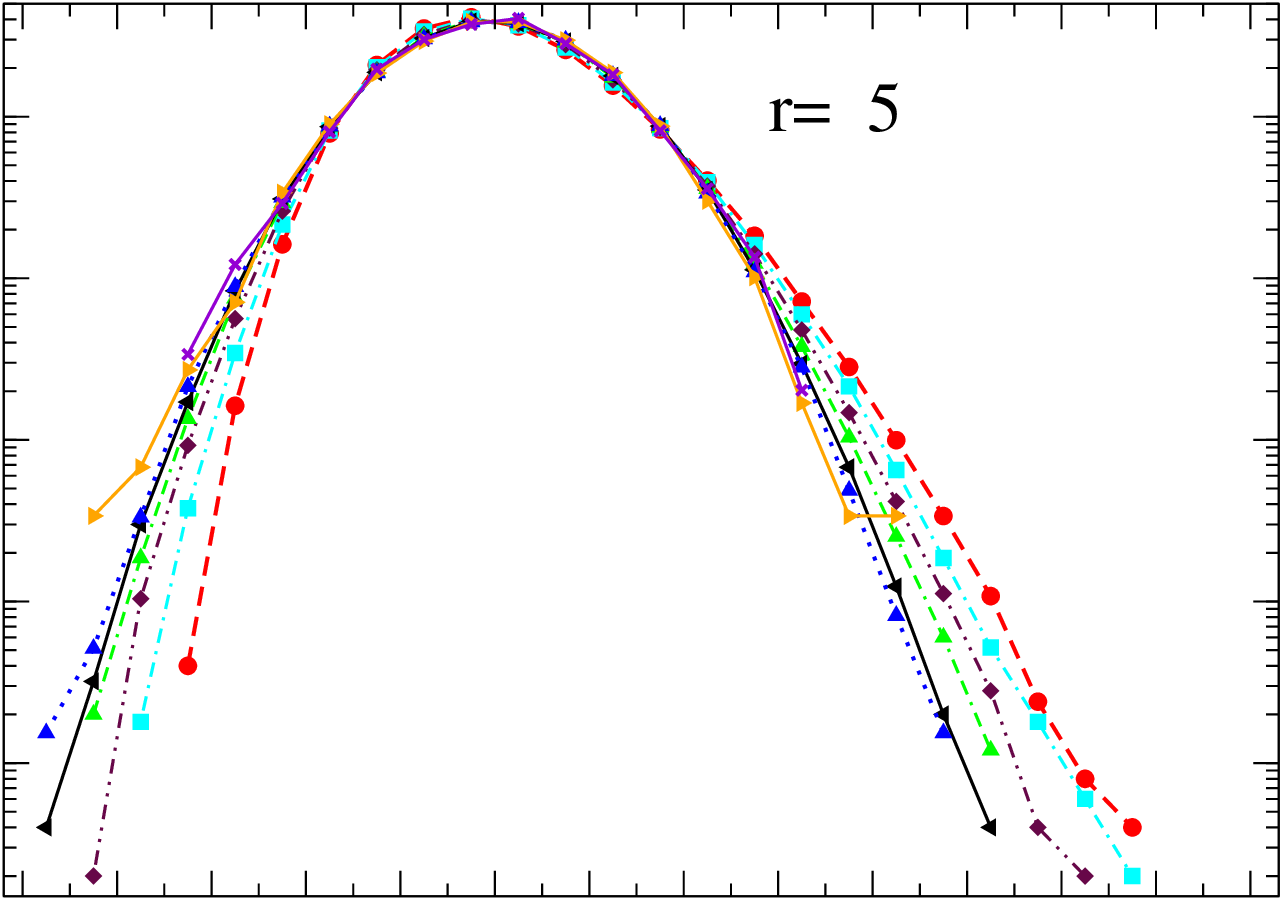} 
\includegraphics{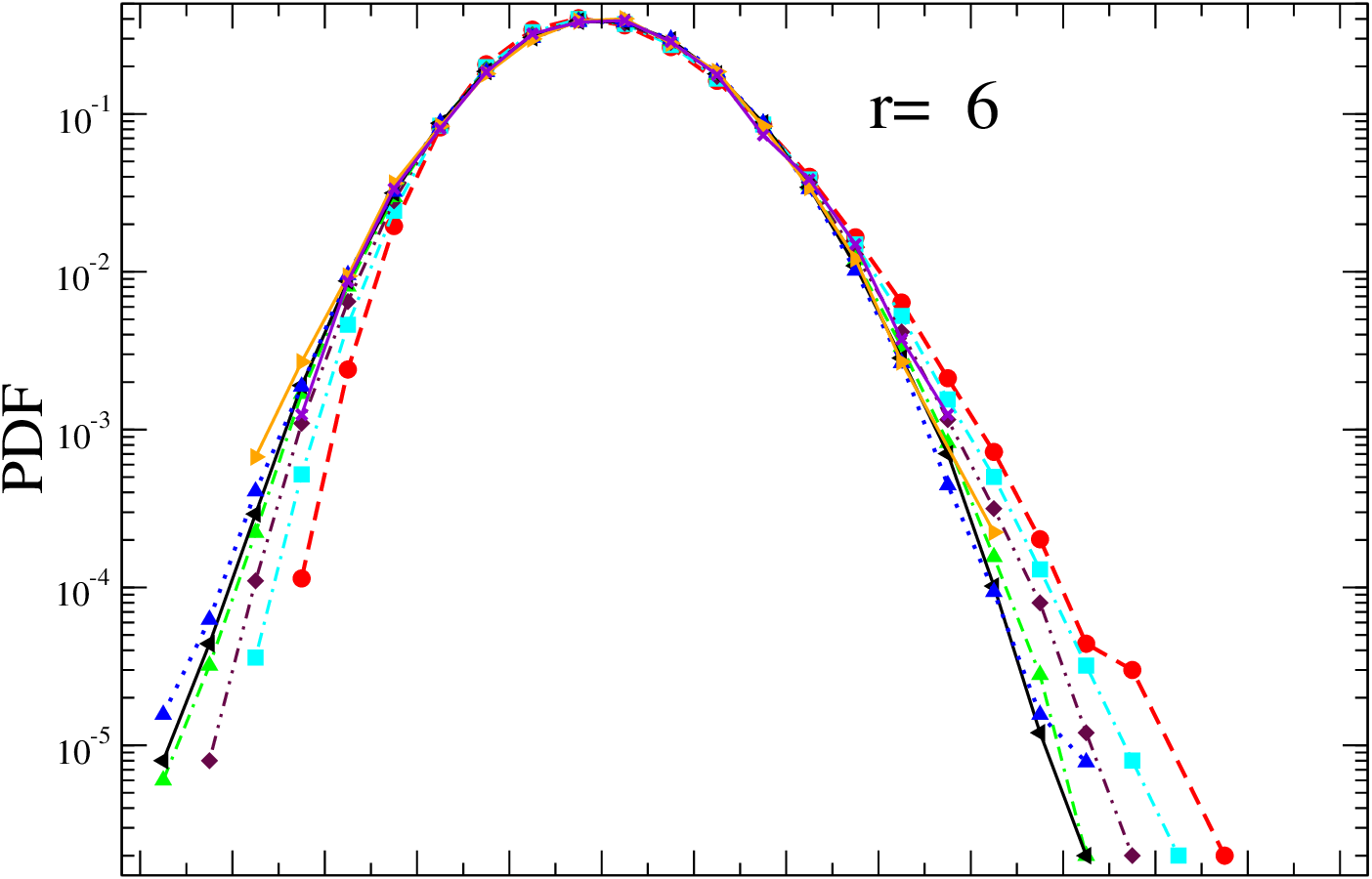} 
\includegraphics{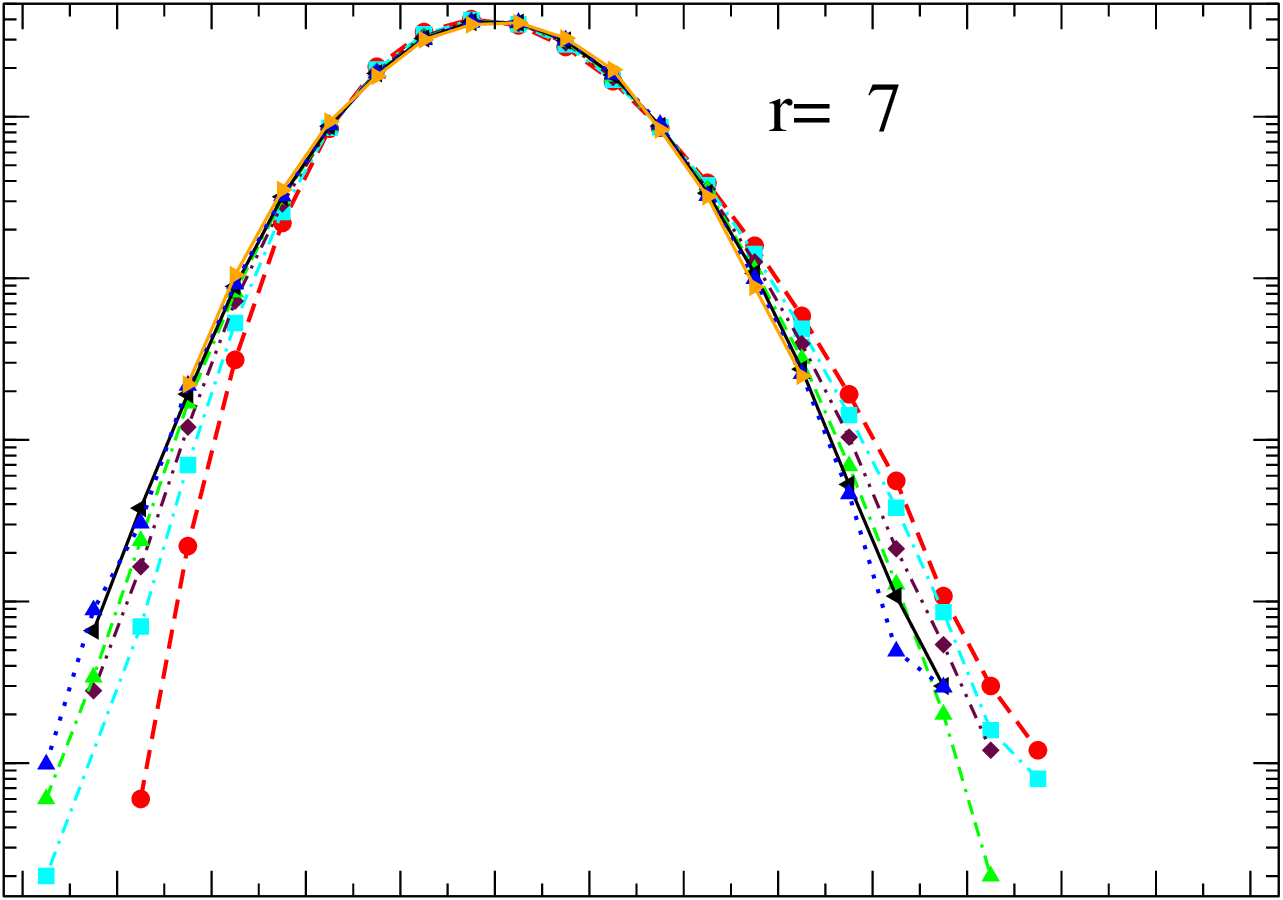} 
\includegraphics{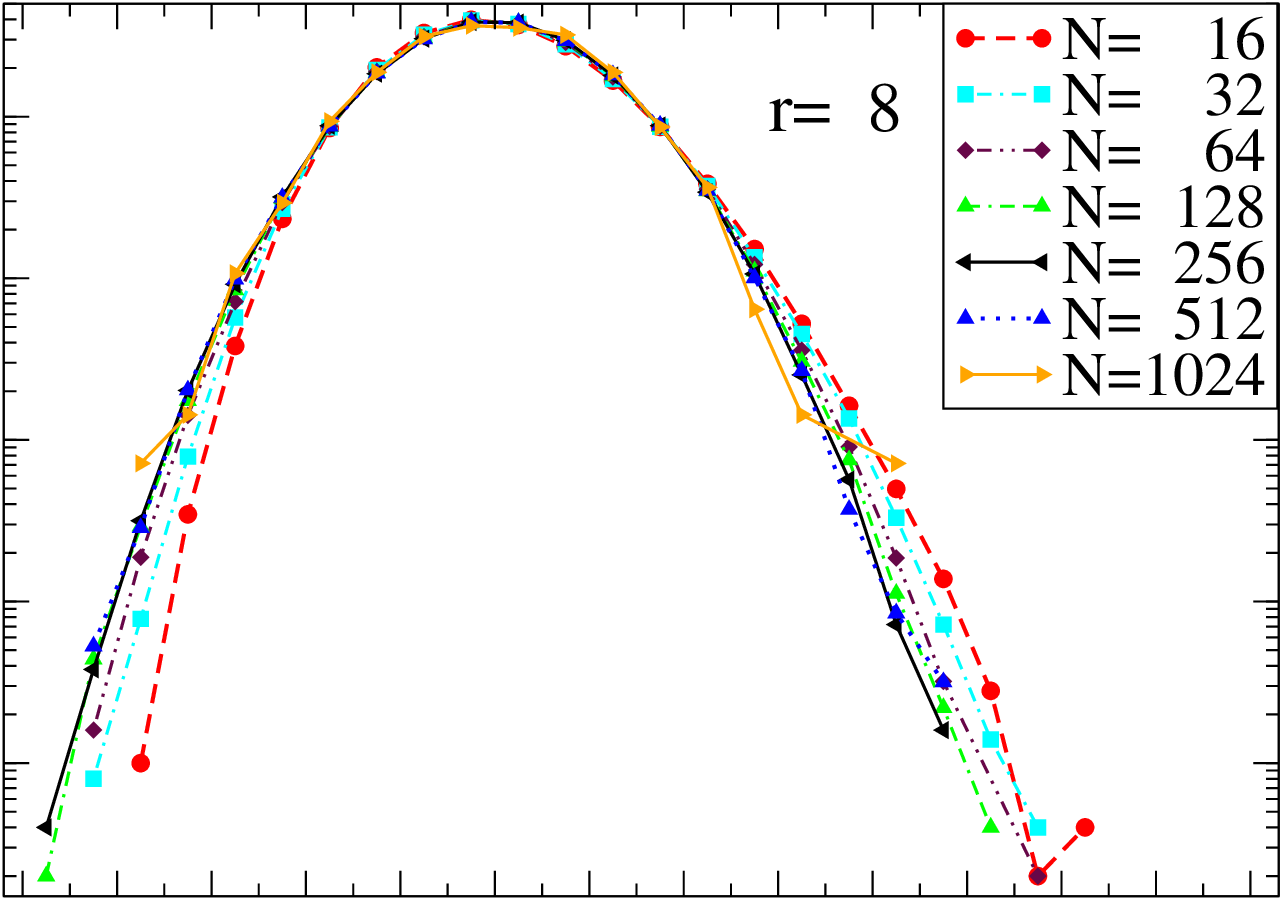} 
\includegraphics{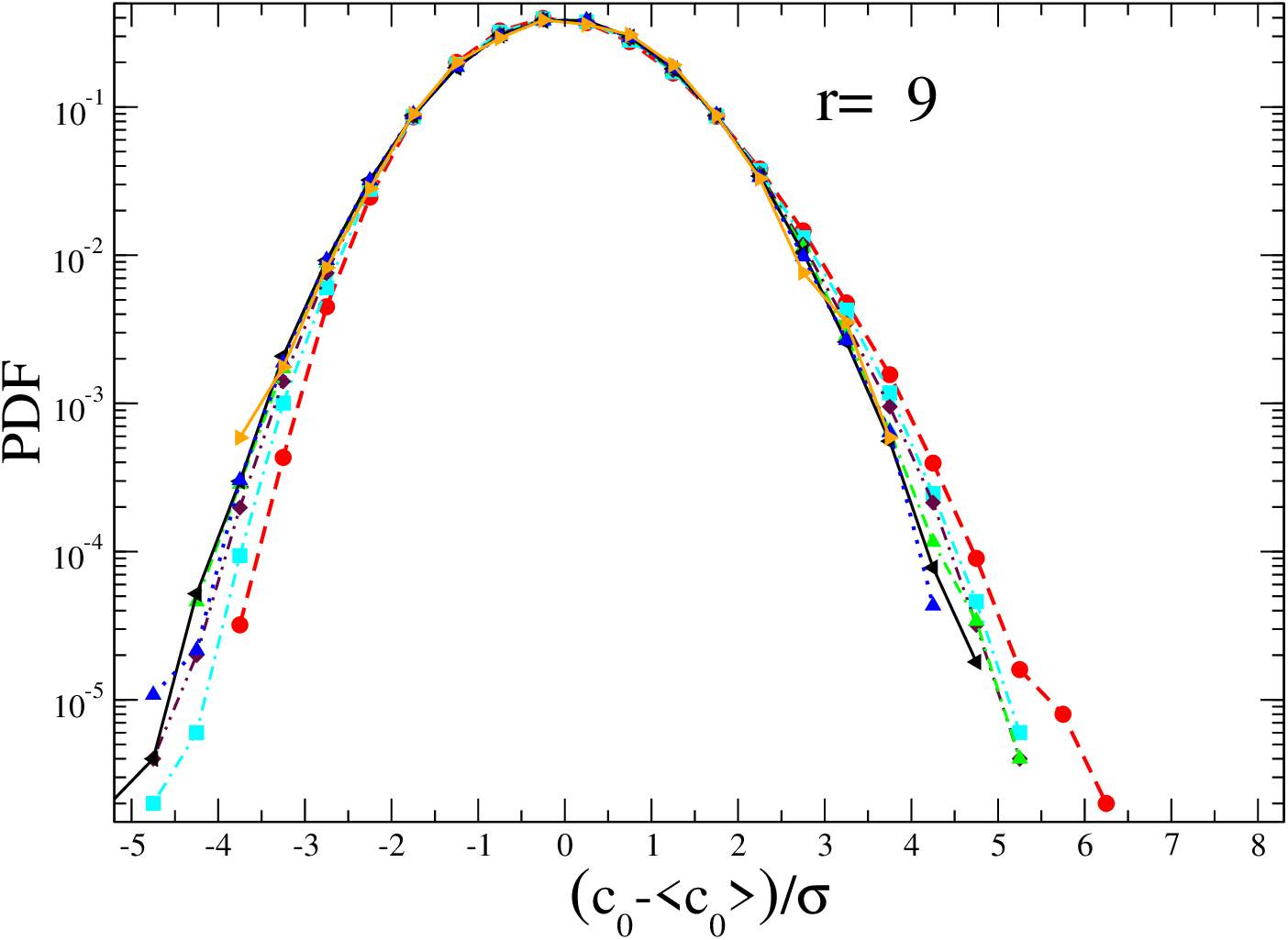} 
\includegraphics{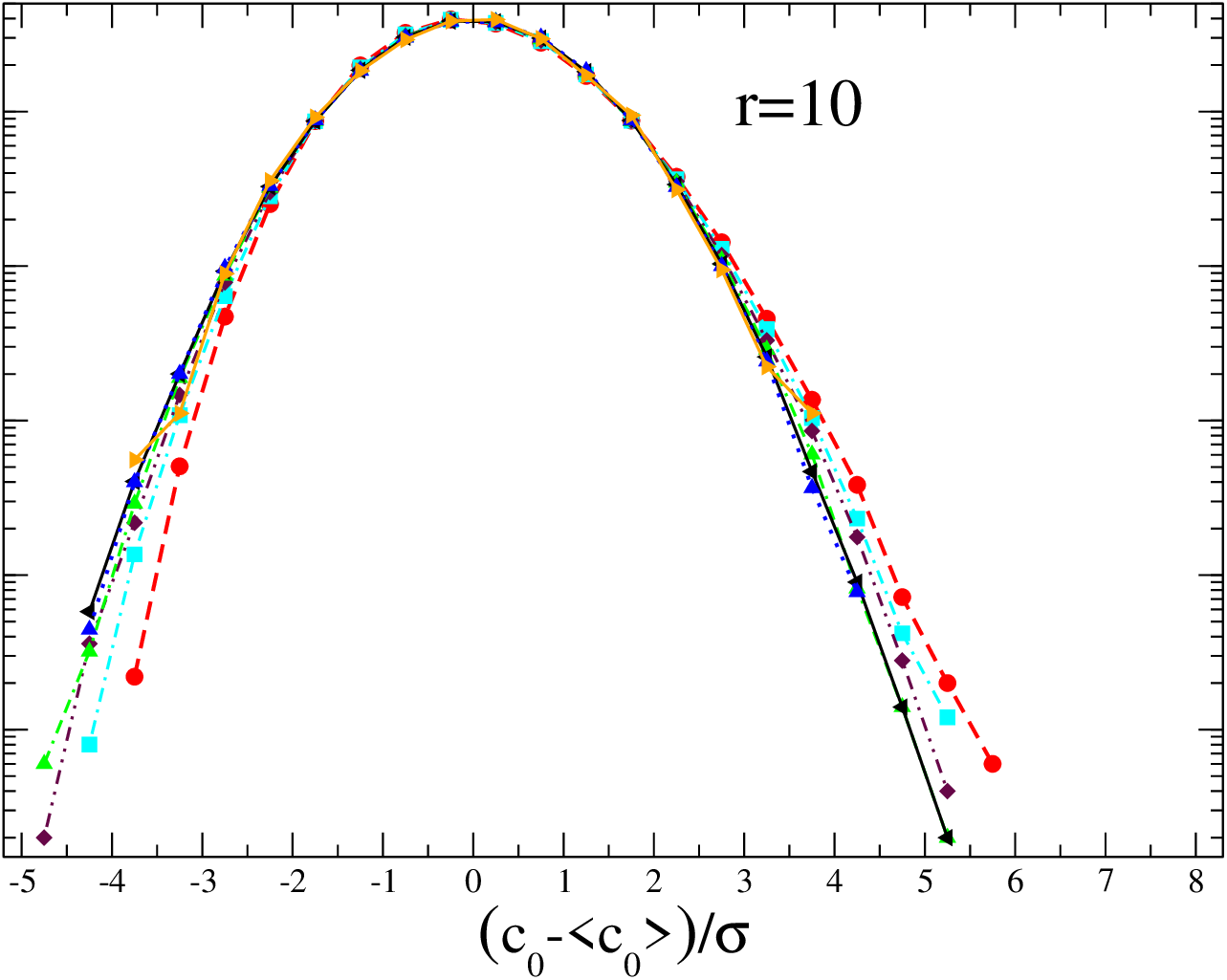}
\includegraphics{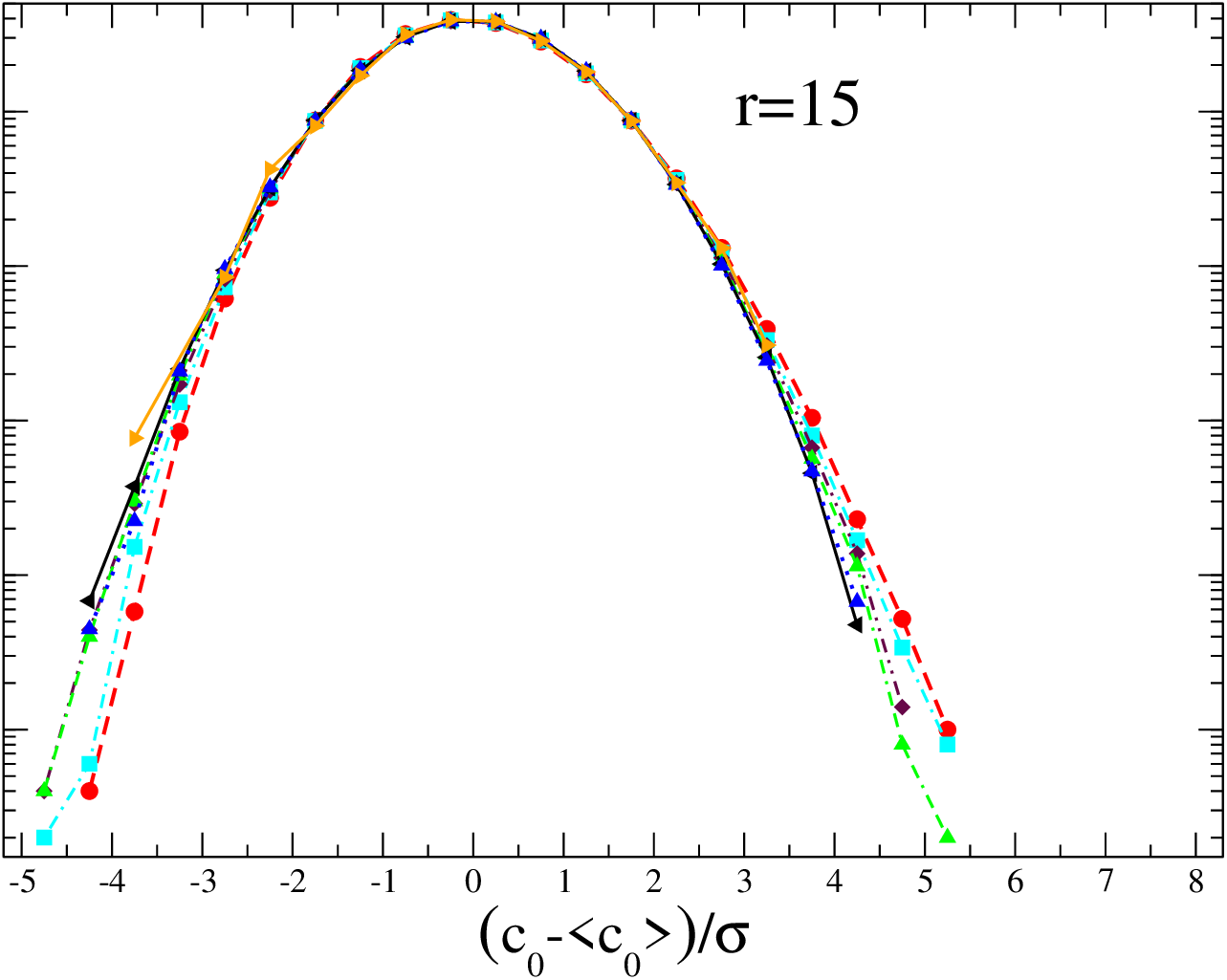} 
\caption{\label{fig:CostFlucGaussian}
Plot of the PDFs for the fluctuations of the ground state costs per
spin for a $p=2$-spin glass with a Gaussian bond distribution on Bethe
lattices of degree $r=3,4,\ldots,10,15$.  For smaller $r$ and sizes
$N$, the PDFs are right-skewed, just as in Fig.~\ref{fig:PDF_p3Gauss},
but especially for larger $r$ those PDFs appear to approach a normal
distribution for large sizes $N$. }
\end{figure*}

With the rapid decay of the width for discrete bonds, it is not
surprising that the PDF for its energy fluctuations has a somewhat
rugged appearance: only few, discrete energy values can be taken on
left or right of the mean. As shown in Fig.~\ref{fig:PDF_p3discrete},
the PDF otherwise skews similarly to all previous cases. It comes as a
surprise then, that the corresponding PDF for Gaussian bonds in
Fig.~\ref{fig:PDF_p3Gauss} skews exactly in the opposite
direction. Although it is plotted for the cost fluctuations instead --
the energy fluctuations in the inset are purely normal --, this does
not explain the difference in the skewness, since cost and energy are
linearly related {[}as in Eq.~(\ref{eq:cost_energy}){]}. Lower cost
correlates with lower energy and vice versa. We speculate that the
skewness here has a rather trivial origin: As indicated by the very
low average ground state cost per spin, $\left\langle
c_{0}\right\rangle _{\infty}=0.0167(4)$ (about $1/4$ of that for
discrete bonds), below-average-cost instances may be hard to find due
to the proximity of entirely cost-free, {}``perfect'' solutions
(although we have not actually generated a single solvable instance
during our study).  On the other hand, higher-cost instances are
likely produced by the addition of isolated ``defects'' in the
quenched bonds that can not be gauge-transformed away as easily as in
SK~\citep{Boettcher05e} (or probably other problems), where many such
frustrated plaquettes overlap.  The smooth and well-converged appearance
of the PDF in Fig.~\ref{fig:PDF_p3Gauss} may provide some confidence
that the lack of scaling in the respective deviations $\sigma$ in
Fig.~\ref{fig:sigma_Bethe34_Gauss} is not due to a systematic error in
the sampling. It is the nature of heuristic optimization that any bias
at large system sizes (larger than can be verified by exact methods)
can never be exclude. But as the examples in the previous sections
demonstrate, strong competition between correction terms may be
equally well a cause in precluding asymptotic scaling.

We can address the puzzling skewness for the Gaussian case further by
comparing with previous results with Gaussian bonds on $p=2$-spin
glasses on Bethe lattices. While we have experienced this case already
for discrete bonds at $r=3$ above in
Sec.~\ref{sec:Spin-Glasses-on-Bethe} without any qualitative
difference in behavior compared to the SK model, its Gaussian version
presents a very odd pattern. In Ref.~\citep{Boettcher10a} we have
already remarked on the unusual finite-size corrections to the
thermodynamic average ground state cost or energy and the confusing
trend that already beset the cost deviations
$\sigma\left(c_{0}\right)$.  We therefore missed the even more
surprising evolution with $r$ and $N$ of the skewness in the
corresponding PDFs. Aside from some finite-size effects, all PDFs for
the energies are approximately normal distributions.  But the PDFs for
cost fluctuations shown in Fig.~\ref{fig:CostFlucGaussian} range from
those strongly skewed, comparable to Fig.~\ref{fig:PDF_p3Gauss}, for
small $r$ to those with only mild skewness at larger $r$. For all
cases, but most drastically for larger $r$, increasing system size $N$
symmetrizes the PDFs towards an apparent normal shape. Note that for
the SK limit $r\to\infty$, these distribution should approach a normal
form for the cost. While energy fluctuations should become the
non-trivial PDF in the SK limit, $r=15$ is apparently not large enough
for this effect to be discernible.

\section{Conclusions\label{sec:Conclusions}}

To elucidate the origin of the unusual ground state deviations
observed in the SK model, we have parsed over a range of different
spin glass and combinatorial models in an effort to provide context
for any typical or atypical properties. The clearest picture emerges
from graph-bipartitioning (GBP) on Bethe lattices, corresponding to a
ferromagnetic system held at zero magnetization. The scaling exponent
$\rho$ for each degree $r$ of the network possesses a convincing
extrapolation, but to values unlike those proposed for
SK~\citep{crisanti:92,Aspelmeier07,Parisi08,Parisi09,Rizzo09b,Rizzo09},
suggesting a very different origin for the fluctuations in GBP. At
best, they come closest to the value of $\rho=\frac{4}{5}$ recently
found for the $m$-vector spin glass~\citep{Aspelmeier10}. The
situation for the SK is less clear, apparently due to very strong
higher-order corrections in the finite-size behavior. Although far
from a scaling with the theoretically favored value of such a behavior
for the SK becomes plausible with very close $1/N$ corrections, a
scenario further supported by the Bethe lattice spin glass at degree
$r=3$ (with discrete bonds) exhibiting very similar behavior but
inherently closer to $\rho=\frac{5}{6}$.  As the example of the vector
model demonstrates, entirely distinct values are conceivable for any
of these models, but since in the limit $r\to\infty$ the Bethe
lattices approaches SK, such a distinction seems implausible. A look
deep into the tails for SK and the Bethe lattices, for which extensive
results have been generated, would suggest that the scaling of the
deviations $\sigma$ with the exponent $\rho$ that is dominated by
near-typical fluctuations may be disconnected from the extremely
atypical fluctuations deep in the tails, which is more consistent with
$\frac{3}{4}$-scaling for both models. But despite the massive amount
of data obtained here, any true asymptotic scaling for the tails may
still be elusive.

The situation might be simpler for models in which more than two spins
interact. For instance, we find for a $p=3$-spin glass model with
discrete bonds on a $r=4$-regular Bethe lattices that fluctuations
scale about with $1/N$, i.~e. $\rho=1$, which would already attain the
result for the REM ($p\to\infty$)~\citep{derrida:80} in
Ref.~\citep{Parisi09}. On the other hand, the same model with Gaussian
bonds demonstrates the fragility of the phenomenon:
energy-fluctuations get overwhelmed by trivial (normal) fluctuations
in the continuous bond weights, while the more pertinent
cost-fluctuations are skewed exactly in the opposite direction from
those from SK, or even those on the same graph with discrete bonds. A
same effect is found for $p=2$ on Bethe lattices at low $r$. In
summary, the large variation in behaviors of fluctuations not only
between models, but even within models for different bond
distributions, hints at the strong dependence on minute details of the
underlying graph geometry and variability in bond weights. Either can
impact whether and how frustrated plaquettes correlate to ease the
cost when an instance possesses more or less of those than the
average.

\section*{Acknowledgments\label{sec:Acknowledgements}}

I would like to thank F. Ricci-Tersenghi for numerous helpful correspondences
during the preparation of this manuscript, and I gratefully acknowledge
support from the Fulbright Kommission and from the U. S. National
Science Foundation through grant number DMR-0812204. 

\bibliographystyle{apsrev}
\bibliography{/Users/stb/Boettcher}
\end{document}